\documentclass[journal,twocolumns, 12px]{IEEEtran}

\ifCLASSINFOpdf
  
\else

\fi
\usepackage{float}
\usepackage{cellspace}
\usepackage{url}
\usepackage{cite}
\usepackage{balance}
\usepackage{array}
\usepackage{booktabs,adjustbox}
\usepackage{amssymb}
\usepackage{multirow}
\usepackage{tabularx}
\usepackage{tikz}
\usepackage{acronym}
\usepackage[none]{hyphenat}

\def\checkmark{\tikz\fill[scale=0.4](0,.35) -- (.25,0) -- (1,.7) -- (.25,.15) -- cycle;}
\usepackage{array,booktabs,arydshln,xcolor}

\usepackage[colorlinks = true,
bookmarks = false,
linkcolor=blue,
citecolor=blue,
filecolor=blue,
urlcolor=blue,
anchorcolor = blue]{hyperref}

\begin{document}

\title{Metaverse Survey \& Tutorial: Exploring Key Requirements, Technologies, Standards, Applications, Challenges, and Perspectives}


\author{\IEEEauthorblockN{Danda B. Rawat, Hassan El alami and Desta Haileselassie Hagos}\\
\IEEEauthorblockA{Howard University, Washington, DC, USA \\ \{danda.rawat, hassan.elalami, desta.hagos\}@howard.edu} 
}

\maketitle

\begin{abstract}
In this paper, we present a comprehensive survey of the metaverse, envisioned as a transformative dimension of next-generation Internet technologies. This study not only outlines the structural components of our survey but also makes a substantial scientific contribution by elucidating the foundational concepts underlying the emergence of the metaverse. We analyze its architecture by defining key characteristics and requirements, thereby illuminating the nascent reality set to revolutionize digital interactions. Our analysis emphasizes the importance of collaborative efforts in developing metaverse standards, thereby fostering a unified understanding among industry stakeholders, organizations, and regulatory bodies. We extend our scrutiny to critical technologies integral to the metaverse, including interactive experiences, communication technologies, ubiquitous computing, digital twins, artificial intelligence, and cybersecurity measures. For each technological domain, we rigorously assess current contributions, principal techniques, and representative use cases, providing a nuanced perspective on their potential impacts. Furthermore, we delve into the metaverse's diverse applications across education, healthcare, business, social interactions, industrial sectors, defense, and mission-critical operations, highlighting its extensive utility. Each application is thoroughly analyzed, demonstrating its value and addressing associated challenges. The survey concludes with an overview of persistent challenges and future directions, offering insights into essential considerations and strategies necessary to harness the full potential of the metaverse. Through this detailed investigation, our goal is to articulate the scientific contributions of this survey paper, transcending a mere structural overview to highlight the transformative implications of the metaverse.
\end{abstract}


\begin{table}[!t]
\footnotesize
\section*{List of Acronyms}
\begin{tabular}{ll}
AI & Artificial Intelligence \\
API & Application Programming Interface \\
AR & Augmented Reality \\
CAD & Computer-Aided Design \\
CGs & Community Groups \\
DRL & Deep Reinforcement Learning \\
De-Fi & Decentralized Finance\\
DEX & Decentralized Exchange \\
DL & Deep Learning \\
DoS & Denied of Service \\
DDoS & Distributed Denied of Service \\
DT & Digital Twin \\
FL & Federated Learning \\
GAI & Generative Artificial Intelligence \\
GPU & Graphics Processing Unit \\
ICP & Industrial Cyber Physical\\
IEC & International Electrotechnical Commission \\
IEEE SA & IEEE Standards Association \\
IP & Internet Protocol \\
ISO & International Organization for Standardization \\
ISP &Internet Service Provider\\
ITU & International Telecommunication Union \\
IT & Information Technology \\
KPI & Key Performance Indicators \\
LLM & Large Language Model\\
MAR & Mobile Augmented Reality \\
ML & Machine Learning \\
MOOC & Massive Open Online Course \\
MPEG & Moving Picture Experts Group \\
MR & Mixed Reality \\
MSF & Metaverse Standards Forum \\
NFT & Non-Fungible Tokens \\
NLP & Natural Language Processing \\
OTT & Over The Top\\
PoW & Proof-of-Work \\
QoE & Quality of Experience \\
QoL& Quality of Life \\
QoS & Quality of Service \\
RL & Reinforcement Learning \\
SDN & Software-Defined Networking \\
SEG & Standardization Evaluation Groups \\
SPoF & Single Point of Failure \\
TCP & Transmission Control Protocol \\
URLLC & Ultra-Reliable Low Latency Communications \\
UX & User Experience \\
VR & Virtual Reality \\
W3C & World Wide Web Consortium \\
XAI & eXplainable Artificial Intelligence \\
WHO&World Health Organisation\\
XR & eXtended Reality \\
xURLLC & eXtreme Ultra-Reliable Low Latency Communications \\
\end{tabular}
\end{table}

\section{Introduction}

\begin{figure*}[htbp]
    \centering
        \includegraphics[width=\textwidth]{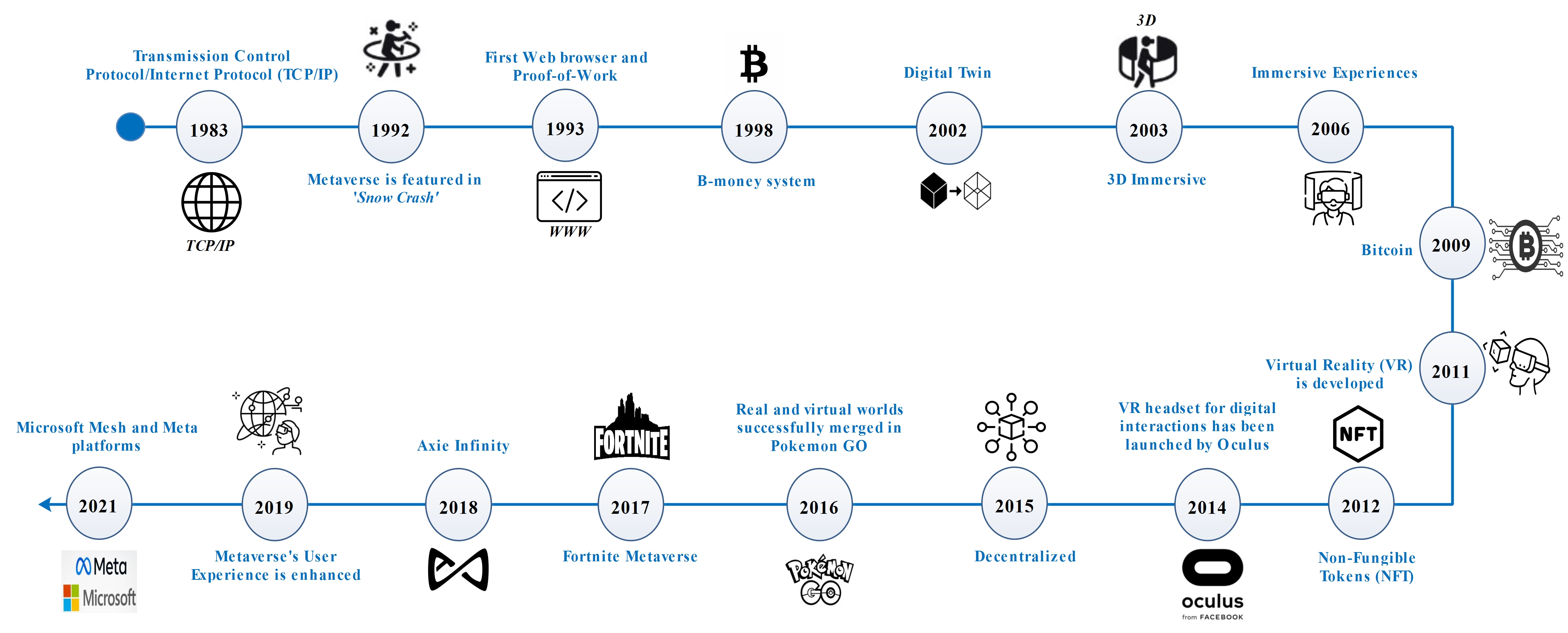}
    \caption{A timeline of Metaverse technologies.}
    \label{Fig: History}
    \vskip-0.45cm
\end{figure*}

\IEEEPARstart{T}{he} metaverse can be defined as a large-scale, persistent, and immersive network of 3D virtual environments facilitated by a convergence of technologies like Augmented Reality (AR), Mixed Reality (MR), Virtual Reality (VR), or eXtended Reality (XR), Artificial Intelligence (AI), and blockchain. These environments allow users, represented by avatars, to interact with each other and their surroundings in real-time through various modalities, simulating real-world experiences and enabling entirely new ones\cite{rawat2023metaverse,ieee2048standards}. It is a digital universe that exists parallel to our physical world, often accessed through the internet, where users can interact with a computer-generated environment and other users in real-time\cite{rawat2023metaverse}. The metaverse is characterized by its immersive nature, allowing individuals to engage in various activities, including work, socializing, entertainment, and business, using virtual avatars or representations. The key idea is the creation of a persistent, interconnected, and interactive digital space that goes beyond traditional online experiences. In addition, the metaverse is envisioned as a virtual world that allows seamless integration between physical and virtual experiences, offering users the ability to live, work, and play in a digital environment\cite{mystakidis2022metaverse}. Despite the ongoing development of the metaverse, it has already shown its potential to transform the way humans interact with technology and each other\cite{sparkes2021metaverse}. According to a survey of experts in\cite{pewresearchcenter2023}, 54\% expect the metaverse to be a part of daily life for half a billion or more individuals by 2040. Although several platforms are working to create immersive virtual worlds and experiences\cite{de2023physical}, including notable ones such as Second Life~\cite{secondlife2023}, Fortnite\cite{fortnite2023}, Sandbox\cite{sandbox2023}, Decentraland\cite{decentraland2023}, Roblox\cite{roblox2023}, VRChat\cite{vrchat2023}, and Somnium Space\cite{somniumspace2023}. Additionally, due to the COVID-19 pandemic, the metaverse has experienced accelerated adoption and development\cite{lee2021all}. As social distancing measures and stay-at-home orders became prevalent, the need for virtual social interactions significantly increased. In response, the metaverse became a platform for virtual events and gatherings during the pandemic, demonstrating its potential for social interaction and entertainment\cite{onggirawan2023systematic}. The pandemic also highlighted the concepts of decentralization and virtual ownership\cite{koohang2023shaping}, as virtual experiences and assets became increasingly valuable when physical events and experiences were limited\cite{tucker2023digital}.\\

As shown in Figure~\ref{Fig: History}, the Metaverse timeline began in 1983 when the Internet was introduced, and TCP/IP was released, enabling computer communication\cite{huynh2023artificial}. The term~\textit{metaverse} was coined by Neal Stephenson in 1992 with the publication of his book~\emph{Snow Crash}, depicting a dystopian future where affluent individuals escape into a connected 3D alternate world\cite{mcwilliam2023metaverse}. In 1993, the first web browser, Mesh, was released, and PoW was invented\cite{internet2023}. PoW is used to verify cryptocurrency transactions by expending energy\cite{farell2015analysis}. The B-money system, introduced in 1998, was the first to publicly broadcast transactions while maintaining anonymity\cite{bonneau2015sok}. In 2002, the concept of the DT was introduced as a means to design, test, manufacture, and support products in a virtual world\cite{fuller2020digital}. Second Life, a 3D immersive environment for social networking and interaction, was launched in 2003\cite{time2023}. Second Life can be considered an early example of a large-scale metaverse. Roblox paltform is introduced in 2006, provided immersive experiences, persistent avatars, and a digital economic component\cite{danielsisson2023}. In 2009, Bitcoin became the world's first cryptocurrency, and its blockchain code was released\cite{julie2023}. Bitcoin's blockchain code is open source, allowing anyone to review and modify it, fostering trust and transparency\cite{karame2016security}. In 2011, Ernest Cline's science fiction novel, Ready Player One, served as inspiration for the development of VR\cite{john2023}. The NFT market was introduced in 2012\cite{wang2021non}, and in 2014, Facebook acquired Oculus, a VR headset aimed at revolutionizing digital social interactions. Decentraland, the metaverse with its own economy and Ethereum blockchain, was introduced in 2015\cite{dannen2017introducing, guidi2022social}. Pokemon GO successfully merged the real and virtual worlds in 2016\cite{paavilainen2017pokemon}, and Epic Games released Fortnite in 2017, resulting in greater connectivity across its games\cite{games2017fortnite}. The virtual world~\textit{Axie Infinity} is launched in 2018, is a blockchain-based war game featuring complex player-owned economies and online rewards\cite{delic2022profiling}. Significant improvements in User Experience (UX) were achieved in 2019 due to advancements in communication technologies and ML\cite{tang2022roadmap}. Furthermore, the reduction in communication overhead for VR devices played a vital role in enhancing the overall UX. In 2021, Microsoft achieved a significant milestone with the release of Microsoft Mesh, an MR platform designed for team collaboration\cite{johnroach2023}. Concurrently, Facebook announced its rebranding as Meta\cite{chang20226g} and initiated research efforts focused on the development of the metaverse. Since these events unfolded, the emergence of the metaverse has ignited a new wave of development. As we enter 2023, the metaverse continues to evolve and progress, fueled by advancements in VR and AR technologies, as well as blockchain and decentralized systems. This emerging concept holds immense potential to revolutionize our interactions with technology and reshape the way we engage with one another.

\subsection{Motivation}
\begin{figure}[!b]
         \includegraphics[width=\linewidth]{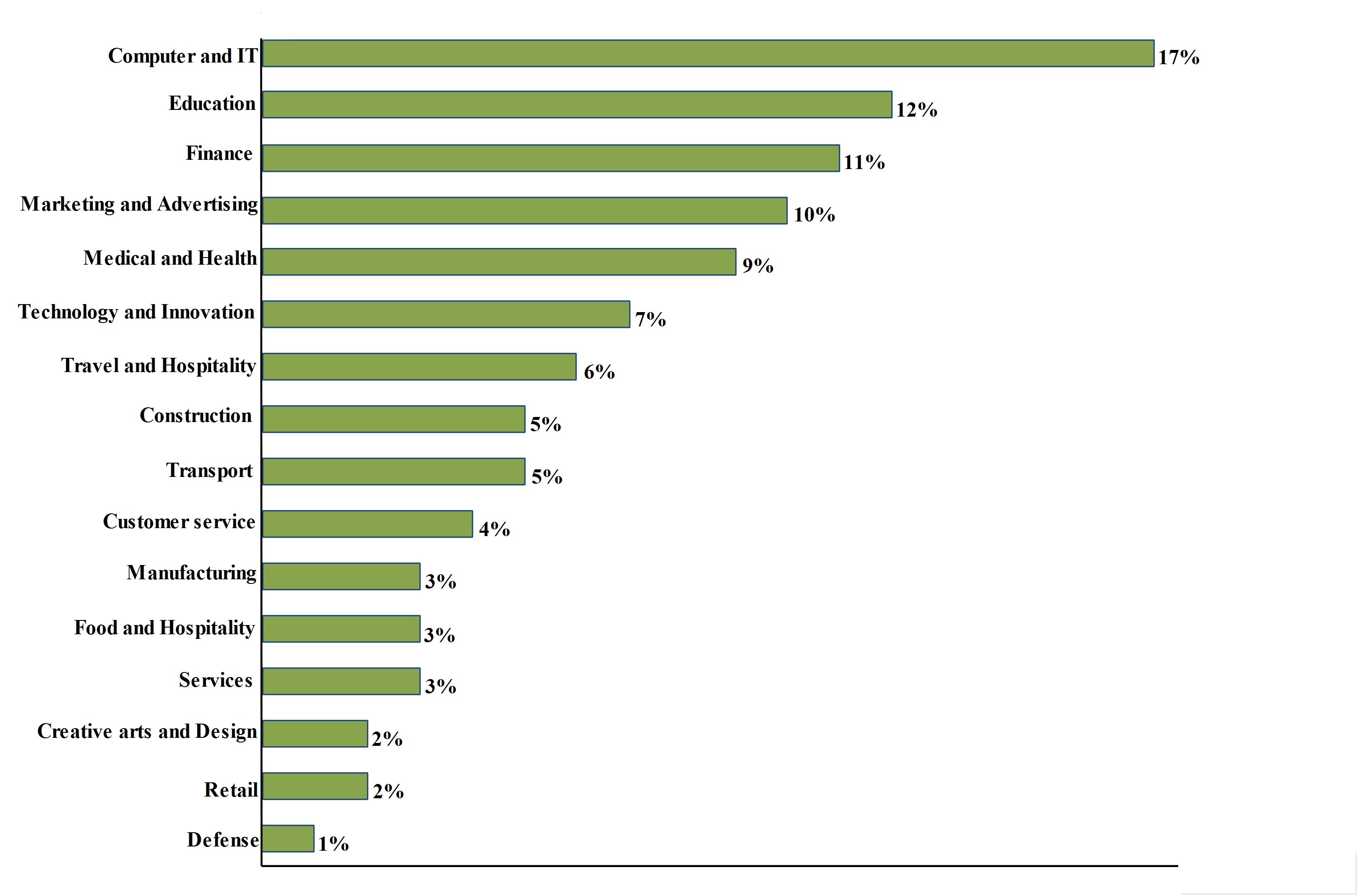}
    \caption{Overview of key global business sectors actively investing in the metaverse as of March 2022.}
    \label{Fig:Statistics}
    \vskip-0.45cm
\end{figure}

The metaverse, a virtual shared space enabling users to engage with computer-generated environments and each other in immersive ways, is poised to reshape numerous sectors. Projections indicate substantial growth in the metaverse market, with a forecasted increase from \$61.8 billion in 2022 to \$426.9 billion by 2027, boasting a compound annual growth rate of 47.2\%, according to a report by~\emph{Markets and Markets}\cite{market2023}. This growth is driven by the increasing adoption of virtual platforms and a growing interest in virtual events and experiences. A survey conducted by \emph{Statista} in March 2022\cite{clement2023} (as depicted in Figure~\ref{Fig:Statistics}) further illuminates this trend, revealing that businesses in the computer and information technology (IT) industries lead in metaverse investments, dedicating 17\% of resources to this transformative technology. Following closely is the education sector, allocating 12\% of its investments.
As industries embrace the metaverse, specific requirements tailored to each sector's unique demands emerge. For instance, the gaming industry prioritizes high-quality graphics and seamless integration with gaming engines to deliver a realistic and immersive experience\cite{njoku2023prospects}. In contrast, the education sector emphasizes interactive learning environments and progress-tracking features~\cite{chen2022exploring}. Similarly, in medicine, the metaverse, accessible through IoT and AR/VR glasses, is envisioned as a tool for advanced healthcare applications\cite{yang2022expert}. The metaverse's architecture plays a pivotal role in meeting these diverse industry needs, demanding scalability and security to support thousands of simultaneous users. Flexibility is crucial for adapting to the varied requirements of different sectors. Ensuring accessibility for all, even those with low-end devices and limited internet access poses a formidable challenge\cite{bhattacharya2023towards}. Additionally, safeguarding user safety, trust, privacy, and security is paramount, given the sensitive nature of personal information and financial transactions within the metaverse\cite{qamar2023systematic}. The motivation behind this manuscript is twofold: First, it aims to offer valuable insights into the complex nature of the metaverse and its impact across various sectors. Second, our survey explores the key factors driving the integration of the metaverse into diverse domains. To achieve widespread success, establishing key standards for the metaverse is essential. These standards will provide a framework for consistency, aiding interoperability with existing technologies and systems. In this survey, we aim to offer a comprehensive overview of the current state and potential impact of the metaverse across various industries. We will delve into the challenges and future research directions associated with its development and highlight essential standards crucial to its success. This manuscript seeks to guide readers through the multifaceted landscape of the metaverse, elucidating its potential and the critical factors underpinning its effective integration into diverse sectors.

\subsection{Surveys on the Metaverse in the Literature}
There has been considerable research interest in the metaverse topic. To date, several surveys and tutorials have been published on various aspects of the metaverse. For instance, the topic of privacy is discussed in\cite{leenes2008privacy} as well as how Second Life regulates privacy. Second Life's current governance mode is analyzed to identify its shortcomings and point out future changes that can be made to it. A metaverse perspective on retailing is presented in\cite{bourlakis2009retail}, along with a study of how metaverses influence the evolution of retailing. Three different, yet intertwined spaces could be used by retailers simultaneously. Traditional retailers, e-retailers, and metaverse retailers are highlighted as well as the key challenges and opportunities. Using 3D virtual spaces as the backdrop, the status work of\cite{dionisio20133d} described what it takes to move from a collection of independent virtual worlds to an integrated network of 3D virtual worlds. This is a compelling alternative to human sociocultural interaction. Additionally, four central features of a viable metaverse, namely realism, ubiquity, interoperability, and scalability are discussed. In\cite{muhanna2015virtual}, the authors conducted a comprehensive literature review focusing on VR and the cave-automated virtual environment (CAVE). The study begins by delivering a concise overview of the historical development of virtual reality. Additionally, the paper highlights essential components of VR systems and introduces a proposed taxonomy that classifies these systems based on the employed technologies and the level of mental immersion they offer. Furthermore, the authors delve into a thorough examination of the CAVE, elucidating its distinctive features, applications, and particularly, the various interaction styles inherent to this environment. In\cite{correia2016computer}, the author presents an overview study that addresses several challenges and opportunities for research on 3D virtual worlds in learning within an open, meta-theoretic framework. Through this research, healthcare contexts, K-12 research, entertainment, security, cultural influence, economic concerns, mobile and multiplayer technologies, open source platforms, gesture recognition, social behaviors, and physical interaction were identified.  This overview study helps researchers better understand the field and previous studies, helping them classify research, identify gaps in the literature, and shape future trends. Considering previous research on privacy and gaming analytics in the metaverse, the authors of\cite{falchuk2018social} propose a survey focusing on technologies that would enable VR participants to maintain a greater level of privacy whilst immersed in social VR. In this context, the term social metaverse describes the above types of virtual realities in which a central objective is socialization and interaction with other avatars, including both players and non-player characters. As presented in\cite{duan2021metaverse}, the authors proposed a survey related to the metaverse for social good.  Using a macro perspective, they proposed a three-layer metaverse architecture consisting of infrastructure, interaction, and ecosystem. Moreover, they developed a timeline with a detailed table of specific attributes and traveled toward both a historical and a novel metaverse. Finally, they illustrated a blockchain-driven metaverse prototype for a university campus and discussed its design and insights. The purpose of the survey presented in\cite{ning2021survey} is to introduce the development status of the metaverse from a technical perspective, covering five aspects, including network infrastructure, management technology, basic common technology, VR object connection, and VR convergence. The proposed survey provides an overview of the technical framework of the metaverse. Moreover, it discussed the first application areas of the metaverse, as well as some of the challenges and problems it may encounter. According to\cite{park2022metaverse}, the authors divided the concepts and essential techniques necessary to realize the metaverse into three components (i.e., hardware, software, and contents) and three approaches (i.e., user interaction, implementation, and application) to conduct a comprehensive analysis rather than utilizing a marketing approach or hardware approach. Additionally, they summarized social influences, constraints, and open challenges for the implementation of the immersive metaverse. In\cite{yang2022fusing}, the authors delved into the metaverse, analyzing the integration of blockchain and AI. They explored the current research landscape encompassing metaverse components, digital currencies, AI applications in virtual environments, and technologies empowered by blockchain. Advancing research and interdisciplinary exploration of the synergy between AI and blockchain for the metaverse necessitates collaborative efforts from academia and industries alike. The authors of\cite{wang2022survey} provide an in-depth analysis of the fundamentals, security, and privacy of the metaverse. Specifically, they investigated the key characteristics of a distributed metaverse architecture with ternary-world interactions. In addition to discussing security and privacy threats, they discussed the critical challenges faced by metaverse systems, as well as reviewed current countermeasures. The review in\cite{al2022review} provided a comprehensive survey of metaverse development, offering a chronological history and highlighting recent technological advances. It covered definitions, properties, architecture, and applications of the metaverse. Notably, the proposed work introduced a framework to address ongoing issues, serving as a guide for future research. The review discusses the challenges faced by researchers, explores pertinent issues, and outlines future trends, making a valuable contribution to the understanding and advancement of the metaverse. According to\cite{venugopalrealm23}, the authors proposed a hypothetical meta-stack framework for gaining a deeper understanding of the various elements within the realm of the metaverse. They provided an in-depth analysis of the most recent developments in the metaverse based on cutting-edge technologies, security vulnerabilities, and prevention measures specifically related to the metaverse as well as research challenges associated with it. The work in\cite{abilkaiyrkyzy2023metaverse} has a two-fold objective. Firstly, it provides an impartial analysis of essential requirements for metaverse platforms, covering criteria such as interoperability, immersiveness, scalability, and security. Secondly, the authors critically evaluate existing metaverse platforms, identifying limitations that must be addressed for fair and trustworthy experiences. The paper emphasizes the need for ongoing research and development, highlighting areas like decentralization, improved security, privacy measures, and integration of emerging technologies such as blockchain and AI as crucial for building a resilient and secure metaverse. The study in\cite{uddin2023unveiling}, highlighted its thorough exploration of fundamental metaverse concepts, intricate details, and advanced technologies. The statement also emphasizes the study's showcase of innovative applications and its valuable insights into challenging research avenues, specifically underlining the significance of establishing an efficient, private, and secure metaverse realm. In\cite{raad2023metaverse}, the authors underscored both existing and anticipated metaverse applications, addressing key concerns and challenges encountered by stakeholders. Additionally, they analyzed the strengths, weaknesses, opportunities, and threats associated with metaverse technology. Lastly, the authors outlined future directions and emphasized crucial recommendations for the development of metaverse systems. In the survey discussed in\cite{khan2024metaverse}, the authors introduced a general architecture for the metaverse in wireless systems. They explored key applications, design trends, and essential enablers of this architecture. Finally, they addressed several open challenges and proposed potential solutions. In summary, Table~\ref{Tab:Com} highlights the distinctions among the aforementioned surveys on the metaverse. Within this table, we conduct a comparative analysis of extant survey papers, focusing on essential aspects such as key requirements, key technologies, applications, challenges, and research directions.
\begin{table*}[htbp]
  \centering
  \caption{Comparison of our Survey with Existing Related Research}
  \label{Tab:Com}
  \begin{tabular}{c|c|c|c|c|c|c}
  \cline{1-7}
    \hline
    \multirow{2}{*}{\textbf{Year}} & \multirow{2}{*}{\textbf{Reference}} & \multicolumn{5}{c}{\textbf{Content Coverage}} \\
    \cline{3-7}
    & & \textbf{Key Requirements} & \textbf{Key Technologies} & \textbf{Applications} & \textbf{Challenges} & \textbf{Future Research Directions} \\
    \hline
    2008 &~\cite{leenes2008privacy}& \checkmark & $\times$ & $\times$ & $\times$ & $\times$ \\
    \hline
    2009 &~\cite{bourlakis2009retail} & $\times$ & $\times$ & \checkmark & $\times$ & \checkmark \\
    \hline
    2013 &~\cite{dionisio20133d} & \checkmark & $\times$ & $\times$ & $\times$ & \checkmark \\
    \hline
    2015 &~\cite{muhanna2015virtual} & \checkmark & $\times$ & $\times$ & \checkmark & \checkmark \\
    \hline
    2016 &~\cite{correia2016computer} & $\times$ & $\times$ & \checkmark & \checkmark & \checkmark \\
    \hline
    2018 &~\cite{falchuk2018social} & \checkmark & $\times$ & \checkmark & $\times$ & $\times$ \\
    \hline
    2021 &~\cite{duan2021metaverse} & $\times$ & \checkmark & \checkmark & $\times$ & $\times$ \\
    \hline
    2021 &~\cite{ning2021survey} & $\times$ & \checkmark & \checkmark & \checkmark & $\times$ \\
    \hline
    2022 &~\cite{park2022metaverse} & $\times$ & $\times$ & \checkmark & \checkmark & $\times$ \\
    \hline
    2022 &~\cite{yang2022fusing} & $\times$ & \checkmark & $\times$ & \checkmark & \checkmark \\
    \hline
    2022 &~\cite{wang2022survey} & $\times$ & \checkmark & \checkmark & $\times$ & \checkmark \\
     \hline
    2022 &~\cite{al2022review} & $\times$ & $\times$ & \checkmark & \checkmark & \checkmark \\
     \hline
    2023 &~\cite{venugopalrealm23} & $\times$ & \checkmark & $\times$ & \checkmark & $\times$ \\
     \hline
    2023 &~\cite{abilkaiyrkyzy2023metaverse} & \checkmark & \checkmark & $\times$ & \checkmark & $\times$ \\
     \hline
    2023 &~\cite{uddin2023unveiling} & $\times$ & \checkmark & \checkmark & \checkmark & $\times$  \\
     \hline
    2023 &~\cite{raad2023metaverse} & $\times$ & $\times$ & \checkmark & \checkmark & \checkmark \\
     \hline
    2024 &~\cite{khan2024metaverse} & $\times$  & \checkmark & $\times$ & $\times$  & \checkmark\\
     \hline
    \textbf{2024} & \textbf{Our survey} &\checkmark & \checkmark &\checkmark & \checkmark & \checkmark \\
     \hline
    \cline{1-7}
  \end{tabular}
\end{table*}

\begin{figure}[htbp]
    \centering
        \includegraphics[width=.50\textwidth]{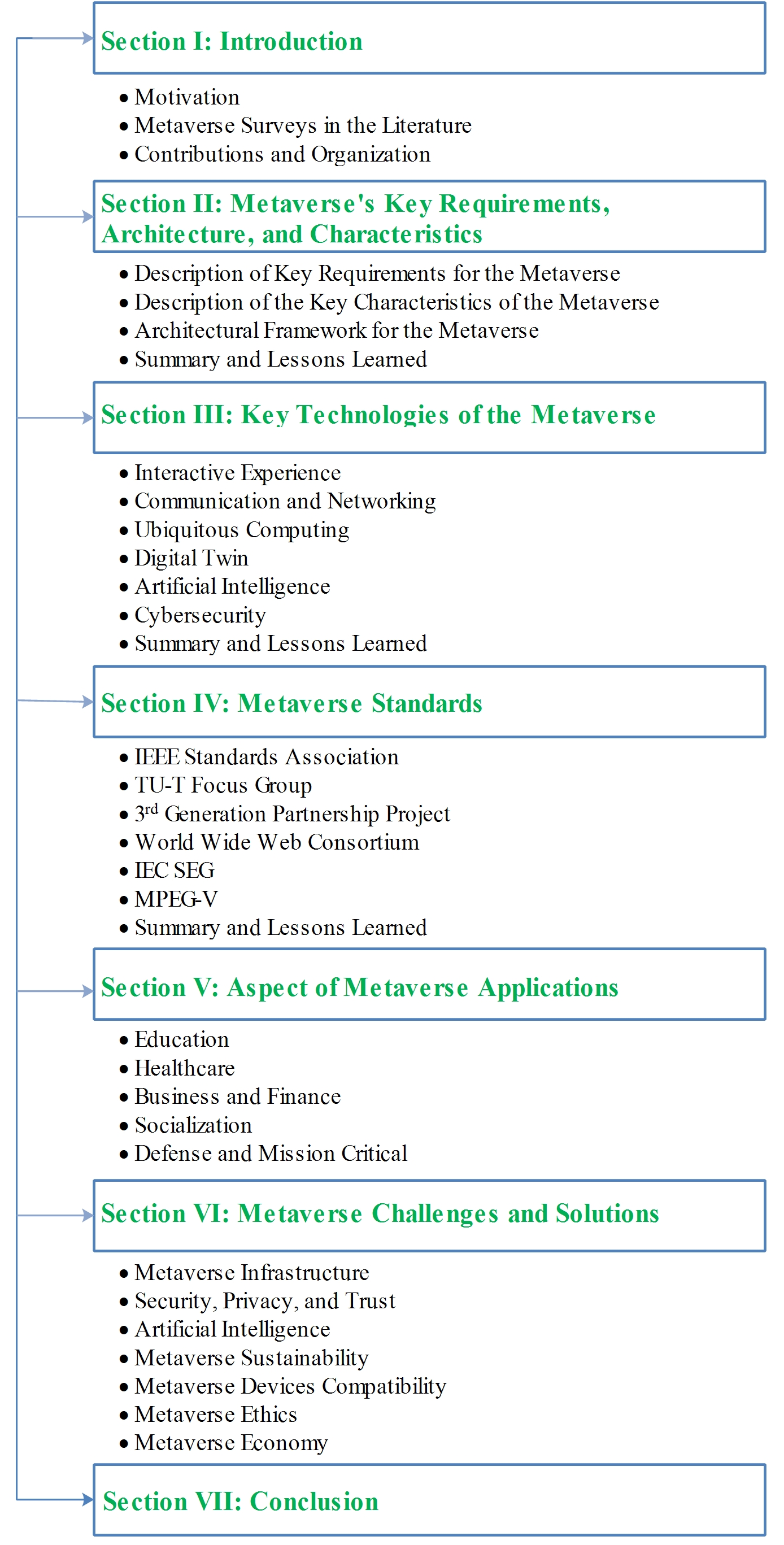}
    \caption{Organization structure of the survey.}
    \label{Fig:Outline}
    \vskip-0.45cm
\end{figure}

\subsection{Organization}
This paper explores various aspects of the metaverse, a virtual world that is replacing the digital realm of the Internet. It covers topics such as requirements, architecture, standards, challenges, and solutions related to the metaverse. The survey provides an in-depth analysis of the necessary architectural framework, requirements, and standards that form the foundation for the development and implementation of the metaverse. Additionally, it presents the current status of the metaverse while discussing the challenges it faces, along with potential solutions and perspectives. In summary, the key contributions of this survey can be outlined as follows:
\begin{itemize}
\item We provide a comprehensive overview of the fundamental aspects of the metaverse, encompassing its key requirements, typical architecture, and essential characteristics.
\item We delve into the essential technologies necessary for constructing the metaverse, encompassing interactive experiences, communication and networking, ubiquitous computing, digital twins, AI, as well as cybersecurity.
\item We provide an outline of the organizations and forums involved in developing standards and guidelines for the metaverse. These include prominent entities such as IEEE SA, ITU, 3GPP, W3C, IEC SEG, and MPEG-V.
  \item We explore the diverse range of applications in various domains where the metaverse can be utilized. These domains include education, healthcare, business, finance, socialization, defense, and mission-critical operations. 
  \item To enhance efficiency within the metaverse realm, we present an overview of the challenges that need to be addressed, potential solutions to overcome these challenges, and recommendations that should be pursued.
\end{itemize}
The survey is structured as follows, following the outline depicted in Figure~\ref{Fig:Outline}. Section II offers an overview of the metaverse, covering key requirements, typical architecture, and the characteristics defining this virtual environment. In Section III, we delve into a discussion of the pivotal technologies essential to the metaverse. Section IV covers the standards and guidelines associated with the metaverse. The survey further advances to Section V, where a comprehensive review of metaverse applications is presented. This section addresses the critical challenges these applications face and provides recommendations for future works. Section VI concentrates on exploring the ongoing challenges, proposing potential solutions, and contemplating future perspectives within the metaverse. Finally, Section VII concludes the survey, summarizing the findings and key takeaways.

\section{Metaverse's Key Requirements, Architecture, and Characteristics}
\label{metaverse_key_characterstics}

In this section, we provide an overview of the metaverse's requirements, typical architecture, and characteristics.
\subsection{Description of Key Requirements for the Metaverse}
\begin{figure}[htbp]
 \includegraphics[width=\linewidth]{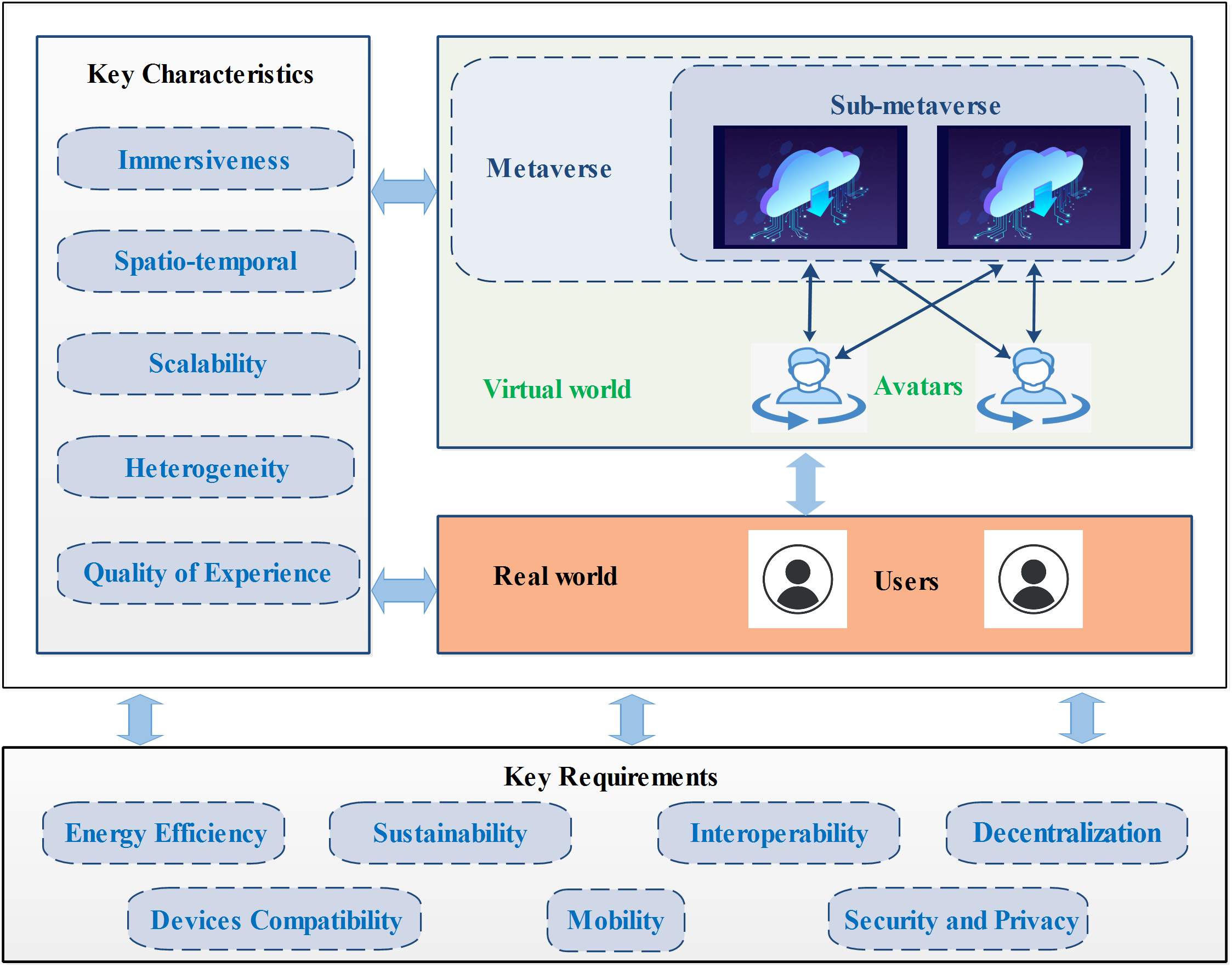}
 \caption{The metaverse's key requirements and characteristics.}
 \label{Fig:Char}
\end{figure}

Metaverse development is guided by fundamental requirements that are shown in Figure~\ref{Fig:Char}. Key facets include energy efficiency, ensuring responsible resource usage, sustainability to support long-term viability, interoperability for seamless interactions, decentralization to enhance security, and prioritized security and privacy measures. The metaverse reflects these requirements collectively, demonstrating its commitment to ethical considerations while also being technologically advanced and inclusive.

\vspace{1.0ex}
\noindent \textbf{Energy Efficiency}. Energy efficiency stands as a foundational and imperative requirement for the metaverse, reflecting a dedication to responsible and sustainable digital practices. As the metaverse continues to expand and integrate into the fabric of our daily experiences, the demand for computational resources and data processing power rises significantly\cite{zhang2023meta}. In this context, optimizing energy consumption becomes not just a preference but an absolute necessity\cite{majid2023optimizing}. The intricate web of servers, data centers, and infrastructure supporting the metaverse requires a precise focus on energy-efficient technologies and operational practices\cite{zhang2023towards}. By adopting measures such as efficient hardware design, optimized algorithms, and environmentally aware data center management, the metaverse endeavors to minimize its overall carbon footprint\cite{liu2023metaverse}. This dedication extends beyond the virtual realm, recognizing the environmental consequences of digital technologies. Highlighting energy efficiency aligns the metaverse with broader sustainability goals, ensuring that its growth is ecologically responsible and contributes positively to global efforts to mitigate climate change\cite{palak2023metaverse}. The requirement for energy efficiency is not solely a technical consideration but a vow to shape a sustainable digital future where the metaverse coexists harmoniously with environmental well-being.

\vspace{1.0ex}
\noindent \textbf{Sustainability}. The sustainability of the metaverse relies on a self-contained economic loop and a consistent value system, coupled with a high degree of independence\cite{jauhiainen2023metaverse}. Ideally, it should embrace openness, continually engaging users in digital content production and fostering open innovation. To ensure resilience, the metaverse should adopt a decentralized architecture, mitigating Single Point of Failure (SPoF) risks and preventing control by a select few powerful entities. In essence, sustainability in this context refers to the ability to maintain and enhance the health and well-being of the virtual world while minimizing its environmental impact. This consideration encompasses social, economic, and environmental aspects\cite{allam2022metaverse}.
Firstly, social sustainability hinges on an inclusive and diverse metaverse community, ensuring accessibility for individuals irrespective of their background, abilities, or income levels. Establishing community policies and standards that promote positive behavior, discourage harassment and bullying, and offer mechanisms for dispute resolution is crucial\cite{arpaci2023investigating}. Secondly, sustainable economic development requires the establishment of a fair, stable, and sustainable virtual economy. Furthermore, it involves advocating for a circular economy approach, which minimizes waste by reusing, repurposing, or recycling materials, and ensures the equitable distribution of resources\cite{huawei2023economic}. Lastly, environmental sustainability entails reducing the metaverse's carbon footprint and limiting its impact on the physical world. Employing sustainable energy solutions to power virtual infrastructure and designing efficient data centers is imperative. Additionally, energy-efficient virtual spaces should be constructed to reduce power consumption by user devices and servers, thereby minimizing waste\cite{zhang2023towards}.

\vspace{1.0ex}
\noindent \textbf{Interoperability}. Metaverse interoperability is defined as the ability of users to move seamlessly across virtual worlds (i.e., sub-metaverses) without interruption of the immersive experience\cite{lee2021all} as well as the ability to exchange digital assets for rendering and reconstructing virtual worlds across multiple platforms\cite{wang2022survey}. Further, the lack of common standards and protocols is one of the greatest challenges to interoperability in the metaverse. Different platforms and virtual worlds may use different programming languages, data formats, and communication protocols, making it difficult to communicate between them. To address this challenge, many companies and organizations are developing common standards and protocols\cite{ghirmai2022self}. For instance, the Open metaverse Interoperability Group is developing a set of standards for interoperability between virtual worlds, including protocols for communication, identity, and asset exchange\cite{li2023metaopera}. Interoperability can also be achieved by creating cross-platform marketplaces where users can buy and sell digital assets that can be used across a variety of virtual worlds and platforms. Users can use their digital assets in different virtual worlds without having to worry about interoperability issues, thereby creating a more seamless experience\cite{chen2022cross}.

\vspace{1.0ex}
\noindent \textbf{Decentralization}. Decentralization emerges as a foundational and transformative requirement for the metaverse, embodying a profound dedication to user empowerment, security, and a resilient distributed framework\cite{karaarslan2023metaverse}. Within the metaverse ecosystem, the decentralization mandate signifies a departure from centralized control, distributing authority and data across a network of nodes. This deliberate design choice not only mitigates the risks associated with a single point of failure but also bolsters the metaverse's resistance to malicious activities and unauthorized access. Beyond its technical implications, decentralization propels the metaverse towards embodying democratic ideals, granting users unprecedented autonomy over their digital identities and interactions\cite{hashash2023towards}. By prioritizing user ownership and control of personal data, this requirement fosters a more inclusive and equitable virtual environment. Additionally, decentralization acts as a safeguard, enhancing the metaverse's resilience against censorship and ensuring a robust, secure, and diverse digital landscape. The emphasis on decentralization underscores a devoted dedication to a metaverse where user trust, security, and governance are distributed, reflecting a vision of a more democratic, secure, and user-centric virtual realm\cite{goldberg2023metaverse}.

\vspace{1.0ex}
\noindent \textbf{Devices Compatibility}. The metaverse's efficacy hinges on its ability to harmonize with the diverse array of devices that users employ in their digital interactions\cite{yaqoob2023metaverse}. This requirement mandates a comprehensive approach to accommodate various technologies such as~\textit{Microsoft HoloLens} and~\textit{Google Glass}. To illustrate, a user seamlessly transitioning from an immersive VR experience on a headset during a virtual meeting to a quick check-in on their smartphone while on the go should encounter consistent and optimized interactions. Developers and platform architects must adhere to standardized practices, ensuring that metaverse applications are designed with adaptability in mind, catering to the unique specifications and capabilities of each device\cite{mystakidis2022metaverse}. By establishing robust device compatibility, the metaverse not only embraces the technological diversity of its user base but also paves the way for a fluid and interconnected digital environment where users can effortlessly navigate between devices, fostering a truly integrated metaverse experience.

\vspace{1.0ex}
\noindent \textbf{Mobility}. In the metaverse, the imperative for mobility stands as a foundational requirement, prominently illustrated by the necessity for instinctive and responsive avatar movements\cite{huang2022mobility}. User expectations center around a cohesive and seamless exploration of virtual environments, emphasizing the evolving nature of mobility beyond conventional avatar navigation. This paradigm extends to the contemplation of avant-garde applications, notably the integration of autonomous systems within the metaverse\cite{deveci2022personal}. Envision a scenario where users traverse vast virtual landscapes through their avatars, augmented by the option to incorporate autonomous vehicles for heightened efficiency and immersive exploration. Such intelligent vehicular entities possess the potential to elevate the overall user experience, providing users with the capacity for fluid, seamless movements and dynamic engagement within the virtual realm. For example, users within the metaverse navigate virtual terrains, approach diverse spatial constructs, and engage with fellow users through authentically responsive avatar movements. Alternatively, users may opt to employ autonomous vehicles to augment their exploration, particularly in scenarios of collaborative work sessions or social gatherings within virtual spaces. The absence of a robust mobility infrastructure, encompassing both avatar movement and the potential integration of autonomous vehicles, may introduce perceptible disjointedness, impeding the natural flow of interactions and diminishing the metaverse's overall allure\cite{wang2023metamobility}. Therefore, the viability and prosperity of the metaverse hinge decisively upon fulfilling this intricate mobility requirement, assuring users of the capacity for genuine, dynamic engagement within the digital milieu and fostering an enduring sense of presence and connection.

\vspace{1.0ex}
\noindent \textbf{Security and Privacy}. The security and privacy requirements for the metaverse stand as a paramount pillar, reflecting an unwavering commitment to safeguarding user information and fostering a secure digital environment\cite{kang2023security}. In the ever-expanding landscape of the metaverse, where user interactions and transactions are increasingly prevalent, ensuring robust security measures and privacy protections is imperative\cite{oh2023secure}. This requirement encompasses the implementation of advanced encryption techniques, secure authentication protocols, and proactive measures against cyber threats. By prioritizing user privacy, the metaverse aims to establish a foundation of trust, assuring users that their personal information is handled with the utmost care and confidentiality\cite{kim2023secure}. Furthermore, the security and privacy requirements align with legal and ethical standards, emphasizing compliance with data protection regulations to instill confidence among users. This resolute focus on security and privacy emphasizes the metaverse's committed efforts to offer a secure and dependable digital environment. In this space, users can confidently engage, communicate, and transact, assured that their data is safeguarded and their privacy is valued.

\subsection{Description of the Key Characteristics of the Metaverse}
During the initial development of Web 1.0, Internet users were merely content consumers, where the content was provided by websites\cite{hidayanto2022designing}. A key characteristic of Web 2.0 (i.e., the mobile Internet) is that users have become both content producers and content consumers, and websites have become platforms for delivering services. The most common platforms are~\textit{LinkedIn} and~\textit{Facebook}\cite{jamil2022role}. Furthermore, Web 3.0, also called the semantic web, is a web that is decentralized, open to everyone, and built on the foundation of blockchain technology. It is characterized by a paradigm known as the metaverse\cite{hackl2022navigating}. Figure~\ref{Fig:Char} illustrates the ability of digital avatars to shuttle seamlessly across various virtual worlds (i.e., sub-metaverses) to experience a digital life, as well as create digital objects and participate in economic transactions, all supported by physical infrastructures and metaverse engines. Particularly, the metaverse exhibits specific characteristics from the following aspects\cite{wang2022survey, rawat2023metaverse}.

\vspace{1.0ex}

\noindent \textbf{Immersiveness}. A virtual environment that has immersiveness means that it is sufficiently realistic to allow users to feel psychologically and emotionally immersed\cite{han2010user}. A similar concept is known as immersive realism\cite{dionisio20133d}. According to the realism perspective, human beings interact with their environment through their senses and their bodily functions. Sensory perceptions, such as sight, sound, touch, temperature, and balance, and expressions, such as gestures, contribute to immersive realism. Further, this immersive experience is achieved through the combination of technology, user interfaces, and design elements that combine to create an immersive experience that is seamless and believable. The use of VR technology is one of the key factors contributing to the immersive nature of the metaverse. As a result of VR technology, users can experience the virtual environment in a fully immersive manner. A VR headset allows users to immerse themselves fully in the metaverse and experience the virtual world as if they were physically present\cite{dincelli2022immersive}. Another factor contributing to the immersiveness of the metaverse is the use of realistic 3D graphics and sound effects. The use of high-quality graphics and sound effects creates a believable and engaging virtual environment that feels like a real-world experience\cite{upadhyay2022metaverse}.

\vspace{1.0ex}
\noindent \textbf{Spatiotemporal}. It is important to remember that there is a finite amount of space in the real world and that time is irreversible. A hyper-spatiotemporal continuum refers to the breaking of limitations of time and space in the metaverse, which is a virtual parallel to our real world\cite{ning2021survey}. This allows the user to freely navigate across different worlds with different spatiotemporal dimensions and experience an alternate life with seamless scene transformation. In addition, spatiotemporal aspects are crucial for creating a believable and immersive experience for its users. This includes the virtual representation of physical space, as well as the passage of time within the virtual environment\cite{yang2023human}. For example, in a metaverse game, the user should be able to navigate through the virtual space, and the objects and characters within that space should respond to the user's movements in real-time. Similarly, the time of day and weather within the virtual world should reflect the corresponding time and weather in the real world. Furthermore, the spatiotemporal aspects of the metaverse can also play a significant role in the development of social interactions within the virtual world\cite{evans2022social}. For example, users can engage in real-time communication and collaboration, and the virtual environment can reflect changes in the user's social network, such as changes in the user's friends list or group memberships.

\vspace{1.0ex}

\vspace{1.0ex}
\noindent \textbf{Scalability}. In terms of type, scope, and range, scalability refers to the capacity of the metaverse to remain efficient with the number of concurrent users/avatars, the complexity of scenes, and the mode of interactions between users and avatars\cite{dionisio20133d}. Essentially, scalability is the capacity of a system to handle an increasing number of users/avatars, transactions, and interactions. This includes the capability of maintaining stable and reliable performance even as the number of users/avatars and interactions increases. For the metaverse to be scalable, technical solutions and architectural design must be combined with a robust and flexible infrastructure that is capable of adapting to changing user needs. For the metaverse to achieve scalability, large amounts of data must be managed and processed. This includes user-generated content, transaction data, as well as interactions between users and virtual objects\cite{cheng2022we}. Several technologies are used by metaverse platforms to address this challenge, including distributed computing, cloud computing, and blockchain technology, to optimize data processing and management\cite{cheng2022will}. As the metaverse scales, it will be able to accommodate increasing numbers of users/avatars without compromising performance or UX. This requires a robust infrastructure capable of handling large volumes of traffic and enabling users to access the virtual environment quickly and seamlessly. Using load balancing techniques, distributed computing, and optimized network architectures, this can be achieved\cite{cheng2022reality}.

\vspace{1.0ex}
\noindent \textbf{Heterogeneity}. Heterogeneity refers to the diversity of technologies, platforms, and experiences that make up the metaverse. It is important to recognize that the metaverse is heterogeneous in that it has heterogeneous virtual spaces, heterogeneous physical devices, heterogeneous data types (such as unstructured and structured), different wireless communication networks (such as cellular and satellites), and a variety of psychological factors. Additionally, the metaverse systems do not interoperate effectively\cite{tang2022roadmap}. One of the key drivers of heterogeneity in the metaverse is the range of technologies and platforms that are used to create and access virtual environments\cite{lim2022realizing}. These include VR and AR technologies, as well as a range of gaming platforms. Each of these technologies has its strengths and limitations, and they are often used in combination to create a more diverse and immersive metaverse experience\cite{han2022dynamic}. In addition to the diversity of technologies, the metaverse is also characterized by a range of different UXs and use cases. For example, some users may use the metaverse primarily for gaming and entertainment, while others may use it for socializing, learning, or even conducting business. This diversity of use cases is reflected in the range of virtual environments and experiences that are available in the metaverse\cite{narang2023mentor}. The diversity of cultures and communities within the metaverse is another important aspect of heterogeneity\cite{sinnappan2023delphi}. Additionally, the metaverse is a global and decentralized space, and it encompasses a range of different communities, each with its own unique culture and identity. These communities may be based on shared interests, languages, or geo-locations, and they contribute to the richness and diversity of the metaverse experience\cite{godwin2023emerging}.

\vspace{1.0ex}
\noindent\textbf{Quality of Experience}. The concept of Quality of Experience (QoE) emerges as a pivotal aspect in the evolving landscape of the metaverse, playing a crucial role in determining user satisfaction and engagement within immersive digital environments. According to\cite{perkis2020qualinet}, QoE is a composite metric that reflects various key components, including user interface design, connectivity reliability, sensory accuracy, and emotional depth in virtual interactions. These elements collectively enhance the user's overall experience in the metaverse. Researchers and service providers must thoroughly understand the complex factors influencing QoE to improve services, maintain user engagement, and ensure loyalty. However, developing a comprehensive QoE model is challenging and resource-intensive. Further, XAI typically enables the use of data-driven QoE modeling to develop generalizable QoE models, while simultaneously allowing us to understand which QoE factors are relevant and how they impact the QoE score. Consequently, The study of\cite{wehner2023explainable} demonstrated the feasibility of using explainable data-driven QoE modeling specifically for video streaming, an area where QoE has been extensively researched. 
In\cite{ahmad2019towards}, the authors investigate the impact of information exchange frequency between Over The Top (OTT) services and Internet Service Providers (ISPs) on user QoE and network resource utilization. They introduced a QoE-aware collaborative management approach, detailing the necessary information exchanges. Their experiments, conducted on an SDN platform, show that while increased information exchange frequency improves network reliability and QoE, exceeding 1/4 Hz does not further enhance QoE. In\cite{murray2022can}, the researchers explore whether the principles from the QoE framework can be applied, adapted, and reimagined to create tools and systems that improve Quality of Life (QoL). According to WHO, health is defined as~\textit{a state of complete physical, mental, and social well-being and not merely the absence of disease or infirmity}\cite{WHO1946}. This definition aligns closely with the concept of QoL, which, despite its widespread use, lacks a precise definition. In a white paper\cite{le2012qualinet} stemming from this initiative, the researchers describe how human, contextual, service and systemic factors affect the quality of experience in multimedia systems, using multimedia quality as a case study. In the same context, meeting the QoE expectations of end-users is essential for the viability of video streaming services. As end-users are the primary audience for streaming content in most operational settings, subjective QoE assessments\cite{bingol2022impact} provide a reliable and straightforward mechanism for evaluating the perceptual quality of video streams. There is also a growing interest in developing objective QoE assessment models, as indicated in\cite{rodriguez2016video,hossfeld2014assessing}. However, many investigations into both subjective and objective QoE have tended to overlook the energy consumption implications of video streaming. In study\cite{bingol2023analysis}, the authors investigate how various elements of the video streaming infrastructure, including data centers, transmission networks, end-devices, and user behaviors, contribute to energy consumption and CO2 emissions. This analysis establishes the foundation for an objective methodology to determine the most appropriate video bitrate, balancing QoE with environmental sustainability. Understanding the relationship between QoE in the metaverse and gaming is crucial, particularly as gaming's established QoS criteria, such as low latency and high frame rates, also influence the metaverse. However, the metaverse introduces additional complexities with varied virtual environments and diverse user interactions\cite{anwar2024moving}.
\begin{figure*}[htbp]
\centering
 \includegraphics[width=0.86\linewidth]{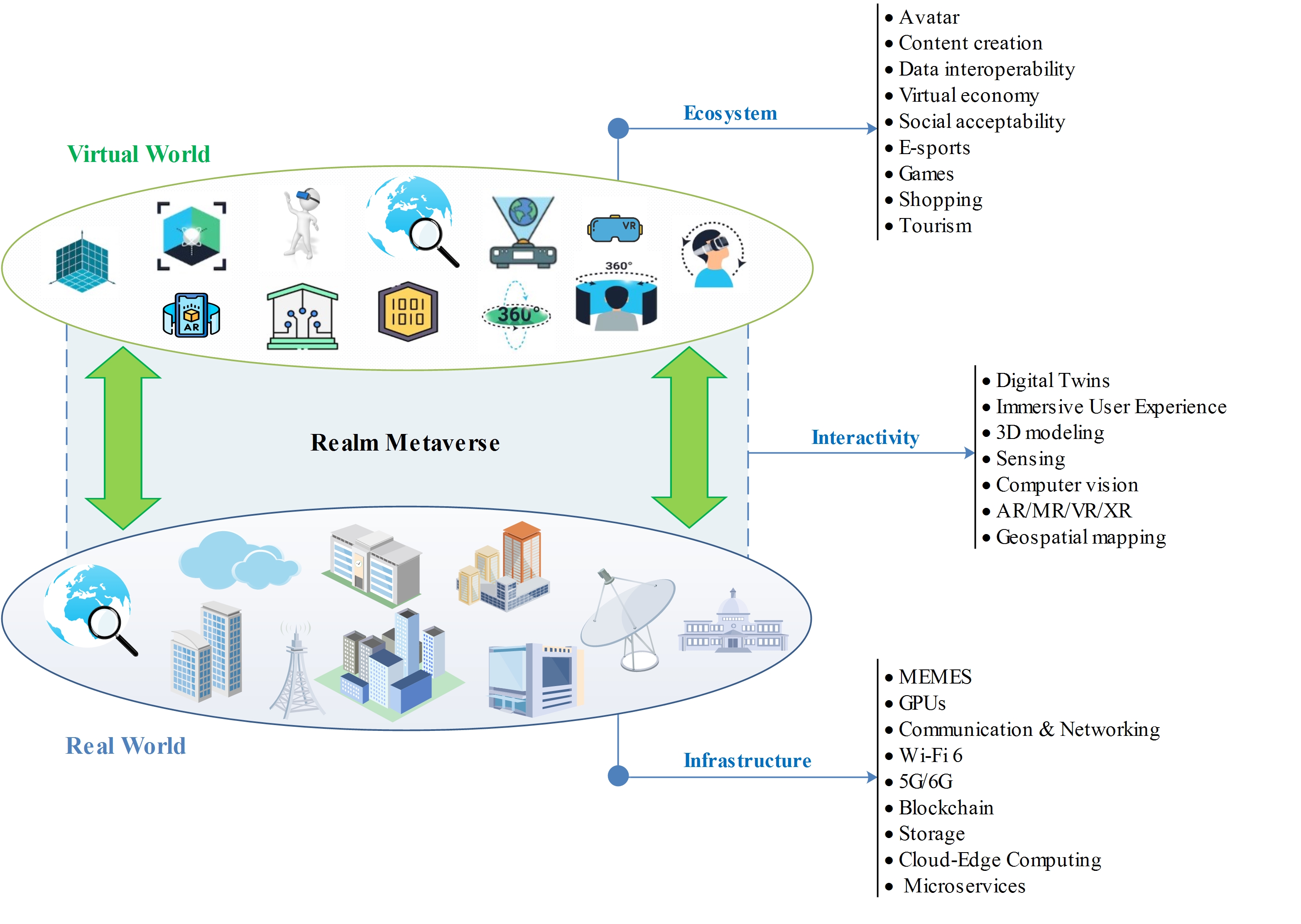}
 \caption{An overview of the typical metaverse architecture.}
 \label{Fig:Arch}
\end{figure*}
\subsection{Architectural Framework for the metaverse}
The metaverse's early-stage development contributes to a divergence in opinions between academia and industry regarding its architecture. The lack of a mature and standardized framework at this nascent stage results in varied interpretations and approaches to building the metaverse. As outlined in\cite{fu2022survey}, stakeholders are navigating through the complexities of integrating real and virtual elements, leading to differing viewpoints on architectural foundations. The proposed architecture, illustrated in Figure~\ref{Fig:Arch}, aims to address this gap by encompassing key components, including infrastructure, interactivity, and ecosystem, to establish common ground for future development and collaboration.
\vspace{1.0ex}

\noindent \textbf{Infrastructure}. A metaverse infrastructure consists of a variety of technologies (including hardware, software, GPU, communication and networking, blockchain, cloud-edge computing, storage, and interoperability protocols) that enable virtual worlds and environments to be created, managed, and operated\cite{seidel2022designing}. The metaverse refers to a collective virtual shared space, where users can interact with each other in real-time, across various devices and platforms. Several layers of technology are involved in the infrastructure of the metaverse, including hardware, software, and networking components\cite{setiawan2022essential}. Hardware components of the metaverse infrastructure include high-performance servers (e.g., GPU), storage systems, and data centers which are capable of processing and storing vast quantities of data generated by virtual environments. The metaverse infrastructure includes a variety of software development tools, programming languages, and frameworks that enable developers to create immersive and interactive virtual environments\cite{zhu2022metaaid}. The environments can range from simple 2D games to complex 3D simulations, depending on the application's requirements. Aside from software and hardware components, the metaverse infrastructure will also require robust networking technologies, such as high-speed broadband networks, 5G/6G wireless networks, and edge computing solutions\cite{zhang2023towards}. To provide real-time, low-latency connectivity between users and virtual environments, these networking technologies are crucial\cite{beshley2023emerging}. Blockchain technology is unequivocally established as a foundational pillar within the intricate framework of the metaverse infrastructure. At its core, blockchain serves as a decentralized and secure platform, offering a robust foundation for the management of virtual assets and transactions in the dynamic realm of the metaverse\cite{truong2023blockchain}. This decentralized nature ensures that no single entity possesses control over the entire network, enhancing security and transparency in virtual asset interactions. Within the metaverse, users heavily rely on blockchain technology to navigate the complex landscape of ownership and exchange of virtual assets and currencies. The decentralized nature of blockchain empowers users with a sense of autonomy and security, fostering a trustless environment where transactions can occur without the need for intermediaries\cite{xiao2023blockchain}. This autonomy extends across multiple virtual worlds, illustrating the interoperability and universality of blockchain solutions within the diverse metaverse ecosystem. Moreover, the integration of blockchain technology introduces a level of transparency and immutability to transactions, mitigating concerns related to fraudulent activities and ensuring the integrity of virtual asset ownership records\cite{mourtzis2023blockchain}. As users traverse through various virtual environments, the decentralized ledger provided by blockchain serves as an immutable record, guaranteeing the legitimacy and provenance of virtual assets\cite{truong2023blockchain}. In essence, the pivotal role of blockchain technology in the metaverse infrastructure lies not only in its technical prowess but also in the transformative impact it brings to user experiences. By providing a secure, transparent, and interoperable foundation, blockchain technology lays the groundwork for a metaverse where users can confidently engage in ownership, exchange, and transactions, thereby shaping the evolution of this digital frontier. As part of the metaverse infrastructure, interoperability protocols must be developed to enable users to seamlessly navigate between different virtual environments and platforms\cite{rawal2022rise}. These protocols ensure that users can maintain a consistent identity and virtual asset ownership across various applications and platforms in the metaverse.

\vspace{1.0ex}
\noindent \textbf{Interactivity}. This is an interface between the virtual and real worlds that provides a connection between the real world and the metaverse. The technology includes DTs, immersive UXs, 3D modeling, sensors, computer vision, geospatial mapping, and AR/MR/VR/XR\cite{rawat2023metaverse}. In the same context, interactivity is achieved through various design elements and technologies. As an example, metaverse environments typically include a wide range of interactive objects and elements, as well as detailed 3D models with realistic lighting and audio effects\cite{he2023engineering}. A key design principle of the metaverse is user-centeredness, which refers to the concept of enhancing virtual environments by prioritizing user needs and preferences. It involves creating intuitive and easy-to-use environments with clear visual cues and interactive elements that react to user input\cite{reinhard2013virtual}. In addition, the use of social features and tools plays a critical role in the architectural design of the metaverse. Social features such as chat and messaging systems are often included in metaverse environments, enabling users to communicate and interact in real-time. A sense of community and social interaction within the virtual environment can be fostered using these social tools\cite{idrees2023interactive}. A variety of interactive technologies are also incorporated into the architecture of the metaverse, including haptic feedback systems, motion tracking, and voice recognition. A virtual environment can be interacted with more naturally and intuitively using these technologies, providing a more immersive and rewarding experience for users\cite{venugopalrealm23}. Many tools and platforms allow users and developers to create their virtual environments and experiences within the metaverse. A variety of interactive experiences can be obtained through this technology, including games and simulations as well as educational and training applications\cite{almarzouqi2022prediction}.

\vspace{1.0ex}
\noindent \textbf{Ecosystem}. The architecture of the metaverse is highly dependent on the concept of the ecosystem. An ecosystem is a complex system of living organisms, their environment, and the interactions between them. The ecosystem facilitates the operation of whole metaverse systems including virtual entities, objects, buildings, environments, and their interactions, avatars, content creation, data interoperability, social acceptance, e-sports, gaming, shopping, tourism, and virtual economy\cite{wei2022gemiverse}. The ecosystem in the architecture of the metaverse is designed to enable users to interact with each other and their virtual surroundings naturally and intuitively. Furthermore, the ecosystem includes not only the software and hardware infrastructure that supports the virtual world but also the people, organizations, and communities that inhabit it. At its core, the metaverse is a shared space where users can interact with each other, create and exchange digital assets, and engage in various activities. As such, the ecosystem in the metaverse encompasses a broad range of interconnected elements, such as virtual real estate, digital currencies, social networks, marketplaces, gaming platforms, and many others\cite{guidi2022social}. In the virtual economy, one of the critical aspects of the metaverse ecosystem is the creation and distribution of digital assets. These assets can take various forms, such as virtual goods, NFTs, and cryptocurrencies\cite{hammi2023non}. Users can buy, sell, or trade these assets, which can have significant economic value\cite{vidal2023illusion}. Therefore, a well-functioning economy is an essential aspect of the metaverse ecosystem, with various actors playing different roles, such as creators, investors, traders, and consumers. Another crucial element of the metaverse ecosystem is the social network\cite{rospigliosi2022metaverse}. Users can connect, form communities, and collaborate on various projects. These communities can be based on shared interests, values, or goals, and can play a vital role in shaping the culture and identity of the metaverse. The metaverse ecosystem also includes the technical infrastructure enabling the virtual world to function\cite{ramadan2023marketing}. This infrastructure includes servers, databases, APIs, protocols, and other software and hardware components that support the creation, distribution, and interaction of digital assets and users\cite{bhattacharya2023towards}.

\begin{figure}[htbp]
 \includegraphics[width=\linewidth]{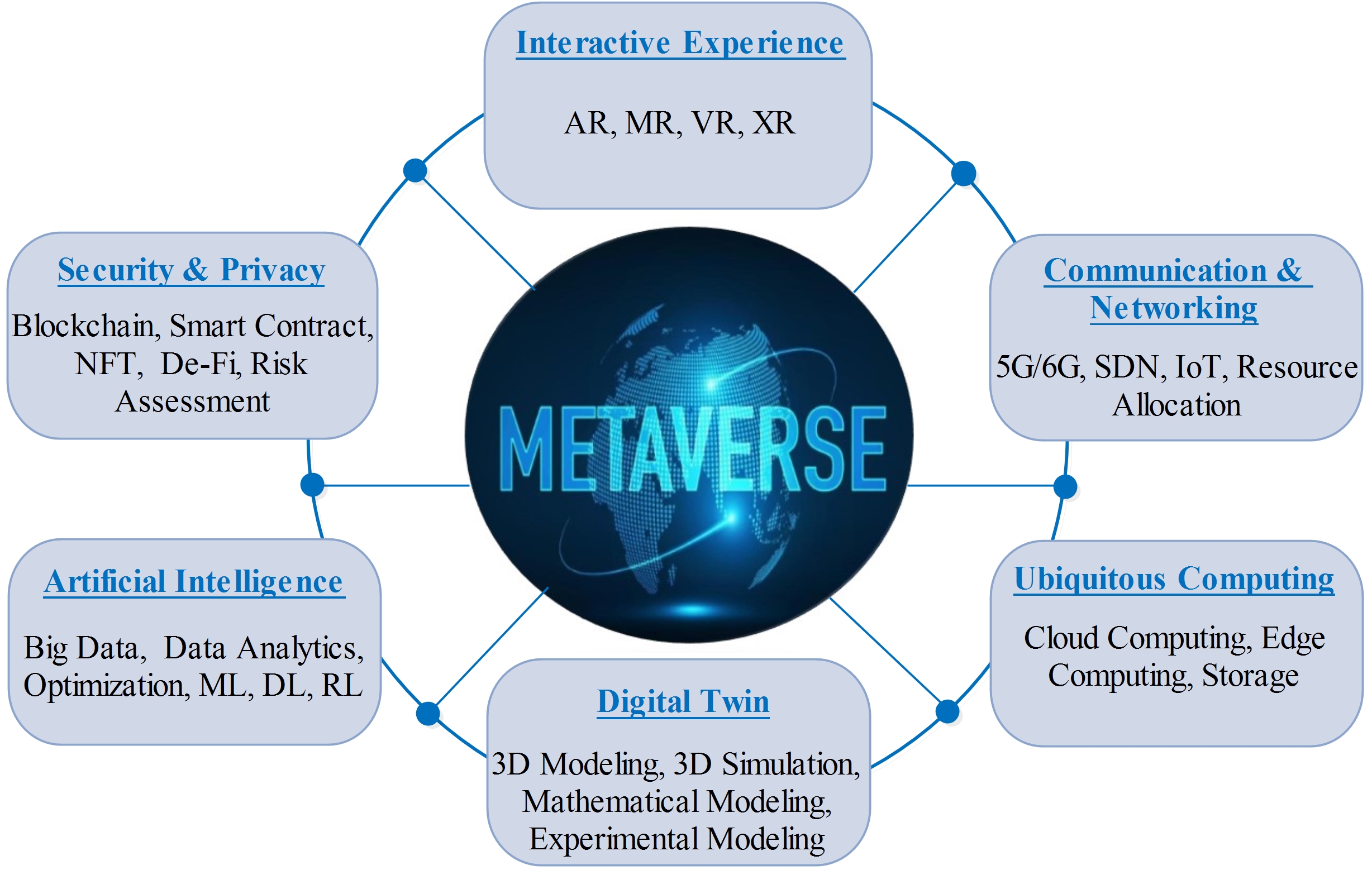}
 \caption{The metaverse's key technologies.}
 \label{Fig:Key}
\end{figure}

\subsection{Summary and Lessons Learned}
Based on an analysis of the metaverse's key requirements, architecture, and characteristics, a series of profound lessons have been uncovered. The elucidation of key requirements, encompassing energy efficiency, sustainability, interoperability, decentralization, and the critical realms of security and privacy, underscores the foundational considerations pivotal to the metaverse's robust development. These prerequisites form the cornerstone of fostering a sustainable, secure, and inclusive digital ecosystem. Delving into the key characteristics of the metaverse, spanning immersiveness, spatiotemporal dynamics, scalability, heterogeneity, and the essence of QoE, reveals a nuanced understanding of user engagement within this expansive digital landscape. As the metaverse evolves, it becomes evident that it is more than just a technological construct, but rather a dynamic fusion of immersive experiences and functional complexities. Moreover, the architectural framework, encapsulating infrastructure, interactivity, and ecosystem components, stands as the linchpin supporting the realization of these requirements and characteristics. It serves as a testament to the delicate equilibrium required between technological innovation and user-centric design, shaping a metaverse that is both robust and user-friendly. As a result of these findings, we believe that it is essential to develop a comprehensive strategy for navigating the metaverse. Recognizing the intricate interplay between technological innovation, design considerations, and ethical dimensions is not only crucial but lies at the core of steering the responsible and effective evolution of these insights act as a guiding compass for plotting the path forward, underscoring the importance of harmonizing with the diverse and evolving needs of metaverse users. This alignment proves pivotal in sculpting a digital frontier that transcends mere technological advancements, placing a strong emphasis on profound user consideration and an ethical foundation. By acknowledging and addressing these multifaceted dimensions, the metaverse has the potential to transform into a digital realm that not only pushes technological boundaries but also prioritizes the well-being, inclusivity, and ethical integrity of its user community.
\section{Key Technologies of the Metaverse}
In this section, we analyze essential technologies that are crucial for the construction of the metaverse, integrating insights obtained from existing literature. We meticulously outline key technological prerequisites, based on scholarly research, to establish the fundamental imperatives necessary for the systematic development of the metaverse. Figure~\ref{Fig:Key} illustrates the diverse technologies that comprise the metaverse's fundamental framework. This illustration carefully outlines the critical role each technology plays in structuring and shaping the dynamic digital landscape of the metaverse.

\subsection{Interactive Experience}
In the metaverse, users are offered the possibility of seamlessly engaging with their digital avatars and navigating a virtual environment with ease. A paramount consideration in constructing the metaverse is ensuring that the interactive experience is not only intuitive but also conducive to exploration and seamless interaction with other users. Virtual Reality (VR) technology stands as a foundational pillar in supporting the metaverse, employing VR headsets and controllers that empower users to fully immerse themselves in a digital environment. This technology facilitates natural interactions with digital avatars, replicating real-life movements and gestures with remarkable fidelity\cite{han2022virtual}. Complementing VR, Augmented Reality (AR) technology enriches the metaverse construction process by creating a Mixed Reality (MR) experience, overlaying digital elements seamlessly onto the real world. AR enables users to see and interact with their digital avatars within their actual surroundings, introducing an additional layer of immersion to the overall experience. The integration of Virtual Reality (VR), Augmented Reality (AR), and Mixed Reality (MR) into eXtended Reality (XR) further expands the horizon of possibilities. XR not only provides immersive encounters but also augments experiences and facilitates real-time interactions among users, avatars, and environments. Front-projected holographic displays, sophisticated human-computer interaction, and large-scale 3D modeling contribute to the creation of a dynamic and engaging metaverse\cite{wang2022metaverse}. However, the realization of the metaverse's full potential necessitates a keen focus on its social dimension. To foster a sense of community and connection among users with shared interests, online forums, chat rooms, and virtual events have become indispensable tools. These platforms serve as virtual gathering spaces, enabling users to interact, collaborate, and build meaningful connections in the digital realm\cite{oh2023social, jovanovic2022vortex}. As the metaverse evolves, the harmonious integration of technology and sociability will be pivotal in shaping a truly immersive and socially vibrant digital universe. In Table~\ref{Tab:Exp}, we present the current body of work on interactive experiences within the metaverse.

\begin{table*}[htbp]
  \centering
\caption{Existing Works of Interactive Experience for the Metaverse}
\label{Tab:Exp}
  \begin{tabular}{@{}l|l|l|l@{}}
    \cline{1-4}
    \textbf{Existing works} & \textbf{Summary of Contributions} & \textbf{Key techniques} & \textbf{Use cases}\\
    \hline
    \begin{tabular}[t]{@{}l@{}}
~\cite{popescu2022augmented}
    \end{tabular}
    & 
    \begin{tabular}[t]{@{}l@{}}
        \textbullet\ It analyzed interactive shopping experiences, retail business analytics, and\\machine vision.\\
        \textbullet\ Visualization tools, sentiment analytics, and ambient scene detection have\\been used to optimize customer engagement on live-streaming shopping platforms.
    \end{tabular}
    & 
    \begin{tabular}[t]{@{}l@{}}
        \textbullet\ Augmented Reality
    \end{tabular}
    & 
    \begin{tabular}[t]{@{}l@{}}
        \textbullet\ Shopping
    \end{tabular}
    \\
    \hline
    \begin{tabular}[t]{@{}c@{}}
~\cite{pranoto2023augmented}
    \end{tabular}
    & 
    \begin{tabular}[t]{@{}l@{}}
        \textbullet\ It suggested using interactive experience technology to enrich tourism at a\\historical site and enhance its preservation.
    \end{tabular}
    & 
    \begin{tabular}[t]{@{}l@{}}
        \textbullet\ Augmented Reality
    \end{tabular}
    & 
    \begin{tabular}[t]{@{}l@{}}
        \textbullet\ Tourism
    \end{tabular}
    \\
    \hline
    \begin{tabular}[t]{@{}l@{}}
~\cite{siyaev2021towards}
    \end{tabular}
    & 
    \begin{tabular}[t]{@{}l@{}}
        \textbullet\ It focused on providing Mixed Reality education and training for Boeing 737\\aircraft maintenance using smart glasses. It incorporates a DL speech\\interaction module to enable the hands-free operation of both physical and virtual\\assets through speech commands for trainee engineers.
    \end{tabular}
    & 
    \begin{tabular}[t]{@{}l@{}}
        \textbullet\ Mixed Reality

    \end{tabular}
    & 
    \begin{tabular}[t]{@{}l@{}}
        \textbullet\ Education \\
        \textbullet\ Training
    \end{tabular}
    \\
    \hline
    \begin{tabular}[t]{@{}l@{}}
~\cite{lee2022virtual}
    \end{tabular}
    & 
    \begin{tabular}[t]{@{}l@{}}
        \textbullet\ The approach explored how young individuals engage with media to understand\\interactive experience settings. It provides insights into potential future metaverses,\\envisioning interactive narrative experiences.
    \end{tabular}
    & 
    \begin{tabular}[t]{@{}l@{}}
        \textbullet\ Mixed Reality
    \end{tabular}
    & 
    \begin{tabular}[t]{@{}l@{}}
        \textbullet\ Media
    \end{tabular}
    \\
    \hline
    \begin{tabular}[t]{@{}l@{}}
~\cite{xu2023trustless}
    \end{tabular}
    & 
    \begin{tabular}[t]{@{}l@{}}
        \textbullet\ The utilization of virtual reality formed the basis of the proposed system, which\\was employed to create a simulation for aircraft maintenance.\\
        \textbullet\ A comparative analysis was carried out between the proposed system and the\\video-based training approach.
    \end{tabular}
    & 
    \begin{tabular}[t]{@{}l@{}}
        \textbullet\ Virtual Reality
    \end{tabular}
    & 
    \begin{tabular}[t]{@{}c@{}}
        \textbullet\ Aircraft simulation
    \end{tabular}
    \\
    \hline
    \begin{tabular}[t]{@{}l@{}}
~\cite{mourtzis2022improving}
    \end{tabular}
    & 
    \begin{tabular}[t]{@{}l@{}}
        \textbullet\ Utilizing multi-user features in virtual environments improves interactivity,\\enhancing the online teaching and learning experience for languages.\\
         \textbullet\ It offered suggestions on promoting student interactions within a culture of\\inquiry, aiming to enhance the overall quality of teaching.
    \end{tabular}
    & 
    \begin{tabular}[t]{@{}l@{}}
        \textbullet\ Virtual Reality
    \end{tabular}
    & 
    \begin{tabular}[t]{@{}l@{}}
        \textbullet\ Online language\\learning
    \end{tabular}
    \\
    \hline
        \begin{tabular}[t]{@{}l@{}}
~\cite{plechata2022can}
    \end{tabular}
    & 
    \begin{tabular}[t]{@{}l@{}}
        \textbullet\ Extended reality features, including presence, agency, and embodiment, have\\the potential to positively impact healthy behaviors by targeting users' threat\\and coping appraisals. \\
        \textbullet\ Extended reality can be employed in health communication to tackle persistent\\health challenges.
    \end{tabular}
    & 
    \begin{tabular}[t]{@{}l@{}}
        \textbullet\ Extended Reality
    \end{tabular}
    & 
    \begin{tabular}[t]{@{}l@{}}
        \textbullet\ Healthcare
    \end{tabular}
    \\
    \hline
    \begin{tabular}[t]{@{}l@{}}
~\cite{mudivcka2023digital}
    \end{tabular}
    & 
    \begin{tabular}[t]{@{}l@{}}
        \textbullet\ This study aims to examine the existing possibilities and trends in interactive\\experiences.\\
        \textbullet\ It explores the use of 3D digital heritage for creating interactive multimedia\\exhibitions in museums and galleries.
    \end{tabular}
    & 
    \begin{tabular}[t]{@{}l@{}}
        \textbullet\ Extended Reality
    \end{tabular}
    & 
    \begin{tabular}[t]{@{}l@{}}
        \textbullet\ Digital heritage
    \end{tabular}
    \\
    \hline
        \begin{tabular}[t]{@{}l@{}}
~\cite{venkatesan2021virtual}
    \end{tabular}
    & 
    \begin{tabular}[t]{@{}l@{}}
        \textbullet\ It focuses on using interactive reality in current biomedical applications, featuring\\case studies in cell biology, proteomics, and 3D cardiac models.\\
       \textbullet\ A discussion is presented on the cost comparison of different platforms and\\the emerging challenges linked to interactive experience technologies.
        
    \end{tabular}
    & 
    \begin{tabular}[t]{@{}l@{}}
        \textbullet\ Virtual Reality\\
        \textbullet\ Augmented Reality
    \end{tabular}
    & 
    \begin{tabular}[t]{@{}l@{}}
        \textbullet\ Biomedical
    \end{tabular}
    \\
    \hline
            \begin{tabular}[t]{@{}l@{}}
~\cite{farshid2018go}
    \end{tabular}
    & 
    \begin{tabular}[t]{@{}l@{}}
        \textbullet\ It introduced different types of interactive experiences within a continuum\\encompassing both real and virtual realities.\\
        \textbullet\ Illustrating differences through a common example, it outlines their respective\\business applications.
    \end{tabular}
    & 
    \begin{tabular}[t]{@{}l@{}}
        \textbullet\ Augmented Reality\\
        \textbullet\ Mixed Reality\\
        \textbullet\ Virtual Reality\\
    \end{tabular}
    & 
    \begin{tabular}[t]{@{}l@{}}
        \textbullet\ Business
    \end{tabular}
    \\
    \hline
    \cline{1-4}
\end{tabular}
\end{table*}

\subsection{Communication and Networking}
The metaverse demands advanced communication and networking technologies tailored to specific criteria. 5G technology stands out for its pivotal role, offering support for high data transfer speeds, ultra-low latency, and significantly expanded bandwidth compared to its predecessors. These attributes are indispensable for enabling seamless interaction within the metaverse, where instantaneous communication is paramount\cite{huang2023standard}. The anticipated 6G technology is poised to elevate performance further, promising even faster data transfer speeds and lower latency. This becomes particularly crucial for accommodating an extensive network of interconnected devices and sustaining enhanced mobility within the metaverse\cite{du2022optimal}. The goal of 6G is to enable advanced computing, communication, and storage capabilities for the development of new applications and services, particularly focusing on interactive immersive applications. This includes advancements in VR, AR, and MR technologies, aiming to provide highly intelligent computer processing and efficient transmission of massive amounts of data with minimal delay. The applications of immersive services span various sectors such as healthcare, online learning, and virtual teleportation. However, realizing these immersive services presents challenges, including the need for synchronization among users, managing excessive bandwidth demands, implementing advanced processing models using AI, and optimizing system architecture. To address these challenges, a proposed solution involves a hybrid edge/cloud model for 6G, allowing data processing closer to users, and reducing latency, and bandwidth consumption. The emphasis is on creating a flexible and distributed framework that seamlessly integrates advancements in edge computing, cloud services, and AI paradigms to meet the evolving requirements of immersive applications in the metaverse\cite{aloqaily2023realizing}. On the other hand, SDN addresses the distinctive challenges of the metaverse by allowing the creation and independent management of virtual networks. This technology enhances efficiency in managing network traffic and contributes to improved security through automated and dynamic network configurations\cite{tuan2022blockchain}. The IoT plays a pivotal role in crafting immersive experiences in the metaverse. Smart sensors, for instance, provide real-time data on physical environments with precise metrics, such as temperature, light levels, and sound. These data enables dynamic adjustments in the virtual environment, ensuring a more realistic and engaging user experience\cite{guan2022extended,li2022internet}. Efficient resource allocation is critical for optimal metaverse performance. Metaslicing technology, an emerging concept, enables dynamic allocation of resources such as bandwidth, processing power, and storage based on real-time demand. This ensures that the metaverse efficiently utilizes its resources, adapting to the evolving needs of different services\cite{chu2023metaslicing}. Overall, the metaverse's communication and networking landscape involves specific and quantifiable requirements. Technologies like 5G and the upcoming 6G promise concrete improvements in data transfer speeds and latency. SDN and IoT address the unique demands of the metaverse, while innovations like meta-slicing tackle the challenge of efficient resource allocation. Table~\ref{Tab:Net} provides an overview of the present developments in communication and networking for the metaverse.
\begin{table*}[htbp]
  \centering
\caption{Existing Works of Communication and Networking for the Metaverse}
\label{Tab:Net}
  \begin{tabular}{@{}l|l|l|l@{}}
    \cline{1-4}
    \textbf{Existing works} & \textbf{Summary of Contributions} & \textbf{Key techniques} & \textbf{Use cases}\\
    \hline
            \begin{tabular}[t]{@{}l@{}}
~\cite{han2022dynamic}
    \end{tabular}
    & 
    \begin{tabular}[t]{@{}l@{}}
        \textbullet\ It aims to create a prototype VR streaming system, investigating\\the feasibility of dynamically managing bandwidth through SDN.
    \end{tabular}
    & 
    \begin{tabular}[t]{@{}l@{}}
        \textbullet\ SDN
    \end{tabular}
    & 
    \begin{tabular}[t]{@{}l@{}}
        \textbullet\ Interactive medical\\
        \textbullet\ VR streaming service
    \end{tabular}
    \\
    \hline
        \begin{tabular}[t]{@{}c@{}}
~\cite{huang2023standard}
    \end{tabular}
    & 
    \begin{tabular}[t]{@{}l@{}}
        \textbullet\ It introduced a 5G system architecture for XR and explored 3GPP\\standards addressing network architecture enhancements for XR. These\\enhancements cover network-assisted service optimization, improved\\QoS for XR transmission, and power-saving features.\\
        \textbullet\ It explored challenges and potential research directions for the metaverse\\within the context of 5G and future mobile networks.
    \end{tabular}
    & 
    \begin{tabular}[t]{@{}l@{}}
        \textbullet\ 5G
    \end{tabular}
    & 
    \begin{tabular}[t]{@{}l@{}}
        \textbullet\ Metaverse applications
    \end{tabular}
    \\
    \hline
    \begin{tabular}[t]{@{}l@{}}
~\cite{njoku2022role}
    \end{tabular}
    & 
    \begin{tabular}[t]{@{}l@{}}
        \textbullet\ It devised a dynamic hierarchical framework tailored for virtual service\\providers to address synchronization challenges within the metaverse.\\
        \textbullet\ It introduced a temporal value decay dynamics approach to measure the\\values of digital twins and evaluate their sensitivity to synchronization\\strategies employed by virtual service providers.
    \end{tabular}
    & 
    \begin{tabular}[t]{@{}l@{}}
        \textbullet\ IoT
    \end{tabular}
    & 
    \begin{tabular}[t]{@{}l@{}}
        \textbullet\ Virtual Service\\Providers
    \end{tabular}
    \\
    \hline

    \begin{tabular}[t]{@{}l@{}}
~\cite{meng2024task}
    \end{tabular}
    & 
    \begin{tabular}[t]{@{}l@{}}
        \textbullet\ The study delves into metaverse features and explores the role of 5G\\standards in realizing the metaverse.\\
    \end{tabular}
    & 
    \begin{tabular}[t]{@{}l@{}}
        \textbullet\ 5G

    \end{tabular}
    & 
    \begin{tabular}[t]{@{}l@{}}
        \textbullet\ Online gaming\\platform
    \end{tabular}
    \\
    \hline
    \begin{tabular}[t]{@{}l@{}}
~\cite{peng20226g}
    \end{tabular}
    & 
    \begin{tabular}[t]{@{}l@{}}
        \textbullet\ This work aims to explore the significance of wireless communication\\in metaverse applications and services.\\
        \textbullet\ 6G is pivotal for the metaverse, supporting cross-platform integration,\\high-speed data connectivity, and improved user interaction.
    \end{tabular}
    & 
    \begin{tabular}[t]{@{}l@{}}
        \textbullet\ 6G
    \end{tabular}
    & 
    \begin{tabular}[t]{@{}l@{}}
        \textbullet\ Metaverse applications
    \end{tabular}
    \\
    \hline
    \begin{tabular}[t]{@{}l@{}}
~\cite{qi2022low}
    \end{tabular}
    & 
    \begin{tabular}[t]{@{}l@{}}
        \textbullet\ It provides a comprehensive overview of metaverse applications and\\investigates key technologies related to 6G in the metaverse context.\\
        \textbullet\ Several outstanding issues were explored, and suggestions for future\\research were put forth.
    \end{tabular}
    & 
    \begin{tabular}[t]{@{}l@{}}
        \textbullet\ 6G
    \end{tabular}
    & 
    \begin{tabular}[t]{@{}c@{}}
        \textbullet\ Metaverse applications
    \end{tabular}
    \\
    \hline
    \begin{tabular}[t]{@{}l@{}}
~\cite{wu2020interactive}
    \end{tabular}
    & 
    \begin{tabular}[t]{@{}l@{}}
        \textbullet\ This work aims to optimize control latency in wireless communication\\by strategically placing controllers and establishing a switch-controller\\mapping.\\
         \textbullet\ The proposed work distinguishes the impact of switches on control\\operations by analyzing flow programmability and critical flows in SDN.
    \end{tabular}
    & 
    \begin{tabular}[t]{@{}l@{}}
        \textbullet\ SDN
    \end{tabular}
    & 
    \begin{tabular}[t]{@{}l@{}}
        \textbullet\ Real backbone\\topology
    \end{tabular}
    \\
    \hline
    \begin{tabular}[t]{@{}l@{}}
~\cite{si2022resource}
    \end{tabular}
    & 
    \begin{tabular}[t]{@{}l@{}}
        \textbullet\ For optimal utility in the metaverse, the proposed approach suggests\\a communications resource allocation algorithm based on outer\\approximation.\\
    \end{tabular}
    & 
    \begin{tabular}[t]{@{}l@{}}
        \textbullet\ Resource allocation
    \end{tabular}
    & 
    \begin{tabular}[t]{@{}l@{}}
        \textbullet\ Interaction with real\\world 
    \end{tabular}
    \\
    \hline
        \begin{tabular}[t]{@{}l@{}}
~\cite{du2023attention}
    \end{tabular}
    & 
    \begin{tabular}[t]{@{}l@{}}
        \textbullet\ Meta-Immersion is an innovative scheme put forth to model the\\quality of experience (QoE) from the perspective of metaverse users.\\
       \textbullet\ The proposed Meta-Immersion scheme includes a new xURLLC\\approach that predicts users' attention to virtual objects using historical\\sparse attention records.
    \end{tabular}
    & 
    \begin{tabular}[t]{@{}l@{}}
        \textbullet\ Resource allocation\\
        \textbullet\ URLLC
    \end{tabular}
    & 
    \begin{tabular}[t]{@{}l@{}}
        \textbullet\ QoE
    \end{tabular}
    \\
    \hline
    \cline{1-4}
\end{tabular}
\end{table*}

\subsection{Ubiquitous Computing}
Ubiquitous computing, representing the seamless integration of technology into our daily lives, holds paramount importance in crafting an immersive metaverse experience accessible anytime and anywhere\cite{monteiro2018evaluating}. Expanding on this notion, various aspects of ubiquitous computing, including cloud computing, edge computing, and storage technologies, significantly contribute to enhancing the metaverse landscape. In the metaverse's dynamic environment, cloud computing emerges as a pivotal player, offering the necessary infrastructure to sustain an extensive network of virtual environments and experiences\cite{jiang2022reliable}. Leveraging cloud-based solutions, applications in virtual and augmented reality can tap into processing power and storage capabilities, ensuring the delivery of high-quality, immersive experiences. Cloud services not only support scalability but also provide the flexibility needed to adapt to evolving user preferences and requirements. As the metaverse continues to grow, cloud-based solutions are instrumental in maintaining responsiveness and accessibility. The integration of edge computing is equally transformative, particularly in delivering real-time, immersive encounters for metaverse users\cite{wang2022survey}. By processing data and running applications in proximity to users, edge computing reduces latency and enhances overall performance. This approach becomes especially advantageous in ensuring high-quality VR and AR experiences even for users with limited internet bandwidth, making the metaverse more inclusive and accessible\cite{shen2022metaverse,aung2023edge,van2022edge}. Storage emerges as a critical component as the metaverse evolves into a vast and intricate digital space\cite{ersoy2023blockchain}. With a substantial amount of data generated by users, including images, videos, and various media forms, robust storage solutions are imperative. Cloud-based storage options like Amazon S3 and Google Cloud Storage offer scalability and flexibility, accommodating the evolving storage needs of the metaverse\cite{amazonS32023,googlecloud2023,maccallum2019teacher}. As the metaverse landscape expands, ensuring efficient storage solutions becomes essential to managing the ever-growing volume of data and supporting the diverse needs of its users\cite{ryskeldiev2018distributed}. In Table~\ref{Tab:Comp}, we showcase the present research on the application of ubiquitous computing within the metaverse.
\begin{table*}[htbp]
  \centering
\caption{Existing Works of Ubiquitous Computing for the Metaverse}
\label{Tab:Comp}
  \begin{tabular}{@{}l|l|l|l@{}}
    \cline{1-4}
    \textbf{Existing works} & \textbf{Summary of Contributions} & \textbf{Key techniques} & \textbf{Use cases}\\
    \hline
    \begin{tabular}[t]{@{}l@{}}
~\cite{oliver2013metaverse}
    \end{tabular}
    & 
    \begin{tabular}[t]{@{}l@{}}
        \textbullet\ It explores the deployment of immersive educational environments in the cloud\\and provides performance results for hosting virtual worlds in a cloud-based setting.
    \end{tabular}
    & 
    \begin{tabular}[t]{@{}l@{}}
        \textbullet\ Cloud computing
    \end{tabular}
    & 
    \begin{tabular}[t]{@{}l@{}}
        \textbullet\ MOOCs
    \end{tabular}
    \\
    \hline
    \begin{tabular}[t]{@{}c@{}}
~\cite{cai2022compute}
    \end{tabular}
    & 
    \begin{tabular}[t]{@{}l@{}}
        \textbullet\ It presented metaverse applications leveraging cloud networks, examining their\\practicality, versatility, and the diverse requirements they encompass.\\
        \textbullet\ A comprehensive explanation of the compute and data-intensive infrastructure for\\the metaverse is presented.
    \end{tabular}
    & 
    \begin{tabular}[t]{@{}l@{}}
        \textbullet\ Cloud networks
    \end{tabular}
    & 
    \begin{tabular}[t]{@{}l@{}}
        \textbullet\ Social\\
        \textbullet\ Gaming\\
        \textbullet\ Healthcare\\
        \textbullet\ Education
    \end{tabular}
    \\
    \hline
    \begin{tabular}[t]{@{}l@{}}
~\cite{chua2023play}
     \end{tabular}
    & 
    \begin{tabular}[t]{@{}l@{}}
        \textbullet\ It investigated a play-to-learn mobile edge computing framework for the metaverse\\and introduces an optimized approach to minimize latency in in-game graphics data\\transmission.\\
        \textbullet\ It proposed a multi-agent loss-sharing algorithm for addressing joint optimization\\problems with different variable weightings, utilizing joint optimization weighting\\analyses.
    \end{tabular}
    & 
    \begin{tabular}[t]{@{}l@{}}
        \textbullet\ Edge computing\\
        \textbullet\ Mobile edge\\computing
    \end{tabular}
    & 
    \begin{tabular}[t]{@{}l@{}}
        \textbullet\ Gaming
    \end{tabular}
    \\
    \hline
    \begin{tabular}[t]{@{}l@{}}
~\cite{cong2022edge}
    \end{tabular}
    & 
    \begin{tabular}[t]{@{}l@{}}
        \textbullet\ It introduced a scheme for edge computing trading within the framework of a\\metaverse enabled by semantic communication.\\
        \textbullet\ The proposed scheme aims to maximize revenue for edge computing providers\\while ensuring crucial economic properties, including individual rationality and\\incentive compatibility.
    \end{tabular}
    & 
    \begin{tabular}[t]{@{}l@{}}
        \textbullet\ Edge computing
    \end{tabular}
    & 
    \begin{tabular}[t]{@{}l@{}}
        \textbullet\ Economy
    \end{tabular}
    \\
    \hline
    \begin{tabular}[t]{@{}l@{}}
~\cite{zhang2023multi}
    \end{tabular}
    & 
    \begin{tabular}[t]{@{}l@{}}
        \textbullet\ It proposed an efficient encryption-based healthcare data-sharing scheme for the\\metaverse, utilizing a feature-rich multi-server architecture.\\
        \textbullet\ The scheme incorporates metaverse healthcare ciphertext validity and equivalence\\detection to mitigate the presence of invalid ciphertexts.
    \end{tabular}
    & 
    \begin{tabular}[t]{@{}l@{}}
        \textbullet\ Storage
    \end{tabular}
    & 
    \begin{tabular}[t]{@{}c@{}}
        \textbullet\ Patient\\diagnosis
    \end{tabular}
    \\
    \hline
    \begin{tabular}[t]{@{}l@{}}
~\cite{dilibal2022implementation}
    \end{tabular}
    & 
    \begin{tabular}[t]{@{}l@{}}
        \textbullet\ The study provides a comprehensive examination of the implementation process\\of an embedded smart healthcare system within the metaverse architecture, with\\a foundation on cloud technology.\\
        \textbullet\ It aims to evaluate the feasibility of implementing a smart healthcare system in the\\metaverse. It investigates potential concerns and limitations during the implementation\\process, ultimately seeking to clarify the system's viability.
    \end{tabular}
    & 
    \begin{tabular}[t]{@{}l@{}}
        \textbullet\ Edge computing\\
        \textbullet\ Data storage
    \end{tabular}
    & 
    \begin{tabular}[t]{@{}l@{}}
        \textbullet\ Telemedicine\\
        \textbullet\ Telehealth
    \end{tabular}
    \\
    \hline
    \cline{1-4}
\end{tabular}
\end{table*}
\subsection{Digital Twin} 
The digital twin (DT) serves as a powerful tool in the metaverse, acting as a virtual counterpart to physical objects, systems, or processes. Its application enhances the construction and maintenance of a virtual world that is not only accurate but also captivating for users. In essence, the digital twin is a faithful and conscious representation of the real world\cite{wu2021digital}. Various modeling techniques contribute to the development of digital twins for the metaverse, each offering unique advantages. 3D modeling, a prominent technique, involves the creation of detailed representations of objects or environments, such as buildings and landscapes, providing a visual foundation for the metaverse\cite{lo2022design}. Computer-Aided Design (CAD) software facilitates the creation of intricate 3D models, seamlessly integrated into the metaverse\cite{jaimini2022imetaversekg}. In 3D simulation, computer software generates interactive virtual environments, enabling real-time interactions within the metaverse. This technique is employed to create virtual games or simulations, fostering immersive experiences where users can explore and interact using game engines like Unity and Unreal Engine\cite{zauskova2022visual,chia2022metaverse}. Mathematical modeling, utilizing equations and algorithms, provides a means to create virtual representations of physical systems and processes in the metaverse. This approach is instrumental in generating realistic simulations of physics and natural processes, employing tools such as computational fluid dynamics and finite element analysis for accurate behavioral representations\cite{moser2021mechanistic}. Experimental modeling in the metaverse is geared towards simulating user behavior and interactions within virtual environments. Through user studies and experiments, valuable data is collected to enhance the UX in the virtual world\cite{tcha2018towards}. Overall, the integration of digital twins and their associated modeling techniques enriches the metaverse by providing a diverse range of tools to create, simulate, and refine virtual environments with high fidelity and interactivity. We provide a summary of the utilization of DT technology for the metaverse, including its applications, as presented in Table~\ref{Tab:DT}.
\begin{table*}[htbp]
  \centering
\caption{Existing Works of Digital Twins for the Metaverse}
\label{Tab:DT}
  \begin{tabular}{@{}l|l|l|l@{}}
    \cline{1-4}
    \textbf{Existing works} & \textbf{Summary of Contributions} & \textbf{Key techniques} & \textbf{Use cases}\\
    \hline
    \begin{tabular}[t]{@{}l@{}}
~\cite{afrooz2019immersive}
    \end{tabular}
    & 
    \begin{tabular}[t]{@{}l@{}}
        \textbullet\ This work explores virtual learning environments within the context of\\architecture, urban planning, and design.\\
        \textbullet\ It aims to conduct a critical evaluation of the capacity of virtual\\environments in supporting experiential online learning.\\
        \textbullet\ To contribute to the literature, a virtual world is implemented in both\\undergraduate and graduate-built environment courses.
        
    \end{tabular}
    & 
    \begin{tabular}[t]{@{}l@{}}
        \textbullet\ 3D Modeling
    \end{tabular}
    & 
    \begin{tabular}[t]{@{}l@{}}
        \textbullet\ Education
    \end{tabular}
    \\
    \hline
    \begin{tabular}[t]{@{}c@{}}
~\cite{ammanuel2019creating}
    \end{tabular}
    & 
    \begin{tabular}[t]{@{}l@{}}
        \textbullet\ The proposed study aims to demonstrate the transformation of 2D\\radiologic images into affordable 3D models using thresholding and\\segmentation, integrated with a VR interface.\\
        \textbullet\ Four anatomy modules have been developed, featuring inputs that~govern\\the rotation and translation of 3D models within the virtual environment.
    \end{tabular}
    & 
    \begin{tabular}[t]{@{}l@{}}
        \textbullet\ 3D modeling
    \end{tabular}
    & 
    \begin{tabular}[t]{@{}l@{}}
        \textbullet\ Medical education\\
        \textbullet\ Radiologic

    \end{tabular}
    \\
    \hline
    \begin{tabular}[t]{@{}l@{}}
~\cite{szalai2020mixed}
     \end{tabular}
    & 
    \begin{tabular}[t]{@{}l@{}}
        \textbullet\ This study introduced a digital twin simulation environment that\\seamlessly integrates a real test car into a virtual setting. This allows\\the car to navigate real and virtual obstacles, conduct traffic simulations,\\and provide visualizations of the entire process.\\
        \textbullet\ Digital twin simulation environments play a crucial role in advancing\\autonomous vehicle development, serving as a basis for establishing\\validation procedures for these systems.
    \end{tabular}
    & 
    \begin{tabular}[t]{@{}l@{}}
        \textbullet\ 3D simulation
    \end{tabular}
    & 
    \begin{tabular}[t]{@{}l@{}}
        \textbullet\ Autonomous Vehicles
    \end{tabular}
    \\
    \hline
    \begin{tabular}[t]{@{}l@{}}
~\cite{carey2022simulation}
    \end{tabular}
    & 
    \begin{tabular}[t]{@{}l@{}}
        \textbullet\ The proposed study centers on smart healthcare within the metaverse.\\
        \textbullet\ The aim of this study is to compile and summarize prior research\\findings on smart wearable Internet medical devices and the decentralized\\healthcare metaverse.
    \end{tabular}
    & 
    \begin{tabular}[t]{@{}l@{}}
        \textbullet\ 3D simulation
    \end{tabular}
    & 
    \begin{tabular}[t]{@{}l@{}}
        \textbullet\ Healthcare
    \end{tabular}
    \\
    \hline
    \begin{tabular}[t]{@{}l@{}}
~\cite{bandyopadhyay2023game}
    \end{tabular}
    & 
    \begin{tabular}[t]{@{}l@{}}
        \textbullet\ The work presents a dynamic and distributed framework for high-quality\\reconstructions using sensor data streams and optimal allocation of\\virtual service providers.\\
        \textbullet\ A simultaneous differential game is employed to model the optimal\\synchronization intensity control between the feeder network and virtual\\service providers.\\
        \textbullet\ A model based on preference-based game theory is utilized to represent\\the allocation of virtual service providers to users.
    \end{tabular}
    & 
    \begin{tabular}[t]{@{}l@{}}
        \textbullet\ Mathematical modeling
    \end{tabular}
    & 
    \begin{tabular}[t]{@{}c@{}}
        \textbullet\ Virtual service provider
    \end{tabular}
    \\
    \hline
    \begin{tabular}[t]{@{}l@{}}
~\cite{van2011modeling}
    \end{tabular}
    & 
    \begin{tabular}[t]{@{}l@{}}
        \textbullet\ This work aims to introduce a theoretical model for effective teamwork\\within 3D virtual environments.\\
        \textbullet\ The presented model seeks to enhance the comprehension of capabilities\\that contribute to successful collaborations within 3D virtual teams.\\
        \textbullet\ As per the proposed model, 3D virtual worlds can facilitate this form\\of teamwork.
    \end{tabular}
    & 
    \begin{tabular}[t]{@{}l@{}}
        \textbullet\ 3D modeling\\
        \textbullet\ Mathematical modeling
    \end{tabular}
    & 
    \begin{tabular}[t]{@{}l@{}}
        \textbullet\ Team collaboration
    \end{tabular}
    \\
    \hline
    \begin{tabular}[t]{@{}l@{}}
    ~\cite{guo2022metaverse}
    \end{tabular}
    & 
    \begin{tabular}[t]{@{}l@{}}
        \textbullet\ The study establishes an experimental English-teaching scenario.\\
        \textbullet\ AI models are developed to detect students' emotions using\\electroencephalograms, considering time, frequency, and spatial domains.\\
    \end{tabular}
    & 
    \begin{tabular}[t]{@{}l@{}}
        \textbullet\ Experimental modeling
    \end{tabular}
    & 
    \begin{tabular}[t]{@{}l@{}}
        \textbullet\ Teaching
    \end{tabular}
    \\
    \hline
       \begin{tabular}[t]{@{}l@{}}
    ~\cite{yang2023efficacy}
    \end{tabular}
    & 
    \begin{tabular}[t]{@{}l@{}}
        \textbullet\ It created a metaverse-based nursing simulation program for early-onset\\schizophrenia, drawing from person-centered therapy. It aims to improve\\therapeutic communication skills among nursing students.\\
        \textbullet\ The simulation program exhibited positive effects, indicating its potential\\as an alternative for clinical psychiatry education without limitations\\in space and time.
    \end{tabular}
    & 
    \begin{tabular}[t]{@{}l@{}}
        \textbullet\ 3D simulation\\
        \textbullet\ Experimental modeling
    \end{tabular}
    & 
    \begin{tabular}[t]{@{}l@{}}
        \textbullet\ Nursing education
    \end{tabular}
    \\
    \hline
    \cline{1-4}
\end{tabular}
\end{table*}
\subsection{Artificial Intelligence}
Realizing the full potential of the metaverse necessitates the integration of artificial intelligence (AI) technologies, serving as the backbone for the efficient, effective, and secure operation of this virtual realm\cite{huynh2023artificial}. Within the metaverse, AI plays a multifaceted role, employing various AI techniques and models for optimization, big data analytics, natural language processing (NLP), and more. Big data analytics, powered by AI, becomes a cornerstone in processing and deciphering the vast datasets generated through metaverse interactions, transactions, and other activities. This AI-driven analytics not only enhances UX but also contributes valuable insights for informed decision-making\cite{zhang2023survey,zhou2022evolution}. NLP emerges as a key facilitator of natural and intuitive communication between users and the metaverse. By leveraging AI-powered NLP technologies, users can interact seamlessly with the metaverse using natural language, voice commands, and other forms of human-like communication, enriching the overall user experience and paving the way for innovative VR applications, including virtual assistants and customer service\cite{de2023nlp}. Optimization algorithms, underpinned by AI, play a crucial role in maintaining the metaverse's efficiency by optimizing network traffic and resource allocation, ensuring stability and uninterrupted availability to users and avatars\cite{zou2017collision}. The impact of AI/ML algorithms on the metaverse is widespread, spanning areas such as computer vision, tactical autonomy, object detection, fraud detection, trust, privacy, and security. These algorithms encompass ML, DL, RL, and FL. Over time, these algorithms continuously learn from user/avatar behavior and other data sources, progressively improving in accuracy and effectiveness\cite{rospigliosi2022adopting}. Explainable AI (XAI) aspects contribute to transparency and accountability in AI systems, enabling researchers and users to better understand the decision-making processes\cite{adnan2022earliest}. Generative AI (GAI) models add a layer of creativity to the metaverse, facilitating the automatic generation of realistic virtual environments, characters, and objects. Leveraging GAI models not only saves time and resources but also elevates the overall quality of the virtual world within the metaverse, demonstrating the potential for AI to shape and enhance the immersive experiences offered\cite{ratican2023proposed}. In addition to these AI components, the integration of Large Language Models (LLMs) further enhances the metaverse's interactive capabilities\cite{lopez2023medical}. LLMs, such as GPT-3.5 and GPT-4 , contribute to advanced natural language understanding and generation, enabling nuanced conversations, context-aware responses, and dynamic storytelling within the metaverse. Users can engage in rich linguistic interactions, fostering a more immersive and intelligent virtual environment. LLMs also act as language mediators, facilitating communication across diverse linguistic backgrounds and fostering a globally inclusive virtual space. Ethical considerations, including data privacy and responsible AI usage, become integral in ensuring the ethical deployment of LLMs within the evolving metaverse landscape. In Table~\ref{Tab:AI}, we present a variety of applications of AI within the context of the metaverse.
\begin{table*}[htbp]
  \centering
\caption{Existing Works of Artificial Intelligence for the Metaverse}
\label{Tab:AI}
  \begin{tabular}{@{}l|l|l|l@{}}
    \cline{1-4}
    \textbf{Existing works} & \textbf{Summary of Contributions} & \textbf{Key techniques} & \textbf{Use cases}\\
    \hline
    \begin{tabular}[t]{@{}l@{}}
~\cite{siwach2022inferencing}
    \end{tabular}
    & 
    \begin{tabular}[t]{@{}l@{}}
        \textbullet\ The goal of this quantitative research is to present various approaches for\\visualizing big data in the metaverse using VR. \\
        \textbullet\ Through the application of AI/ML algorithms, the authors provided\\a comprehensive implementation of statistical and modeling techniques.\\
        \textbullet\ An inferencing architecture and process in real-time are employed to\\perform statistical analysis on big data within the metaverse.
        
    \end{tabular}
    & 
    \begin{tabular}[t]{@{}l@{}}
        \textbullet\ Big Data\\
        \textbullet\ Machine Learning\\
    \end{tabular}
    & 
    \begin{tabular}[t]{@{}l@{}}
        \textbullet\ Infrastructure level\\security
    \end{tabular}
    \\
    \hline
    \begin{tabular}[t]{@{}c@{}}
~\cite{tyagi2022applications}
    \end{tabular}
    & 
    \begin{tabular}[t]{@{}l@{}}
        \textbullet\ The study provided an analyze of existing literature regarding the\\applications of AI and data science in hospital administration, accompanied\\by the inclusion of various case studies.\\
        \textbullet\ The discussion includes metaverse, AI, and data science in the~context\\of smart healthcare, highlighting unanswered research\\questions and challenges.\\
        \textbullet\ It outlined future research directions and explores the integration of the\\metaverse, AI, and data science in smart healthcare, offering valuable\\insights for researchers in the field.
    \end{tabular}
    & 
    \begin{tabular}[t]{@{}l@{}}
        \textbullet\ Big data analytic
    \end{tabular}
    & 
    \begin{tabular}[t]{@{}l@{}}
        \textbullet\ Smart healthcare

    \end{tabular}
    \\
    \hline
    \begin{tabular}[t]{@{}l@{}}
~\cite{rathore2023digital}
     \end{tabular}
    & 
    \begin{tabular}[t]{@{}l@{}}
        \textbullet\ The objective of the proposed study is to assess the potential impact of\\integrating AI with the metaverse on the fashion industry.\\
        \textbullet\ Recommendations have been made for future research
    \end{tabular}
    & 
    \begin{tabular}[t]{@{}l@{}}
        \textbullet\ Machine Learning\\
        \textbullet\ Deep Learning\\
    \end{tabular}
    & 
    \begin{tabular}[t]{@{}l@{}}
        \textbullet\ Marketing sector
    \end{tabular}
    \\
    \hline
    \begin{tabular}[t]{@{}l@{}}
~\cite{ali2023metaverse}
    \end{tabular}
    & 
    \begin{tabular}[t]{@{}l@{}}
        \textbullet\ The proposed architecture includes platforms for the doctor, the patient,\\and the metaverse.\\
        \textbullet\ XAI contributes to the trust, explanation, interpretability, and transparency\\in the diagnosis and prediction of diseases.\\
        \textbullet\ The proposed architecture ensures transparency and trust in patient data\\security and disease diagnosis.
    \end{tabular}
    & 
    \begin{tabular}[t]{@{}l@{}}
        \textbullet\ eXplainable AI
    \end{tabular}
    & 
    \begin{tabular}[t]{@{}l@{}}
        \textbullet\ Patients data\\security
    \end{tabular}
    \\
    \hline
    \begin{tabular}[t]{@{}l@{}}
~\cite{li2023optimization}
    \end{tabular}
    & 
    \begin{tabular}[t]{@{}l@{}}
        \textbullet\ The study investigates VR technology, utilizing visual direction with\\optimization algorithms and considering aspects like perception, existence,\\and interactive operation. It aims to propose preservation methods for folk\\arts within the VR technology domain.
    \end{tabular}
    & 
    \begin{tabular}[t]{@{}l@{}}
         \textbullet\ Deep Learning\\
         \textbullet\ Optimization algorithms\\
    \end{tabular}
    & 
    \begin{tabular}[t]{@{}c@{}}
        \textbullet\ Folk arts
    \end{tabular}
    \\
    \hline
    \begin{tabular}[t]{@{}l@{}}
~\cite{gu2023metaverse}
    \end{tabular}
    & 
    \begin{tabular}[t]{@{}l@{}}
        \textbullet\ The study proposes a deep reinforcement earning (DRL) model within\\the metaverse to solve the efficient evacuation problem. It showcases\\a training system within the metaverse that allows evacuees to choose the\\most efficient route for a swift evacuation.\\
        \textbullet\ To dynamically guide evaluation, cloud servers process the collected data\\using a DRL model.
    \end{tabular}
    & 
    \begin{tabular}[t]{@{}l@{}}
        \textbullet\ Deep Reinforcement\\Learning\\
        \textbullet\ Optimization algorithms
    \end{tabular}
    & 
    \begin{tabular}[t]{@{}l@{}}
        \textbullet\ Teaching building\\evacuation Guidance
    \end{tabular}
    \\
    \hline
    \begin{tabular}[t]{@{}l@{}}
    ~\cite{zhou2022mobile}
    \end{tabular}
    & 
    \begin{tabular}[t]{@{}l@{}}
        \textbullet\ Federated learning (FL) and mobile AR (MAR) synergize in the metaverse\\to protect user privacy while making use of the computing resources on\\mobile devices.\\
        \textbullet\ This work provides an explanation and a comprehensive list of promising\\technologies that enable the implementation of FL-MAR systems within the\\metaverse.
    \end{tabular}
    & 
    \begin{tabular}[t]{@{}l@{}}
        \textbullet\ Federated Learning
    \end{tabular}
    & 
    \begin{tabular}[t]{@{}l@{}}
        \textbullet\ Education\\
        \textbullet\ Autonomous driving\\
        \textbullet\ Shopping
    \end{tabular}
    \\
    \hline
       \begin{tabular}[t]{@{}l@{}}
    ~\cite{xu2023generative}
    \end{tabular}
    & 
    \begin{tabular}[t]{@{}l@{}}
        \textbullet\ An autonomous driving architecture based on Generative AI (GAI) is\\introduced within the framework of the metaverse.\\
        \textbullet\ A developed multi-task digital twin offloading model is implemented to\\ensure the reliable execution of autonomous vehicle digital twin tasks.\\
        \textbullet\ The authors used GAI to create diverse and conditional driving simulation\\datasets for offline training of autonomous vehicles.
    \end{tabular}
    & 
    \begin{tabular}[t]{@{}l@{}}
        \textbullet\ Generative AI
    \end{tabular}
    & 
    \begin{tabular}[t]{@{}l@{}}
        \textbullet\ Autonomous vehicles
    \end{tabular}
    \\
    \hline
           \begin{tabular}[t]{@{}l@{}}
    ~\cite{cho2022dave}
    \end{tabular}
    & 
    \begin{tabular}[t]{@{}l@{}}
        \textbullet\ The goal of this work is to design a platform-optimized interface using DL\\in an asymmetric virtual environment, fostering interaction between VR and\\AR participants.\\
        \textbullet\ The authors proposed a DL-based asymmetric virtual environment to\\enhance immersive experiential metaverse content.
    \end{tabular}
    & 
    \begin{tabular}[t]{@{}l@{}}
        \textbullet\ Deep Learning
    \end{tabular}
    & 
    \begin{tabular}[t]{@{}l@{}}
        \textbullet\ Gesture interface\\
        \textbullet\ Tex interface\\
    \end{tabular}
    \\
    \hline
    \cline{1-4}
\end{tabular}
\end{table*}

\subsection{Cybersecurity}
In the expansive application of the metaverse across various facets of modern life, cybersecurity emerges as a critical technology. Its significance lies in establishing trust within the metaverse, ensuring secure transactions, safeguarding user data, and maintaining privacy in a virtual space where personal information is openly shared. As highlighted by recent research\cite{chow2022visualization}, the metaverse faces diverse threats and potential attacks. Addressing these challenges necessitates a comprehensive cyber-defense strategy, encompassing detection mechanisms, and countermeasures, as well as leveraging blockchain, Non-Fungible Tokens (NFT), and Decentralized Finance (De-Fi). Blockchain technology, in particular, proves instrumental in fortifying metaverse security. Its decentralized and transparent ledger enables meticulous tracking of digital assets and transactions. This capability ensures the confidentiality, integrity, and availability of the metaverse by recording and verifying all changes and transactions, presenting a formidable barrier against malicious alterations\cite{sathya2023blockchain}. Moreover, blockchain can facilitate secure and trustless transactions between users in the metaverse. In this context, blockchain technology enables users to transact with each other without the need for a third-party intermediary, thereby reducing the risk of fraud and theft. Developers and users in the metaverse must ensure that they implement appropriate security measures, such as multi-factor authentication and encryption, to mitigate these risks\cite{swati2023innovations}. An NFT represents ownership of a virtual item or digital content. NFTs have become increasingly popular in the metaverse, with users buying and selling them for millions of dollars. It is important to note, however, that NFTs also present privacy concerns\cite{zhang2024privacy}. To avoid identity theft or other malicious activities, users should exercise caution when sharing information about their NFTs and their ownership of them\cite{zelenyanszki2023privacy}. Decentralized finance (De-Fi) is a financial system based on blockchain technology that allows peer-to-peer financial transactions using financial instruments without intermediaries\cite{qiao2023time}. Metaverse, on the other hand, is a virtual world where users can interact with each other. Integrating De-Fi into the metaverse could revolutionize how we interact with digital assets. With blockchain technology, users could gain greater control over their digital assets, eliminating the need for intermediaries such as banks and brokers. Furthermore, smart contracts could enable automated financial transactions within the metaverse\cite{duan2021metaverse,chao2022study}. In addition, De-Fi could be applied to the creation of a Decentralized EXchange (DEX) for virtual assets in the metaverse\cite{mackenzie2022criminology}. Users of such an exchange could trade virtual assets such as virtual estate, digital collectibles, and other digital assets peer-to-peer. By eliminating centralized exchanges, which are often hacked and compromised, security breaches could be prevented\cite{christodoulou2022nfts}. In the metaverse, De-Fi may also be used to create decentralized lending and borrowing platforms. By using these platforms, users would be able to borrow and lend virtual assets, with interest rates and collateral determined by smart contracts. A virtual asset account would allow users to earn interest on their assets while providing them with increased access to capital. On the other hand, the metaverse is not immune to attacks, and it is important to implement effective techniques to detect and prevent these malicious activities. These attacks can be categorized into social engineering threats and traditional IT attacks. Social engineering attacks manipulate the user psychologically so that he makes security mistakes or discloses sensitive data\cite{falchuk2018social}. The fraudsters will have a more difficult time concealing their identities. A criminal may attempt to penetrate the metaverse by posing as a company, provider, family member, or friend. It is also possible that law enforcement agencies may have difficulty intercepting crimes and criminals in the metaverse. Furthermore, metaverse technologies are based on IT standards, making them susceptible to IT attacks. As far as IT threat scenarios are concerned, such as API attacks, ransomware, etc., these scenarios will continue to occur throughout the metaverse\cite{canbay2022privacy}. For example, the release of APIs for Metaverse applications may result in the creation of malicious code or phishing attempts. Behavioral analysis, AI, transaction monitoring, user verification, and security audits are some techniques that can be used to identify fraudulent activities in the metaverse.  AI-based solutions employ ML/DL to detect attacks and fraud. Additionally, transaction monitoring oversees all transactions and movements of virtual assets. User verification is implemented to thwart the creation of fake accounts and to verify identities before granting access. Furthermore, security audits are conducted to identify vulnerabilities and prevent attacks\cite{pooyandeh2022cybersecurity}. Risk assessment methods involve identifying potential risks and vulnerabilities in the Metaverse, and then evaluating their potential impact on users and the metaverse as a whole. Additionally, there are many potential risks associated with the metaverse, including cyberattacks, smart contract vulnerabilities, identity theft, and malicious actors. In Table~\ref{Tab:Sec}, we present the current body of research focused on cybersecurity within the metaverse.
\begin{table*}
  \centering
\caption{Existing Works of Cybersecurity for the metaverse}
\label{Tab:Sec}
  \begin{tabular}{@{}l|l|l|l@{}}
    \cline{1-4}
    \textbf{Existing works} & \textbf{Summary of Contributions} & \textbf{Key techniques} & \textbf{Use cases}\\
        \hline
            \begin{tabular}[t]{@{}l@{}}
~\cite{xu2023trustless}
    \end{tabular}
    & 
    \begin{tabular}[t]{@{}l@{}}
        \textbullet\ It described the performance and dynamics of the new crypto niche,\\which is characterized by continuous growth driven by gaming, trading,\\and corporate interest.\\
        \textbullet\ Key findings demonstrate that play-to-earn and metaverse tokens are\\associated with positive long-run performance.
    \end{tabular}
    & 
    \begin{tabular}[t]{@{}l@{}}
        \textbullet\ Non-fungible tokens
    \end{tabular}
    & 
    \begin{tabular}[t]{@{}c@{}}
        \textbullet\ Gaming
    \end{tabular}
    \\
    \hline
           \begin{tabular}[t]{@{}l@{}}
~\cite{chao2022study}
    \end{tabular}
    & 
    \begin{tabular}[t]{@{}l@{}}
        \textbullet\ It presented a framework for future metaverse applications composed\\of multiple synchronized data flows from multiple operators through\\multiple wearable devices.\\
        \textbullet\ Based on the ability to customize utility functions for each of\\the individual data flows, a new QoS model is proposed.\\
        \textbullet\ Using a distributed network based on NFTs, the proposed approach\\enables dynamic fine-grained allocation and selection of data flow\\between users and operators.
    \end{tabular}
    & 
    \begin{tabular}[t]{@{}l@{}}
        \textbullet\ Blockchain\\
        \textbullet\ Non-fungible tokens
    \end{tabular}
    & 
    \begin{tabular}[t]{@{}l@{}}
        \textbullet\ Industry 5.0
    \end{tabular}
    \\
        \hline
    \begin{tabular}[t]{@{}l@{}}
~\cite{nguyen2022metachain}
    \end{tabular}
    & 
    \begin{tabular}[t]{@{}l@{}}
        \textbullet\~An overview is provided of the innovation in virtual communities\\related to decentralized financial services on the blockchain. \\
        \textbullet\~Financial supervision issues related to blockchain are within a tolerable\\
        range of risk and affect both the virtual and real words.\\
    \end{tabular}
    & 
    \begin{tabular}[t]{@{}l@{}}
        \textbullet\ Non-fungible tokens\\
        \textbullet\ Decentralized finance\\
        \textbullet\ Decentralized exchange 
    \end{tabular}
    & 
    \begin{tabular}[t]{@{}l@{}}
        \textbullet\ Financial supervision
    \end{tabular}
    \\
    \hline
    \begin{tabular}[t]{@{}c@{}}
~\cite{maksymyuk2022blockchain}
    \end{tabular}
    & 
    \begin{tabular}[t]{@{}l@{}}
        \textbullet\ The proposed study considered the possibility of enhancing the risk\\assessment method by applying the method previously developing\\a multiple risk communicator for metaverse a risk assessment technique.

    \end{tabular}
    & 
    \begin{tabular}[t]{@{}l@{}}
        \textbullet\ Risk assessment technique
    \end{tabular}
    & 
    \begin{tabular}[t]{@{}l@{}}
        \textbullet\ Gaming\\
        \textbullet\ Shopping
    \end{tabular}
    \\
    \hline
    \begin{tabular}[t]{@{}l@{}}
~\cite{vidal2022new}
    \end{tabular}
    & 
    \begin{tabular}[t]{@{}l@{}}
        \textbullet\ It proposed MetaChain, a blockchain-based framework for managing\\metaverse applications efficiently.\\
        \textbullet\ In the metaverse, MetaChain enables smart and trustful between\\metaverse users and service providers.\\
        \textbullet\ A novel sharding scheme can improve MetaChain performance and\\harvest metaverse users resources for metaverse applications.
    \end{tabular}
    & 
    \begin{tabular}[t]{@{}l@{}}
        \textbullet\ Blockchain \\
        \textbullet\ Smart contract \\

    \end{tabular}
    & 
    \begin{tabular}[t]{@{}l@{}}
        \textbullet\ Meeting \\
        \textbullet\ Gaming \\
        \textbullet\ Concert
    \end{tabular}
    \\
    \hline

    \begin{tabular}[t]{@{}l@{}}
~\cite{kuo2023metaverse}
    \end{tabular}
    & 
    \begin{tabular}[t]{@{}l@{}}
        \textbullet\ Using a blockchain-enabled metaverse, it introduced a novel trustless\\architecture aimed at integrating and allocating resources efficiently.\\
        \textbullet\ A metaverse is a weighted hypergraph with local trust.\\
        \textbullet\ The architecture uses a hypergraph to represent a metaverse, in which\\each hyper-edge links a group of users with certain relationship.\\
    \end{tabular}
    & 
    \begin{tabular}[t]{@{}l@{}}
        \textbullet\ Blockchain\\
        \textbullet\ Non-fungible tokens\\
        \textbullet\ Decentralized finance
    \end{tabular}
    & 
    \begin{tabular}[t]{@{}l@{}}
        \textbullet\ Medical infrastructure
    \end{tabular}
    \\
    \hline
        \begin{tabular}[t]{@{}l@{}}
~\cite{gupta2023ddos}
    \end{tabular}
    & 
    \begin{tabular}[t]{@{}l@{}}
        \textbullet\ A method is proposed to detect wormhole attacks in the metaverse,\\which contains physical and virtual information.\\
        \textbullet\ This method is efficient since it requires no additional hardware or\\complex calculations.\\
        \textbullet\ For dynamically analyzing sequential data, a sequential probability\\ratio test method is selected as the fundamental approach.\\
    \end{tabular}
    & 
    \begin{tabular}[t]{@{}l@{}}
        \textbullet\ Wormhole attacks\\detection technique
    \end{tabular}
    & 
    \begin{tabular}[t]{@{}l@{}}
        \textbullet\ Metaverse-related\\IoT application
    \end{tabular}
    \\
    \hline
    \cline{1-4}
\end{tabular}
\end{table*}
\subsection{Summary and Lessons Learned}
In this section, we discuss the metaverse’s key technologies encompassing six distinct domains: interactive experience, communication and networking, ubiquitous computing, digital twin, AI, and cybersecurity.
\begin{enumerate}
    \item Interactive Experience: In building the metaverse, a paramount focus lies on crafting an interactive and immersive user experience. AR, MR, VR, and XR are foundational technologies for creating a dynamic environment where users can engage with digital content seamlessly. These technologies not only enhance the visual and sensory aspects but also contribute to a more immersive and personalized metaverse experience.
    \item Communication and Networking: The backbone of the metaverse is a robust communication and networking infrastructure. The transition to 5G/6G ensures high-speed, low-latency connectivity, while SDN facilitates efficient network management. Effective resource allocation is critical to sustaining seamless interactions within the metaverse, optimizing data transfer, and supporting the diverse communication needs of users and applications.
    \item Ubiquitous Computing: Ubiquitous computing forms the technological underpinning for a scalable and responsive metaverse. Cloud Computing offers the ability to scale resources dynamically, ensuring flexibility and accessibility. Simultaneously, Edge Computing is employed to minimize latency, facilitating real-time interactions. Efficient storage solutions are imperative to manage the vast amounts of data generated within the metaverse environment.
    \item Digital Twin: Digital Twin technologies, encompassing 3D modeling, simulation, and mathematical modeling, play a pivotal role in creating a lifelike representation of the virtual world. These technologies contribute to the dynamic and accurate portrayal of virtual scenarios within the metaverse. From mimicking physical spaces to simulating complex systems, Digital Twin technologies provide the foundation for a realistic and adaptive virtual environment.
    \item Artificial Intelligence: AI is instrumental in enhancing the intelligence and adaptability of the metaverse. Big data and data analytics provide insights for informed decision-making, while optimization techniques enhance system performance. ML, DL, and RL contribute to adaptive functionalities, allowing the metaverse to learn, evolve, and personalize user experiences over time.
    \item Security and Privacy: Security and privacy are paramount considerations in developing a trustworthy metaverse. Blockchain technology, alongside smart contracts and NFTs, ensures secure transactions and ownership within the virtual space. Implementing De-Fi principles and conducting thorough risk assessments are critical components in safeguarding user data and maintaining the integrity of the metaverse ecosystem. Establishing a robust security framework is essential for fostering user trust and confidence in the virtual realm.
\end{enumerate}
In summary, the integration of essential technological prerequisites forms the cornerstone for nurturing a holistic and revolutionary metaverse ecosystem. This fusion not only cultivates a vibrant and secure virtual environment but also guarantees a user-centric experience. Additionally, we summarize the pros and cons of each key technology used in the metaverse in Table~\ref{Tab:Pro}.
\begin{table*}[htbp]
    \centering
    \caption{Summary of Key Technologies Used in the metaverse: Pros and Cons}
    \label{Tab:Pro}
    \begin{tabular}{@{}l|l|l@{}}
        \cline{1-3}
        \textbf{Key Technology} & \textbf{Pros} & \textbf{Cons} \\
        \hline
        Interactive Experience &
        \begin{tabular}[t]{@{}l@{}}
            \textbullet\ An immersive interactive experience enables users to\\deeply engage with their digital avatars and the virtual\\ environment, heightening the overall sense of presence\\and participation.\\
            \textbullet\ An intuitive interactive experience supports effortless\\exploration in the metaverse, enhancing accessible\\interactions and accessibility.\\
            \textbullet\ Using VR technology with headsets and controllers\\lets users fully immerse in the digital environment,\\enhancing natural interaction with avatars by replicating\\real-life movements.
        \end{tabular} &
        \begin{tabular}[t]{@{}l@{}}
            \textbullet\ Relying on VR and AR technologies may\\present barriers for users without access to\\compatible devices or facing challenges\\in adopting and adapting to these technologies.\\
            \textbullet\ The intricacy of immersive interactions and the\\learning curve in navigating virtual environments\\may discourage certain users, especially those\\unfamiliar with VR, AR, or MR technologies.\\
             \textbullet\ In the metaverse, extensive social interactions\\raise privacy concerns as users may be cautious\\about personal data collection and usage.
        \end{tabular} \\
        \hline
        Communication and Networking &
        \begin{tabular}[t]{@{}l@{}}
            \textbullet\ Improved data transfer speeds with 5G and 6G ensure\\seamless and high-speed communication in the metaverse.\\
            \textbullet\ 6G technology contributes to low latency, ensuring near-\\instantaneous communication and interaction, crucial for\\immersive experiences in the metaverse.\\
            \textbullet\ 5G/6G technology provides expanded bandwidth,\\supporting an extensive network of interconnected\\devices and enhancing mobility in the metaverse.\\
            \textbullet\ Metaslicing enables dynamic resource allocation for\\optimal and adaptable metaverse performance.\\
            \textbullet\ SDN enhances security by enabling independent\\management of virtual networks, improving security \\hrough automated and dynamic configurations.
        \end{tabular} &
        \begin{tabular}[t]{@{}l@{}}
            \textbullet\ Excessive bandwidth demands for immersive\\ applications with VR, AR, and MR technologies\\pose challenges for effective communication.\\
            \textbullet\ Implementing AI-driven processing for immersive\\applications introduces complexity and demands\\significant computational resources in the metaverse.\\
            \textbullet\ Creating a flexible and distributed framework for 6G,\\ integrating advancements in edge computing,\\cloud services, and AI paradigms, poses challenges in\\terms of system architecture optimization.\\
            \textbullet\ Social interactions in the metaverse may raise\\privacy concerns as users are cautious about personal\\data collection and use.
        \end{tabular} \\
        \hline
        Ubiquitous Computing &
        \begin{tabular}[t]{@{}l@{}}
            \textbullet\ Ubiquitous computing, especially through cloud solutions,\\provides vital infrastructure support for a vast network\\of virtual environments in the metaverse.\\
            \textbullet\ Cloud computing empowers VR and AR with ample\\processing power and storage for immersive experiences.\\
            \textbullet\ Cloud services enhance metaverse scalability and flexibility,\\accommodating evolving user needs.\\
            \textbullet\ Edge computing is vital for real-time, immersive\\experiences, processing data closer to users to minimize\\latency and enhance performance.
        \end{tabular} &
        \begin{tabular}[t]{@{}l@{}}
            \textbullet\ Cloud integration in the metaverse raises security\\and privacy concerns, demanding robust measures\\to address potential risks.\\
            \textbullet\ Cloud-based solutions may require substantial\\ computational resources, raising environmental\\concerns if not managed sustainably.\\
           \textbullet\ Ubiquitous computing, especially edge computing,\\ relies on internet connectivity, presenting challenges\\for users with limited access to high-speed internet.\\
           \textbullet\ Cloud integration in the metaverse raises security\\and privacy concerns, demanding robust measures to\\mitigate potential risks.
        \end{tabular} \\
        \hline
        Digital Twin &
        \begin{tabular}[t]{@{}l@{}}
            \textbullet\ Digital twins offer a high-fidelity representation, boosting\\accuracy and realism in the virtual world.\\
            \textbullet\ Various modeling techniques, such as 3D modeling,\\simulation, mathematical modeling, and experimental \\modeling, provide versatility for creating and interacting\\with virtual content in the metaverse.\\
            \textbullet\ Mathematical modeling and simulations enhance metaverse\\with realistic physics and natural process simulations,\\fostering a more immersive experience.
        \end{tabular} &
        \begin{tabular}[t]{@{}l@{}}
            \textbullet\ Creating and maintaining detailed digital twins may\\require substantial computational resources, posing\\potential challenges in processing power and storage.\\
            \textbullet\ The effectiveness of digital twins in the metaverse\\depends on seamless internet connectivity, posing\\challenges for users with limited access to high-\\speed internet.\\
            \textbullet\ Ethical concerns may arise with digital twins\\simulating real-world environments, especially\\regarding privacy, data security, and potential\\misuse of user behavior data.
        \end{tabular} \\
        \hline
        Artificial Intelligence &
        \begin{tabular}[t]{@{}l@{}}
            \textbullet\ AI, through optimization algorithms and big data analytics,\\enhances user experiences in the metaverse by processing \\vast data, providing valuable insights for improved UX.\\
            \textbullet\ GAI models in the metaverse automatically generate realistic\\virtual content, saving time and resources while improving\\overall quality.
        \end{tabular} &
        \begin{tabular}[t]{@{}l@{}}
            \textbullet\ AI in the metaverse, particularly in data analytics\\and ML, poses privacy risks with user behavior\\analysis, leading to potential misuse.\\
            \textbullet\ As AI models learn from user/avatar behavior,\\ ethical considerations about fairness, accountability,\\and transparency arise, emphasizing the need for\\ethical AI practices.
        \end{tabular} \\
                \hline
        Cybersecurity &
        \begin{tabular}[t]{@{}l@{}}
            \textbullet\ Cybersecurity is crucial in the metaverse, ensuring trust by\\safeguarding user data, securing transactions, and preserving\\privacy in a space where personal information is openly shared.\\
            \textbullet\ Blockchain in the metaverse ensures a secure, transparent\\ledger, preserving confidentiality, integrity, and availability.\\It acts as a robust barrier against alterations and enables\\secure, trustless transactions among users.\\
            \textbullet\ Cybersecurity safeguards the metaverse using detection\\mechanisms, including AI and behavioral analysis, to identify\\and prevent fraudulent activities through transaction monitoring,\\user verification, and security audits.
        \end{tabular} &
        \begin{tabular}[t]{@{}l@{}}
            \textbullet\ NFTs' popularity in the metaverse raises privacy\\concerns, necessitating user caution in sharing\\information to prevent identity theft and malicious\\activities.\\
            \textbullet\ Integrating De-Fi in the metaverse brings new\\pssibilities and risks, requiring users to adopt\\security measures such as multi-factor authentication\\and encryption for risk mitigation.\\
            \textbullet\ The metaverse faces social engineering threats and\\IT attacks, necessitating techniques like behavioral\\analysis, AI, and transaction monitoring for identifi-\\cationand prevention of malicious activities.
        \end{tabular} \\
        \hline
        \cline{1-3}
    \end{tabular}
\end{table*}
\section{Metaverse Standards}
Designing a metaverse requires careful consideration of both technical and social factors, involving coordination among international standards organizations to establish a globally pervasive, open, and inclusive environment. To translate metaverse concepts into practical applications and stimulate new markets, these organizations must develop a broad range of standards. The metaverse combines the web’s connectivity with advanced technologies like spatial computing, XR, and AI, supported by 5G/6G networks, necessitating novel integrations for interoperability. Standardized access to these technologies facilitates market entry by reducing friction and ensuring universal access. Open standards further enable technology adoption across diverse market segments, fostering healthy competition and broad application. This section provides an overview of the standardization bodies involved in enhancing metaverse interoperability, highlighting the industry's push towards ethical and responsible development to maximize its social and economic potential.

\subsection{IEEE Standards Association}
Whether it is device and technology developers, service providers, metaverse designers, content creators, academics, government agencies, or other stakeholders, the IEEE Standards Association (IEEE SA) is at the forefront of developing a wide range of metaverse-related standards and resources. To define, develop, and deploy technology, applications, and governance practices needed to make metaverse concepts a reality and help drive new markets, the overarching goal is to help turn metaverse concepts into practical realities. The metaverse is, by nature, a converged space that integrates a variety of technologies to create new applications. With its wide and deep range of experience with the necessary technologies, IEEE SA is uniquely positioned to lead the standardization efforts required to meet the evolving needs of new markets and applications. As an illustration, we provide the following examples of standards of metaverse components:

\begin{itemize}
\item \textbf{Taxonomy, terminologies, and definitions in the Metaverse}. 

The IEEE P2048 standard\cite{ieee2048standards} addresses the challenges and ambiguities surrounding the emerging metaverse industry by defining a comprehensive vocabulary, categories, and levels for this evolving concept. Recognizing the infancy of the metaverse and the potential for confusion due to the lack of consensus on fundamental terms, the standard aims to establish a shared language. The standard's development is driven by the need for a common ground that fosters meaningful discussions, encourages sustainable development, and supports healthy market growth within the metaverse ecosystem. By clarifying key aspects such as metaverse categories (Immersive World, Mirror World, and Hybrid World), metaverse levels (ranging from basic virtual environments to seamless integration with the physical world), and additional key definitions (such as Avatar, Virtual Object, metaverse Platform, and metaverse Economy), the IEEE P2048 standard provides a framework that benefits both users and developers. The ongoing evolution of the standard reflects the dynamic nature of the metaverse itself, ensuring relevance and adaptability as the industry progresses.

\vspace{1.0ex}
\item \textbf{AR learning experience model}. In the metaverse, AR is likely to play an important role, and IEEE standards are directly relevant. For example, the IEEE SA board approved the IEEE 1589-2020 standard for AR learning experiences\cite{ieee15892020standards}, which presents an overarching integrated conceptual model, data specifications, and suggestions for how AR-enhanced learning activities can be represented in a standardized interchange format to represent activities, learning contexts, specific environments, and potentially other components. Integrating physical world interactions with web applications, using connected objects and computer vision, lowers entry barriers for users.

\vspace{1.0ex}
\item \textbf{Motion learning framework for MR}. As part of the metaverse, MR can enhance immersion, enable social interaction, and facilitate practical applications. This technology has the potential to revolutionize the way people interact with virtual content and interact with each other in the metaverse, making it an essential part of this ever-changing world of digital content. As an example, IEEE P3079.2\cite{ieee30792standards} defines a framework for MR content aimed at motion learning. The standard includes terms and definitions, requirements, and data formats. It is defined as a mechanism for synchronizing the motion sensor and projector coordinate systems. A description of motion acquisition methods, APIs, and user interfaces is provided.

\vspace{1.0ex}
\item \textbf{DT maturity model and methodology of assessment}. metaverse technology enables the integration of physical objects and environments, enhances functionality and interconnectivity, facilitates virtual commerce and transactions, and facilitates remote collaboration through DT technology. DT has the potential to create a rich, dynamic, and interconnected virtual world within the metaverse, as well as to provide users with a diverse and interactive experience. A proposed IEEE P3144 standard\cite{ieeeP3144standards} defines DT maturity models for industry, including DT capability domains and corresponding subdomains. Additionally, the standard defines assessment methodologies, including assessment content, assessment processes, and assessment maturity levels.

\vspace{1.0ex}
\item \textbf{Human-factors-based DL assessment of visual experience}. Measuring Quality of Experience (QoE) involves examining the human, system, and contextual factors that contribute to the perceptual experience of a user. QoE should be estimated by investigating the mechanism of human visual perception since QoE is a function of human interaction with various devices. QoE measurement remains a challenging endeavor. As of February 2022, IEEE 3333.1.3-2022\cite{ieee3333132022standards} has been approved to categorize factors to be considered when developing DL models for a variety of QoE assessments, as well as providing a reliable subjective test methodology and a process to construct databases.

\vspace{1.0ex}
\item \textbf{Guidelines for metaverse ethics}. To design and utilize the metaverse effectively, ethical guidelines must be established from the outset. The IEEE SA has published a series of white papers\cite{ieeeethical2022} on the impact of XR on user privacy, safety, identity, and inclusion, as well as its ethical implications for education, medicine, finance, and other fields. The IEEE P7016 standard\cite{ieee7016standards} provides an overview of the technological and social aspects of Metaverse systems, as well as a methodology for assessing their ethical implications. Specifically, it guides Metaverse developers to help them prioritize ethically aligned system design; defines ethical system content about accessibility and functional safety; and guidance on how to increase human well-being and environmental sustainability by promoting ethically aligned values and robust public engagement in the research, implementation, and proliferation of metaverse systems.
\end{itemize}

\subsection{ITU-T Focus Group for Metaverse}
A Focus Group on metaverse has been established by the International Telecommunication Union (ITU) in December 2022. The goal of this group is to produce a technical report based on the results of discussions about the analysis of metaverse standardization technology trends and the development of strategies for determining metaverse standardization items using those technologies\cite{hyun2023study}. A focus group is being convened to support pre-standardization activities, including:

\begin{itemize}
\item Building a community of practitioners and experts to unify concepts, and develop common understandings, so that it will benefit not only the ITU industry but also the global community.
\item Identifying and studying the technologies that enable standardization, their evolution, and key tasks, including multimedia, networking optimization, connectivity, interoperability of services and applications, security, protection of personally identifiable information, quality (including bandwidth), digital assets, IoT, accessibility, a DT, and sustainability of the environment.
\item Studying and gathering information to develop a pre-standardization roadmap.
\item Studying terminology, concepts, visions, and ecosystems.
\end{itemize}

This group will also facilitate collaboration between ITU-T and stakeholders and enable non-members to participate in technical pre-standardization work through a collaborative platform. Use cases will be identified to enrich Focus Group work\cite{iutt2023}.

\subsection{3rd Generation Partnership Project}
Since Release 15, the 3rd Generation Partnership Project (3GPP) has been developing and improving its specifications for XR, a core technology for the metaverse. 5G New Radio developed by 3GPP supports emerging XR use cases requiring such Key Performance Indicators (KPI). 5G NR benefits XR, but 5G enhancements and balanced KPIs require more optimization from end to end. XR relies on low latency, high reliability, low power consumption, and high capacity as key service requirements. Releases 15 and 16 provide a solid foundation for XR, but they have not been designed or optimized specifically for XR\cite{salehi2022enhancing}. To enhance the latency and power efficiency of XR devices, 3GPP is developing 5G Advanced on the road to 6G. 
A 5G-enabled system will support end-to-end XR with benefits such as low latency, higher reliability, increased rates, and reduced device computation through a distributed computing architecture. Enhancements in XR, including awareness, power optimizations, and capacity improvements, are achieved from Release 17 to Release 18. Release 19 explores localized mobile metaverse services through 3GPP SA1, involving coordination of input and output data across various devices to support applications\cite{petrov2022extended}. The purpose of this study is to identify specific use cases and service requirements for 5GS support of enhanced XR-based services. metaverse services considered in this study include XR-based services, as well as other functionality to offer shared and interactive UXs of local content and services, either by proximity or remotely. A particular focus will be on the different areas. For example, the sharing of interactive XR media among multiple users in the same location, as well as the acquisition, use, and exposure of local (physical and virtual) information to enable metaverse services.

\subsection{ World Wide Web Consortium for Metaverse}
The World Wide Web Consortium (W3C) is an international organization that develops standards for the World Wide Web. metaverse has become increasingly popular as the web continues to develop\cite{cruz2023towards}. The W3C, as a leader in the development of web technology standards, has been actively involved in discussions and efforts related to the metaverse. The mission of the W3C is to ensure that the web remains an open, accessible, and interoperable platform, which extends to the metaverse as well. Currently, the W3C has three metaverse-related Community Groups (CGs). To ensure interoperability, accessibility, privacy, and security in the metaverse, W3C is developing standards and best practices\cite{ning2021survey}. W3C is focusing on developing standards for VR and AR technologies, which are crucial components of the metaverse. At present, the VR website and metaverse CG, both established in 2015, are almost inactive. Additionally, Galaxy metaverse CG, which was established in January 2022, plans to discuss virtual world infrastructure, virtual land management, marketing, avatar communication, and goods. Although it is difficult to see that the three W3C CGs are actively standardizing at the moment, it is expected to become one of the most powerful standardization organizations when the web-based metaverse becomes fully functional\cite{hyun2023study}. Apart from the VR website, the W3C also develops standards for 3D graphics, spatial audio, and social interactions within virtual worlds. These standards aim to provide a consistent and interoperable framework for developers so that they can create metaverse experiences that can be accessed and experienced seamlessly across a wide range of platforms and devices\cite{smart2007cross}.

\subsection{International Electrotechnical Commission on Metaverse}
A variety of products and services are being offered by the industry in the metaverse, but progress has been slow as different technical solutions and products are not interoperable. There are also ethical and governance issues to be addressed, as risk management, safety, and data protection issues are dealt with. Furthermore, there is a lack of a common vocabulary which hinders the exchange of information. All of these areas could benefit from international standards. The specific technologies used in the metaverse are already subject to international standards designed to ensure their trustworthiness and safety\cite{iecseg2023}. For instance, AI/ML, holography, the IoT, and VR are among these technologies. The metaverse as a whole does not yet have global standards that address it systematically. To address the needs of the metaverse, the International Electrotechnical Commission (IEC) board established a Standardization Evaluation Group (SEG), namely SEG 15 in November 2022. The IEC SEG 15 will develop a common understanding and definition of the metaverse, investigate the need for standardization, and provide recommendations for an initial roadmap for the metaverse. In addition, it will engage with technical committees of the International Organization for Standardization (ISO) and IEC as well as other relevant organizations\cite{narang2023mentor}.

 \subsection{Moving Picture Experts Group for Metaverse}
 As the first standardized framework for networked virtual environments within the metaverse, Moving Picture Experts Group (MPEG-V) provides ISO/IEC 23005, which enables seamless data exchange, simultaneous reactions, and interoperability across virtual worlds\cite{timmerer2009interfacing}. ISO/IEC 23005 is designed to standardize interfaces between the virtual world (simulation, gaming, etc.) and real-world (sensors, actuators, connected objects, etc.), and among virtual worlds. As of 2020, the 4th edition of this standard has been released since its first edition in 2011. Standard ISO/IEC 23005 can be used for a range of business services related to the metaverse, such as avatars and virtual items, where the association of audiovisual data and rendered sensory effects can enhance interaction between the virtual world and real world\cite{osistandard2023,bricken1991virtual}.

 \subsection{Metaverse Standards Forum}
 In June 2022, the metaverse Standards Forum (MSF) was established; it brought together leading standards organizations and companies to develop interoperability standards for the open metaverse\cite{metaverseforum2022}. A key objective of the MSF is to identify where interoperability is hindering the deployment of metaverse and how standards-development organizations can coordinate and accelerate their work to define and evolve the standards required. In addition to accelerating the testing and adoption of metaverse standards, the MSF will develop consistent terminology and deployment guidelines to facilitate implementation prototyping, hackathons, plugfests, and open-source tools, which are open to all organizations at no cost. To identify critical metaverse standards requirements, and to demonstrate real-world interoperability through plugfest projects, the MSF provides a venue for discussion and coordination between standards organizations and companies building metaverse-related products. There have already been more than 1800 members of the MSF since it began with just 35 founding members.
 
\subsection{Summary and Lessons Learned}
The exploration of metaverse Standards within this section highlights the integral contributions of key organizations, each playing a pivotal role in shaping standards for diverse aspects of the metaverse. The IEEE SA takes a central position, addressing taxonomy, terminologies, AR learning models, Motion learning for MR, DT maturity models, Human-factors-based DL assessments, and metaverse ethics guidelines. Simultaneously, the ITU-T Focus Group for metaverse, 3GPP, W3C, IEC SEG, and MPEG-V contribute their specialized expertise, collectively forming a comprehensive framework for metaverse standards. Based on this investigation of the metaverse standard, we draw several conclusions, as outlined below:

\begin{itemize}
    \item Diverse Expertise: The involvement of multiple organizations underscores the need for a diverse range of expertise to comprehensively address the complex aspects of metaverse development and standardization.
    \item Global Perspective: The participation of international bodies such as ITU-T and 3GPP emphasizes the global perspective necessary for establishing universal metaverse standards, ensuring inclusivity and interoperability on a global scale.
    \item User-Centric Design: The inclusion of metaverse ethics guidelines highlights a commitment to user-centric design principles, prioritizing responsible and ethical practices within virtual environments for a positive user experience.
    \item Specialized Contributions: Each organization brings specialized knowledge, contributing to specific elements crucial to the development and standardization of the metaverse, ensuring a comprehensive approach.
    \item Foundations for Future Cooperation: While explicit collaboration details may not be outlined, the combined efforts of these organizations establish the groundwork for potential future cooperation, fostering a more unified and cohesive approach to metaverse standards development.
\end{itemize}

The synthesis from these organizations is pivotal in establishing robust standards that promote a user-friendly, ethical, and globally interoperable metaverse ecosystem.
\section{Metaverse Application Aspects}
\begin{figure*}[htbp]
\centering
 \includegraphics[width=0.99\linewidth]{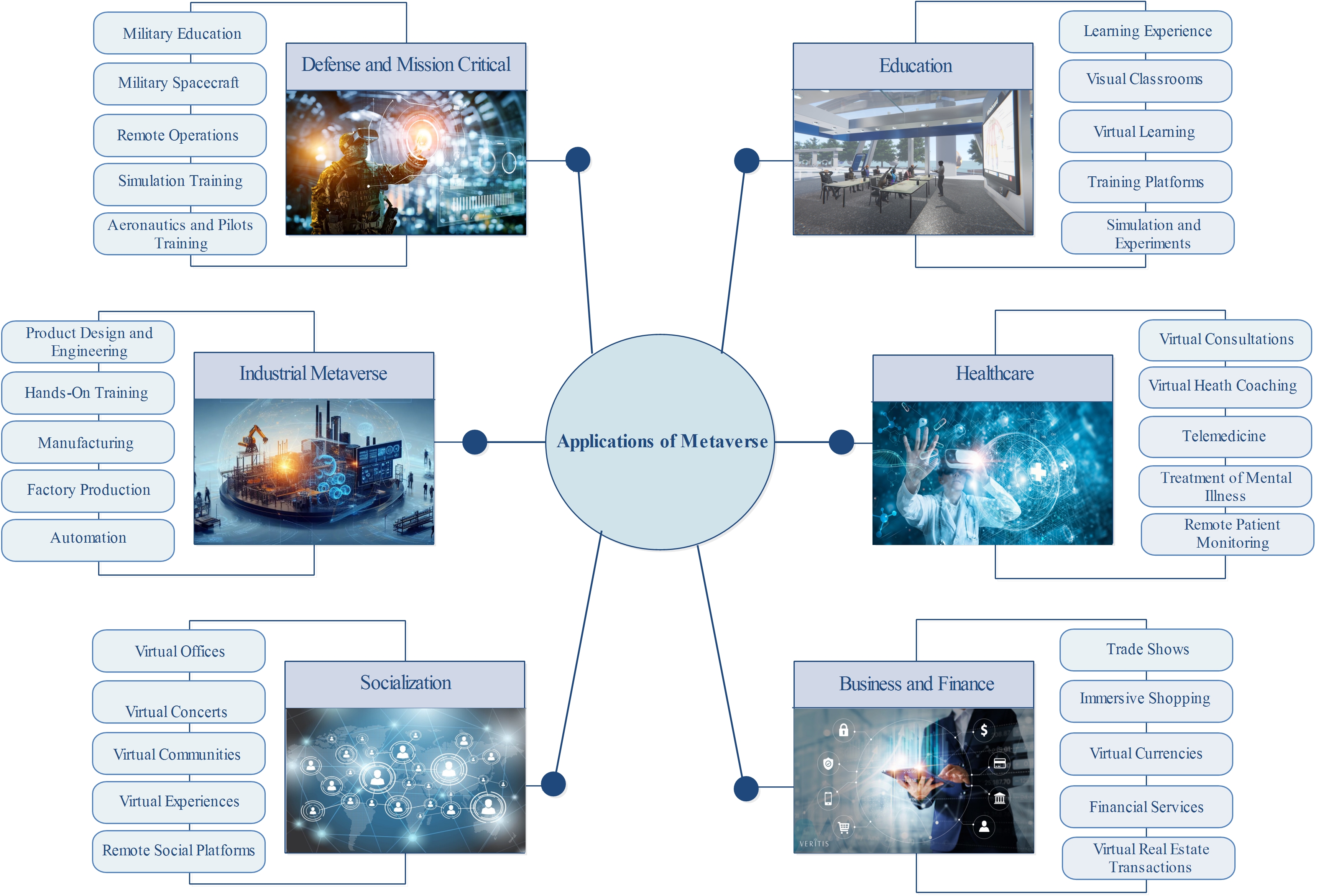}
 \caption{The typical applications of the metaverse include education, healthcare, social, business, finance, industrial metaverse, defense, and mission-critical.}
 \label{Fig:App}
\end{figure*}

As digital technology continues to develop, humans may eventually migrate from the real world to the metaverse. metaverse applications will continue to grow and expand as they evolve until they can be integrated into all aspects of people's lives. In this section, we will discuss five typical applications of the metaverse shown in Figure~\ref{Fig:App}.

\subsection{Education}
Recently, several education methods have shifted from offline to online in response to the COVID-19 epidemic. Through the use of metaverse video in education, more realistic scenes can be presented in an immersive manner, thus making the learning experience more engaging and stimulating. In the metaverse, teachers and coaches can interact with students in the same virtual space without the restrictions of physical distance\cite{abraham2023study}. As part of the teaching process, teachers and students will provide data to improve the quality of the class and learning efficiency. Using virtual classrooms would allow students and teachers to interact in 3D virtual spaces for lectures, discussions, and collaborative activities, regardless of their physical location\cite{suzuki2020virtual}. Simulations and experiments in a controlled virtual environment could provide opportunities for hands-on learning. Through virtual learning, students can perform experiments, analyze data, and learn scientific concepts, increasing their comprehension of complex topics. Also, the metaverse could facilitate virtual field trips where students could explore historical sites, natural wonders, and cultural landmarks, thereby expanding their knowledge of the global environment\cite{buhalis2023metaverse}. As part of the metaverse, students could also access customized educational content and resources tailored to their needs\cite{zhou2022building}. Moreover, the metaverse could be used to foster social interaction and community building among students and create virtual spaces in which students can collaborate and support one another. For the metaverse to be used in education equitably, challenges such as accessibility, privacy, and digital literacy must be addressed. The metaverse has the potential to revolutionize education by providing immersive, interactive, and personalized learning experiences that can enhance student engagement, foster creativity, and expand educational opportunities beyond the physical environment\cite{beck2023educational}. We discussed various works on metaverse applications within the education sector, as outlined in Table~\ref{Tab:Edu}.

\begin{table*}[htbp]
  \centering
\caption{Metaverse Applications in Education}
\label{Tab:Edu}
 \begin{tabular}{@{}l|l|l|l|l@{}}
    \cline{1-5}
    \textbf{Existing works} & \textbf{Summary of Contributions} & \textbf{Key Technologies} & \textbf{Challenges} & \textbf{Recommendations and Future Works}\\
    \hline
    \begin{tabular}[t]{@{}l@{}}

~\cite{yu2022exploration,kye2021educational}
    \end{tabular}
    & 
    \begin{tabular}[t]{@{}l@{}}
        \textbullet\ Exploring the educational opportunities\\provided by four different types of\\metaverse within the realm of\\physical education. \\
        \textbullet\ AR, lifelogging, mirror worlds,\\and virtual worlds are the four\\categories under consideration.
        
    \end{tabular}
        & 
    \begin{tabular}[t]{@{}l@{}}
        \textbullet\ VR\\
        \textbullet\ AR\\
        \textbullet\ IoT
    \end{tabular}
    & 
    \begin{tabular}[t]{@{}l@{}}
        \textbullet\ Privacy risks \\
        \textbullet\ Technologies\\implementation \\
        \textbullet\ Addiction\\
        \textbullet\ Governance challenges\\
        \textbullet\ Inclusiveness
    \end{tabular}
    & 
    \begin{tabular}[t]{@{}l@{}}
        \textbullet\ The integration of the metaverse\\into university physical education\\will enhance educational\\experience by incorporating\\various sports.\\
        \textbullet\ Propose varied developmental\\paths for metaverse-based physical\\education strategies. 
    \end{tabular}
    \\
    \hline
    \begin{tabular}[t]{@{}c@{}}
~\cite{hwang2022definition}
    \end{tabular}
    & 
    \begin{tabular}[t]{@{}l@{}}
        \textbullet\ Define the metaverse in education,\\outline its roles,\\and identify potential research areas.\\
        \textbullet\ The function of AI within metaverse.
    \end{tabular}
      & 
    \begin{tabular}[t]{@{}l@{}}
        \textbullet\ IA\\
        \textbullet\ AR\\
        \textbullet\ VR
    \end{tabular}
    & 
    \begin{tabular}[t]{@{}l@{}}
        \textbullet\ Connecting the metaverse\\
        \textbullet\ Technologies \\implementation\\
        \textbullet\ Ethical issues
    \end{tabular}
    & 
    \begin{tabular}[t]{@{}l@{}}
        \textbullet\ Exploring prospective learning\\and curriculum designs.\\
        \textbullet\ Assessing metaverse learning\\performance and perception.\\
        \textbullet\ Metaverse-based learning strategies.

    \end{tabular}
    \\
    \hline
    \begin{tabular}[t]{@{}l@{}}
~\cite{hare2022hierarchical}
     \end{tabular}
    & 
    \begin{tabular}[t]{@{}l@{}}
        \textbullet\ Develop a metaverse learning framework,\\emphasizing the traits of AI-based non-\\player entities.\\
        \textbullet\ The issues of state explosion and\\convergence time are mitigated through \\the implementation of an innovative\\hierarchical RL approach.
    \end{tabular}
       & 
    \begin{tabular}[t]{@{}l@{}}
        \textbullet\ RL\\
        \textbullet\ Optimization\\
    \end{tabular}
    &
    
    \begin{tabular}[t]{@{}l@{}}
        \textbullet\ Metaverse learning\\development\\
        \textbullet\ Experiments development 
    \end{tabular}
    & 
    \begin{tabular}[t]{@{}l@{}}
        \textbullet\ Creating additional metaverse\\learning applications utilizing\\the suggested approach.\\
        \textbullet\ Examining the fundamentals of\\deduction and induction to enhance\\computational experiments.
    \end{tabular}
    \\
    \hline
    \begin{tabular}[t]{@{}l@{}}
~\cite{wang2022constructing}
    \end{tabular}
    & 
    \begin{tabular}[t]{@{}l@{}}
        \textbullet\ A novel educational metaverse ecosystem\\is suggested by reviewing literature and\\amalgamating best practices in the design of\\ metaverse learning framework.\\
        \textbullet\ The framework comprises four components:\\instructional design, performance technology,\\knowledge, research, and talent training.
    \end{tabular}
       & 
    \begin{tabular}[t]{@{}l@{}}
        \textbullet\ AI\\
        \textbullet\ AR\\
        \textbullet\ XR\\
        \textbullet\ Communication\\
         and networking
    \end{tabular}
    &
    \begin{tabular}[t]{@{}l@{}}
        \textbullet\ Technologies\\implementation\\
        \textbullet\ Metaverse learning\\development\\
        \textbullet\ Security and privacy
    \end{tabular}
    & 
    \begin{tabular}[t]{@{}l@{}}
        \textbullet\ Testing and refining are necessary\\to align the ecosystem with real-\\world practices.\\
        \textbullet\ The proposed ecosystem prioritizes\\learner and teacher support, community\\collaboration, and post-graduation\\assistance.
    \end{tabular}
    \\
    \hline
    \begin{tabular}[t]{@{}l@{}}
~\cite{gim2022metaverse}
    \end{tabular}
    & 
    \begin{tabular}[t]{@{}l@{}}
        \textbullet\ Explore the connection between metaverse\\learning, self-determination, and learner\\satisfaction.\\
        \textbullet\ Analyze learner satisfaction by integrating\\technology acceptance, information systems\\success, and self-determination theory.
    \end{tabular}
        & 
    \begin{tabular}[t]{@{}l@{}}
        \textbullet\ VR\\
        \textbullet\ AR\\
        \textbullet\ Data analysis
    \end{tabular}
    & 
    \begin{tabular}[t]{@{}l@{}}
        \textbullet\ The lack of popularity\\of VR-based content\\
        \textbullet\ Metaverse learning\\development
    \end{tabular}
    & 
    \begin{tabular}[t]{@{}l@{}}
        \textbullet\ Explore the educational impact of\\VR-based learning beyond satisfaction.\\
        \textbullet\ Create a detailed framework for future\\research on assessing learners' academic\\performance.
    \end{tabular}
    \\
    \hline
    \cline{1-4}
\end{tabular}
\end{table*}

\subsection{Healthcare}
Applied to healthcare, the metaverse has the potential to improve patient outcomes and revolutionize healthcare delivery. The metaverse offers enhanced accessibility and convenience to patients, one of the key benefits of its use in healthcare. Telemedicine and remote patient monitoring provide medical care to patients who live in remote or under-served areas, eliminating the need for them to travel long distances for care. Furthermore, patients with disabilities or mobility issues can receive care from the comfort of their homes, reducing the burden on their caregivers. Moreover, the metaverse transcends physical space, and its virtual, real-time nature, and stability facilitate the practice of telemedicine. Digital modeling of the patient's condition in the virtual world can assist doctors in better understanding the patient's condition\cite{mozumder2022overview,chengoden2023metaverse}. By tricking the brain into a virtual experience, the metaverse may also be used to treat mental illness, allowing patients to remain happy or erase as many bad memories as possible. The use of 3D visualizations of organs in medical education allows students to gain a deeper understanding of the body structure as well as acquire systematic knowledge\cite{skalidis2022cardioverse}. Additionally, the metaverse can provide opportunities for patient education and engagement. Virtual environments can simulate medical conditions and procedures, allowing patients to understand their health issues better and make informed decisions about their care. The adoption of the metaverse in healthcare has the potential to enhance efficiency and reduce costs. By facilitating virtual consultations and procedures, it minimizes the dependence on physical infrastructure, equipment, and personnel, thereby lowering healthcare expenses. This can improve the accessibility and affordability of healthcare, especially for uninsured or underinsured patients\cite{wang2022development}. Additionally, virtual health coaching and support groups can offer valuable community and emotional support to patients\cite{cai2022co}. Table~\ref{Tab:HC} presents the current works on metaverse applications in healthcare.

\begin{table*}[htbp]
  \centering
\caption{The use of the metaverse for Healthcare}
\label{Tab:HC}
  \begin{tabular}{@{}l|l|l|l|l@{}}
    \cline{1-5}
    \textbf{Existing works} & \textbf{Summary of Contributions} & \textbf{Key Technologies} & \textbf{Challenges} & \textbf{Recommendations and Future Works}\\
    \hline
    \begin{tabular}[t]{@{}l@{}}
    
~\cite{chengoden2023metaverse}
    \end{tabular}
    & 
    \begin{tabular}[t]{@{}l@{}}
        \textbullet\ Comprehensive review of the metaverse\\in healthcare, covering the current state,\\enabling technologies and applications. \\
        \textbullet\ Future research directions involve\\identifying and addressing issues related\\to adapting the metaverse for healthcare.
        
    \end{tabular}
            & 
    \begin{tabular}[t]{@{}l@{}}
        \textbullet\ Degital twin\\
        \textbullet\ cybersecurity\\
        \textbullet\ AI\\
        \textbullet\ IoT\\
        \textbullet\ VR\\
        \textbullet\ AR
        
    \end{tabular}
    & 
    \begin{tabular}[t]{@{}l@{}}
        \textbullet\ Privacy and security\\
        \textbullet\ Interoperability issues\\
        \textbullet\ High technology costs\\
        \textbullet\ Legal and regulatory
    \end{tabular}
    & 
    \begin{tabular}[t]{@{}l@{}}
        \textbullet\ A metaverse for healthcare with\\enhanced security features\\
        \textbullet\ A synced healthcare metaverse\\on the cloud edge. \\
        \textbullet\ A metaverse for healthcare\\designed to be energy-efficient 
    \end{tabular}
    \\
    \hline
    \begin{tabular}[t]{@{}c@{}}
~\cite{kashif2023survey}
    \end{tabular}
    & 
    \begin{tabular}[t]{@{}l@{}}
        \textbullet\ Distinguishing traditional healthcare from\\the healthcare metaverse.\\
        \textbullet\ FL applications for healthcare metaverse.
    \end{tabular}
            & 
    \begin{tabular}[t]{@{}l@{}}
        \textbullet\ Digital twin\\
        \textbullet\ VR/XR\\
        \textbullet\ IoT\\
        \textbullet\ AI\\
        \textbullet\ FL
    \end{tabular}
    & 
    \begin{tabular}[t]{@{}l@{}}
        \textbullet\ Sync between physical\\and virtual worlds\\
        \textbullet\ Security and privacy\\
        \textbullet\ Lack of explainability\\
        \textbullet\ Fair FL models
    \end{tabular}
    & 
    \begin{tabular}[t]{@{}l@{}}
        \textbullet\ Boost research focus for novel FL\\-enabled metaverse healthcare ideas.\\
        \textbullet\ Discover novel directions and\\propose solutions for\\the discussed challenges.

    \end{tabular}
    \\
    \hline
    \begin{tabular}[t]{@{}l@{}}
~\cite{bansal2022healthcare}
     \end{tabular}
    & 
    \begin{tabular}[t]{@{}l@{}}
        \textbullet\ Current healthcare metaverse application,\\ assessing impact on practices.\\
        \textbullet\ Provides an overview of technical\\challenges and solutions for a self-\\sustaining, persistent, and future-proof\\solution.
    \end{tabular}
            & 
    \begin{tabular}[t]{@{}l@{}}
        \textbullet\ VR\\
        \textbullet\ AR\\
        \textbullet\ IoT\\
        \textbullet\ AI\\
        \textbullet\ Blockchain\\
        \textbullet\ 6G\\
        \textbullet\ Networking
    \end{tabular}
    & 
    \begin{tabular}[t]{@{}l@{}}
        \textbullet\ Hardware, security,\\and privacy\\
        \textbullet\ Identity hacking\\
        \textbullet\ Mental health\\
        \textbullet\ Currency and\\transactions\\
        \textbullet\ Law and jurisdiction
    \end{tabular}
    & 
    \begin{tabular}[t]{@{}l@{}}
        \textbullet\ A comprehensive approach to\\building a medical metaverse.\\
        \textbullet\ Analyze existing healthcare services\\spanning diverse technologies and\\eco-systems to facilitate a more\\expansive discourse. \\
        \textbullet\ Identify pivotal subjects to influence\\the future trajectory of the metaverse.
    \end{tabular}
    \\
    \hline
    \begin{tabular}[t]{@{}l@{}}
~\cite{bhattacharya2022metaverse}
    \end{tabular}
    & 
    \begin{tabular}[t]{@{}l@{}}
        \textbullet\ Examine patient, virtual hospitals,\\and doctor interaction with blockchain\\-based tele-surgical and XAI. \\
        \textbullet\ Showcase benefits of proposed\\architecture over traditional tele-surgery\\experimentally.
    \end{tabular}
            & 
    \begin{tabular}[t]{@{}l@{}}
        \textbullet\ VR/AR\\
        \textbullet\ Blockchain\\
        \textbullet\ AI/DL\\
        \textbullet\ XAI\\
        \end{tabular}
    & 
    \begin{tabular}[t]{@{}l@{}}
        \textbullet\ Technologies\\implementation\\
        \textbullet\ Security and privacy\\
        \textbullet\ Real-time operation\\
        \textbullet\ Accuracy of predictions
    \end{tabular}
    & 
    \begin{tabular}[t]{@{}l@{}}
        \textbullet\ Explore possible resolutions to\\the mentioned challenges.\\
        \textbullet\ Analyze various XAI models and\\predict the optimal use case module\\for telesurgery through metaverse\\setups. 
    \end{tabular}
    \\
    \hline
    \begin{tabular}[t]{@{}l@{}}
~\cite{musamih2022metaverse}
    \end{tabular}
    & 
    \begin{tabular}[t]{@{}l@{}}
        \textbullet\ Demonstrate the applications of the\\metaverse in healthcare.\\
        \textbullet\ Summarize the metaverse, including\\its essential features and enabling\\technologies.\\ 
        \textbullet\ Integrating the metaverse into healthcare\\needs specific components and requirements.
        
    \end{tabular}
            & 
    \begin{tabular}[t]{@{}l@{}}
        \textbullet\ VR/AR/XR\\
        \textbullet\ Cybersecurity\\
        \textbullet\ Cloud/Edge\\
        \textbullet\ IoT\\
        \textbullet\ Communication\\ and networking
    \end{tabular}
    & 
    \begin{tabular}[t]{@{}l@{}}
        \textbullet\ Technological limitations\\
        \textbullet\ Security and privacy\\
        \textbullet\ Socio-ethical\\implications\\
        \textbullet\ Enforcement of laws\\and regulations
    \end{tabular}
    & 
    \begin{tabular}[t]{@{}l@{}}
        \textbullet\ Developing foundational techniques\\to empower the metaverse.\\
        \textbullet\ Develop a strategy for achieving a\\balance between virtual and real-life\\experiences.\\
        \textbullet\ For legitimacy, it must align with\\established regulations, standards,\\and ethical codes.
    \end{tabular}
    \\
    \hline
    \cline{1-4}
\end{tabular}
\end{table*}
\subsection{Business and Finance}
The metaverse presents businesses with numerous opportunities to revolutionize customer engagement through innovative means. Beyond offering a fully immersive environment for product interaction, one of its pivotal applications is the creation of virtual showrooms, particularly beneficial for businesses showcasing large or challenging-to-display items\cite{hopkins2022virtual}. Virtual events such as product launches, conferences, and trade shows find a dynamic platform in the metaverse, enabling real-time interactions between attendees and company representatives, thereby enhancing customer engagement\cite{tahira2023business}. The consumer metaverse serves individual users, providing a space for entertainment, social interaction, and personalized experiences. On the other hand, the enterprise metaverse focuses on business applications, offering tools and environments for collaboration, innovation, and productivity. Businesses navigate these realms strategically, leveraging the consumer, industrial, and enterprise metaverses to shape a future where the metaverse becomes an integral part of daily life and business strategies. This interconnected ecosystem fosters a dynamic digital landscape where individual experiences, industrial processes, and corporate operations converge. The metaverse transforms the retail and e-commerce landscape by providing users with an immersive shopping experience in virtual worlds\cite{jeong2022innovative}. Users can employ digital avatars to try out products, gaining familiarity with specific features. This immersive approach not only facilitates timely and valuable feedback on product improvements but also redefines marketing and advertising methodologies. Merchants can sell and promote virtual counterparts of real goods, exemplified by augmented reality experiences on social media platforms. Moreover, the metaverse extends its influence into the realm of employee collaboration, fostering virtual work environments that may boost productivity and reduce the dependency on physical office spaces\cite{enache2022metaverse}. Simultaneously, the metaverse holds immense potential for transforming the financial sector by introducing novel products, services, and business models\cite{mozumder2023metaverse}. Decentralized Finance (De-Fi) gains a more immersive and engaging environment through metaverse integration, offering users a unique space to access and utilize financial services. The metaverse also opens avenues for the creation of virtual digital currencies backed by real-world assets, traded within or between them\cite{adams2022virtual}. Furthermore, the metaverse can facilitate cross-border transactions, promising faster, cheaper, and more secure transactions compared to existing methods\cite{amalia2023legal}. Table~\ref{Tab:Business} outlines various applications of the metaverse in the domains of business and finance.

\begin{table*}[htbp]
  \centering
\caption{Leveraging the metaverse for business and financial applications}
\label{Tab:Business}
    \begin{tabular}{@{}l|l|l|l|l@{}}
    \cline{1-5}
    \textbf{Existing works} & \textbf{Summary of Contributions} & \textbf{Key Technologies} & \textbf{Challenges} & \textbf{Recommendations and Future Works}\\
    \hline
    \begin{tabular}[t]{@{}l@{}}
~\cite{anshari2022ethical}
    \end{tabular}
    & 
    \begin{tabular}[t]{@{}l@{}}
        \textbullet\ Explore metaverse business ethics\\and promote ethical responsibility\\and sustainability.\\
        \textbullet\ Propose transparent policies for\\businesses' metaverse applications\\to foster an ethical culture.
        
    \end{tabular}
                & 
    \begin{tabular}[t]{@{}l@{}}
        \textbullet\ Optimization\\
        \textbullet\ cybersecurity\\
        \textbullet\ Big data
    \end{tabular}
    & 
    \begin{tabular}[t]{@{}l@{}}
        \textbullet\ Security and privacy\\
        \textbullet\ Customer profiling\\
        \textbullet\ Agressive marketing\\
        \textbullet\ Digital personality\\mining
    \end{tabular}
    & 
    \begin{tabular}[t]{@{}l@{}}
        \textbullet\ The suggested framework will undergo\\testing through the collection of primary\\data.\\
        
        \textbullet\ Explore potential solutions to the raised\\concerns.
    \end{tabular}
    \\
    \hline
    \begin{tabular}[t]{@{}l@{}}
~\cite{gao2022research}
     \end{tabular}
    & 
    \begin{tabular}[t]{@{}l@{}}
        \textbullet\ Explore the innovation of the IoT\\business model within the new\\context of the metaverse.\\
        \textbullet\ Examine the correlation between\\the IoT business and the metaverse.
    \end{tabular}
                & 
    \begin{tabular}[t]{@{}l@{}}
        \textbullet\ IoT\\
        \textbullet\ Blockchain\\
        \textbullet\ AI\\
    \end{tabular}
    & 
    \begin{tabular}[t]{@{}l@{}}
        \textbullet\ Marketing plan\\
        \textbullet\ Personalized, diversified,\\demanding consumer needs
    \end{tabular}
    & 
    \begin{tabular}[t]{@{}l@{}}
        \textbullet\ Improve the accuracy of\\the marketing plan.\\
        \textbullet\ Marketing initiatives should be\\centered around creating value.\\
        \textbullet\ Marketing services need to provide\\more timely responses.\\
        \textbullet\ Decentralization of marketing activities\\is necessary.
    \end{tabular}
    \\
    \hline
    \begin{tabular}[t]{@{}l@{}}
~\cite{chen2022digital}
    \end{tabular}
    & 
    \begin{tabular}[t]{@{}l@{}}
        \textbullet\ Digital economy applications and\\challenges intersect with metaverse.\\
        \textbullet\ Integrate ML, finance, and data\\management advancements in the\\metaverse for enhanced problem\\-solving in the digital economy.
    \end{tabular}
                & 
    \begin{tabular}[t]{@{}l@{}}
        \textbullet\ Blockchain\\
        \textbullet\ De-Fi\\
        \textbullet\ NFT\\
        \textbullet\ AI\\
        \textbullet\ FL\\
    \end{tabular}
    & 
    \begin{tabular}[t]{@{}l@{}}
        \textbullet\ Digital industrialization\\
        \textbullet\ Digital governance\\
        \textbullet\ Data analysis
    \end{tabular}
    & 
    \begin{tabular}[t]{@{}l@{}}
        \textbullet\ Identify potential solutions.\\
        \textbullet\ Developing the ecosystem of the digital\\economy within the context of metaverse. 
    \end{tabular}
    \\
    \hline
    \begin{tabular}[t]{@{}l@{}}
~\cite{wu2023financial}
    \end{tabular}
    & 
    \begin{tabular}[t]{@{}l@{}}
        \textbullet\ Outline the background, foundation,\\and applications of metaverse.\\
        \textbullet\ Examine security risks and financial\\crimes linked to the evolution of the\\decentralized metaverse. 
        
    \end{tabular}
                & 
    \begin{tabular}[t]{@{}l@{}}
        \textbullet\ Data analysis\\
        \textbullet\ Blockchain\\
        \textbullet\ De-Fi\\
        \textbullet\ NFT
    \end{tabular}
    & 
    \begin{tabular}[t]{@{}l@{}}
        \textbullet\ Blockchain challenges\\
        \textbullet\ Smart contract issues\\
        \textbullet\ De-Fi issues
    \end{tabular}
    & 
    \begin{tabular}[t]{@{}l@{}}
        \textbullet\ Analyze cases and apply relevant\\statistical and analytical techniques.\\
        \textbullet\ Examine financial crimes within the\\metaverse employing cutting-edge\\and tools, including AI.\\
        \textbullet\ Conduct research, analysis, and provide\\metaverse development recommendations\\leveraging cross-disciplinary expertise.
    \end{tabular}
    \\
    \hline
    \cline{1-4}
\end{tabular}
\end{table*}
\subsection{Socialization}
Metaverse technology opens new possibilities in human social forms\cite{oh2023social}. It offers a range of potential applications in socialization, providing users with immersive and engaging virtual environments to connect and interact with others. Various forms of social interaction can be provided by the metaverse, which can break down the barriers of time and space\cite{kim2021study}. Virtual offices, virtual dating, and virtual gatherings are examples of higher-level needs that can be met outside of the physical world. Due to the impact of the COVID-19 pandemic, remote social networking has become increasingly important\cite{lyttelton2022telecommuting}. Using the metaverse, remote social platforms can be enhanced further to provide a more authentic social environment, compensating for the limitations of conventional models. The metaverse can be used to facilitate socialization through the creation of virtual communities, where individuals can connect with like-minded individuals and participate in events and activities online\cite{simpson2022liveness}. The metaverse can also offers virtual experiences that simulate real-life social interactions, such as virtual parties and concerts, creating a more immersive and interactive experience\cite{zhao2022metaverse}. Furthermore, the metaverse could offer a platform for various forms of social media. These forms allow users to share content and interact with others in an immersive and interactive manner. Examples include creating and sharing VR videos, or participating in VR live streams. With the metaverse's continued evolution, it has the potential to revolutionize socialization and create new ways to interact and connect. Table~\ref{Tab:Social} highlights various metaverse applications within the socialization domain.
\begin{table*}[htbp]
  \centering
\caption{The utilization of the metaverse for socializing}
\label{Tab:Social}
    \begin{tabular}{@{}l|l|l|l|l@{}}
    \cline{1-5}
    \textbf{Existing works} & \textbf{Summary of Contributions} & \textbf{Key Technologies} & \textbf{Challenges} & \textbf{Recommendations and Future Works}\\
    \hline
    \begin{tabular}[t]{@{}l@{}}
~\cite{jiaxin2022socializing}
    \end{tabular}
    & 
    \begin{tabular}[t]{@{}l@{}}
        \textbullet\ Metaverse communication innovation\\and challenges.\\
        \textbullet\ Analyze the metaverse concept,\\historical context,and current social\\interactions to offer insights for social\\development.
        
    \end{tabular}
                & 
    \begin{tabular}[t]{@{}l@{}}
        \textbullet\ VR\\
        \textbullet\ AR\\
        \textbullet\ 5G\\
        \textbullet\ NLP\\
    \end{tabular}
    & 
    \begin{tabular}[t]{@{}l@{}}
        \textbullet\ Ethical dimension\\
        \textbullet\ Technologies\\implementation
        \end{tabular}
    & 
    \begin{tabular}[t]{@{}l@{}}
        \textbullet\ Analyze the challenges and social\\impacts that social interactions in\\the metaverse will face, providing\\additional reflections.\\
        \textbullet\ Consider potential solutions to\\the challenges that have been raised.
    \end{tabular}
    \\
    \hline
    \begin{tabular}[t]{@{}c@{}}
~\cite{sykownik2022something}
    \end{tabular}
    & 
    \begin{tabular}[t]{@{}l@{}}
        \textbullet\ An investigation into the objectives,\\subjects, and contextual factors of self\\-disclosure in commercial social VR.\\
        \textbullet\ Presenting an online survey to\\investigate how users disclose personal\\information within the social VR.\\
        \textbullet\ Explore the role of self-disclosure in\\sustaining interpersonal relationships in\\commercial social VR.
    \end{tabular}
                & 
    \begin{tabular}[t]{@{}l@{}}
        \textbullet\ VR\\
        \textbullet\ AR\\
        \textbullet\ Data analysis\\
        \textbullet\ Security\\
        \textbullet\ Privacy
    \end{tabular}
    & 
    \begin{tabular}[t]{@{}l@{}}
        \textbullet\ The sample is uniform\\in demographics\\
        \textbullet\ Social VR users defy\\definitive categorization\\
        \textbullet\ The work lacks a full\\grasp of self-disclosure\\in social VR
    \end{tabular}
    & 
    \begin{tabular}[t]{@{}l@{}}
        \textbullet\ Demographic details regarding social\\VR users.\\
        \textbullet\ Assess extensive samples of users\\engaged in social VR.\\
        \textbullet\ Examining the impact of contextual\\factors on self-disclosure decisions\\within specific topic areas.\\
        \textbullet\ Evaluate how user characteristics\\contribute to predicting self-disclosure\\in social VR.

    \end{tabular}
    \\
    \hline
    \begin{tabular}[t]{@{}l@{}}
~\cite{liang2023metaverse}
     \end{tabular}
    & 
    \begin{tabular}[t]{@{}l@{}}
        \textbullet\ Establishing a metaverse social center\\to facilitate communication among the\\elderly during social isolation.\\
        \textbullet\ Develop a framework for establishing\\a virtual social center for the elderly.
    \end{tabular}
                & 
    \begin{tabular}[t]{@{}l@{}}
        \textbullet\ VR\\
        \textbullet\ AR\\
        \textbullet\ Sensors
    \end{tabular}
    & 
    \begin{tabular}[t]{@{}l@{}}
        \textbullet\ The virtual experience\\lacks real-world touch\\sensations for the elderly\\
        \textbullet\ Regulations and law\\enforcement\\
        \textbullet\ Cyberecurity
    \end{tabular}
    & 
    \begin{tabular}[t]{@{}l@{}}
        \textbullet\ Develop potential solutions for the\\identified challenges.\\
        \textbullet\ The optimization of the proposed\\framework will be carried out.
    \end{tabular}
    \\
    \hline
    \begin{tabular}[t]{@{}l@{}}
~\cite{buana2023metaverse}
    \end{tabular}
    & 
    \begin{tabular}[t]{@{}l@{}}
        \textbullet\ Detail the impact of metaverse on\\social life.\\
        \textbullet\ Explore the metaverse's impacts on\\individuals, address its challenges, and\\analyze it sociologically.
    \end{tabular}
                & 
    \begin{tabular}[t]{@{}l@{}}
        \textbullet\ VR\\
        \textbullet\ XR\\
        \textbullet\ MR\\
        \textbullet\ NFTs\\
        \textbullet\ Privacy\\
    \end{tabular}
    & 
    \begin{tabular}[t]{@{}l@{}}
        \textbullet\ Device availability\\for metaverse access\\
        \textbullet\ Security and privacy\\
    \end{tabular}
    & 
    \begin{tabular}[t]{@{}l@{}}
        \textbullet\ Explore emerging technologies\\supporting the metaverse.\\
        \textbullet\ Explore potential solutions\\to the mentioned challenges.
    \end{tabular}
    \\
    \hline
    \begin{tabular}[t]{@{}l@{}}
~\cite{alvarez2022social}
    \end{tabular}
    & 
    \begin{tabular}[t]{@{}l@{}}
        \textbullet\ The study, using social cognitive theory,\\explores the impact of institutional support,\\technological literacy, and self-efficacy on\\engagement in the Facebook metaverse.\\
        \textbullet\  Analyze the influence of institutional\\support, technological literacy, and self-\\efficacy on participants' intent to engage in\\the Feacebook metaverse during COVID-19.
        
    \end{tabular}
                & 
    \begin{tabular}[t]{@{}l@{}}
        \textbullet\ VR\\
        \textbullet\ AR\\
        \textbullet\ XR\\
        \textbullet\ Data Analysis
    \end{tabular}
    & 
    \begin{tabular}[t]{@{}l@{}}
        \textbullet\ Data collection\\
        \textbullet\ Adults aged 18 and\\above are more likely\\to adopt the metaverse\\if receptive to emerging\\technologies
    \end{tabular}
    & 
    \begin{tabular}[t]{@{}l@{}}
        \textbullet\ Understanding participation factors\\is crucial for schools and businesses\\to invest in Facebook metaverse.\\
        \textbullet\ Identify possible solutions.
    \end{tabular}
    \\
    \hline
    \cline{1-4}
\end{tabular}
\end{table*}
\subsection{Industrial Metaverse}
The industrial metaverse, at its core, is a transformative force reshaping various facets of industrial operations\cite{yao2022enhancing}. In the context of product design and engineering, the metaverse comes to life through collaborative virtual environments, offering a space for innovative and cooperative product design processes\cite{jaimini2022imetaversekg}. Hands-on training takes on a new dimension as the industrial metaverse introduces immersive simulations, enabling workers to refine their skills within a virtual realm that is both realistic and risk-free\cite{alpala2022smart}. Manufacturing undergoes a paradigm shift, with the metaverse revolutionizing processes to optimize efficiency and streamline production\cite{zaidan2023uncertainty}. In Factory Production, the integration of virtual models, simulations, and real-time data analytics enhances overall efficiency, offering a dynamic and adaptive approach to contemporary industrial workflows\cite{lv2023digital}. This study in\cite{mourtzis2023blockchain} presented blockchain integration in the industrial metaverse era, acknowledging existing limitations in scalability, flexibility, and cybersecurity for blockchain technologies. It delved into the implications of blockchain for overcoming emerging cybersecurity challenges, specifically in ensuring safe and intelligent manufacturing within Industry 5.0, a subset of the broader concept of Society 5.0. Table~\ref{Tab:Industrial} presents a comprehensive overview of various metaverse applications within the industrial metaverse.
\begin{table*}[htbp]
  \centering
\caption{Leveraging the metaverse for the industrial metaverse.}
\label{Tab:Industrial}
   \begin{tabular}{@{}l|l|l|l|l@{}}
    \cline{1-5}
    \textbf{Existing works} & \textbf{Summary of Contributions} & \textbf{Key Technologies} & \textbf{Challenges} & \textbf{Recommendations and Future Works}\\
    \hline
        \begin{tabular}[t]{@{}l@{}}
~\cite{mourtzis2023blockchain}
    \end{tabular}
    & 
    \begin{tabular}[t]{@{}l@{}}
        \textbullet\ Analyze issues related to the integration of\\blockchain in the industrial metaverse.\\
        \textbullet\ Analyze economic challenges and ethical\\issues for a human-centric in the metaverse.\\
        \textbullet\ Examine how 5G integration transforms\\the industrial metaverse, enhancing\\connectivity and capabilities.
        
    \end{tabular}
                & 
    \begin{tabular}[t]{@{}l@{}}
        \textbullet\ Cybersecurity\\
        \textbullet\ Blockchain\\
        \textbullet\ Storage\\
        \textbullet\ Data Anlaysis
    \end{tabular}
    & 
    \begin{tabular}[t]{@{}l@{}}
        \textbullet\ Data sharing\\
        \textbullet\ Data interoperability\\
        \textbullet\ Data privacy \\
       \textbullet\ Copyright laws\\and regulations\\
       \textbullet\ Regulatory regime\\enforcement mechanisms
    \end{tabular}
    & 
    \begin{tabular}[t]{@{}l@{}}
        \textbullet\ Focus on creating smaller,\\decentralized networks for\\key societal aspects.\\
        \textbullet\ Develop a blockchain of\\blockchains for unified data\\ exchange, enhancing\\decentralization.\\
        \textbullet\ Implement cryptographic\\algorithms to minimize security\\issues in decentralized networks.
    \end{tabular}
    \\
    \hline
    \begin{tabular}[t]{@{}c@{}}
~\cite{alpala2022smart}
    \end{tabular}
    & 
    \begin{tabular}[t]{@{}l@{}}
        \textbullet\ The proposed VR framework facilitates\\modeling and computing of computer-\\generated 3D virtual models.\\
        \textbullet\ Explore the metaverse's potential in an\\industrial context, incorporating multi-\\location teaching-learning with real-world\\sensations, virtual interactions, and real-time\\teaching.
    \end{tabular}
                & 
    \begin{tabular}[t]{@{}l@{}}
        \textbullet\ VR\\
        \textbullet\ 3D modeling\\
        \textbullet\ digital twin
    \end{tabular}
    & 
    \begin{tabular}[t]{@{}l@{}}
        \textbullet\ Resistance\\overcoming\\
        \textbullet\ Technology\\implementation\\
        \textbullet\ UX enhancement
    \end{tabular}
    & 
    \begin{tabular}[t]{@{}l@{}}
        \textbullet\ Conduct practical tests on new\\applications for both the productive\\sector and education.\\
        \textbullet\ Developing VR-based metaverse\\environments for the modular\\design of smart factories.

    \end{tabular}
    \\
    \hline
    \begin{tabular}[t]{@{}l@{}}
~\cite{zaidan2023uncertainty}
     \end{tabular}
    & 
    \begin{tabular}[t]{@{}l@{}}
        \textbullet\ The ICP metaverse is based on a zero\\inconsistency method for calculating weights.\\
        \textbullet\ ICP metaverse smart manufacturing systems\\use an innovative uncertainty decision\\modeling approach based on a decision matrix.
    \end{tabular}
                & 
    \begin{tabular}[t]{@{}l@{}}
        \textbullet\ Digital twin\\
        \textbullet\ Optimization\\
        \textbullet\ 3d modeling
    \end{tabular}
    & 
    \begin{tabular}[t]{@{}l@{}}
        \textbullet\ Experts are not ranked\\based on experience\\
        \textbullet\ Expert decision matrix\\(EDM) created with five\\ linguistic expressions
    \end{tabular}
    & 
    \begin{tabular}[t]{@{}l@{}}
        \textbullet\ Ordering the experts according\\to their experience for more\\reasoned outcomes.\\
        \textbullet\ Exploring varied linguistic\\expressions and numerical scales\\to formulate EDM.\\
        \textbullet\ Using alternative fuzzy sets,\\e.g., spherical hesitant 
         interval\\-valued T-spherical fuzzy sets.
    \end{tabular}
    \\
    \hline
    \begin{tabular}[t]{@{}l@{}}
~\cite{lv2023digital}
    \end{tabular}
    & 
    \begin{tabular}[t]{@{}l@{}}
        \textbullet\ For improving factory production efficiency,\\this study explored the application and DT.\\
        \textbullet\ Explored the architecture and technologies\\of the industrial metaverse, like DT, VR,\\big data, and IoT through a literature review.
    \end{tabular}
                & 
    \begin{tabular}[t]{@{}l@{}}
        \textbullet\ Digital twin\\
        \textbullet\ Blockchain\\
        \textbullet\ IoT\\
        \textbullet\ VR
    \end{tabular}
    & 
    \begin{tabular}[t]{@{}l@{}}
        \textbullet\ DT integration\\
        \textbullet\ DT application \\
        \textbullet\ Visibility Improvement\\
        \textbullet\ Mapping precision\\emphasis
    \end{tabular}
    & 
    \begin{tabular}[t]{@{}l@{}}
        \textbullet\ Optimize production scheduling\\in the industrial metaverse for\\seamless interaction with virtual\\equipment.\\
        \textbullet\ Explore real-world\\implementations for\\integrating into existing\\industrial processes.
    \end{tabular}
    \\
    \hline

    \cline{1-4}
\end{tabular}
\end{table*}
\subsection{Defense and Missions Critical}
The metaverse is an attractive technology for professional military education and aerospace training due to its many applications in defense and mission critical. In addition to providing realistic and immersive training environments, the metaverse enables the simulation of a wide variety of scenarios, including combat, emergencies, and complex missions\cite{training2023}. Developing critical skills and decision-making abilities in a controlled environment can benefit military personnel and other professionals. The immersive and collaborative nature of the metaverse can be leveraged to create realistic training simulations for military and emergency response personnel\cite{hwang2022definition}. These simulations allow individuals to practice and refine their skills in a safe, controlled environment, reducing the risk of injury and equipment damage during live training exercises. In addition, the metaverse can be used for remote operations and surveillance, enabling personnel to remotely control drones, robots, or other devices in hazardous or hard-to-reach locations. Using the metaverse for remote operations can help reduce the risk of personnel exposure to danger and improve situational awareness. In professional military education, the metaverse can be used to train soldiers in a wide range of scenarios, from combat operations to humanitarian missions\cite{ndia2023}. Soldiers can learn and practice complex skills, such as tactical planning, communications, and decision-making, in a realistic and immersive environment. Moreover, in aerospace training, the metaverse can simulate flight environments, including normal and emergency procedures. Pilots can train for various scenarios, such as engine failures, instrument failures, and emergency landings, in a virtual environment that accurately mimics real-world conditions. This can improve their ability to respond quickly and effectively to unexpected events, which can be critical in emergencies. The metaverse can also provide access to virtual simulations of military aircraft and spacecraft, allowing pilots and astronauts to familiarize themselves with the operation and maintenance of complex systems\cite{tas2022bibliometric}. Furthermore, the metaverse can be used for communication and collaboration between different users, platforms, and agencies involved in defense and mission-critical\cite{solly2022unlocking}. This can help improve coordination and information sharing, leading to better decision-making and more efficient use of resources. Table~\ref{Tab:Defense} outlines various applications of the metaverse within the realm of defense and mission-critical operations.

\begin{table*}[htbp]
  \centering
\caption{Leveraging the metaverse for defense and mission-critical purposes.}
\label{Tab:Defense}
    \begin{tabular}{@{}l|l|l|l|l@{}}
    \cline{1-5}
    \textbf{Existing works} & \textbf{Summary of Contributions} & \textbf{Key Technologies} & \textbf{Challenges} & \textbf{Recommendations and Future Works}\\
    \hline
    \begin{tabular}[t]{@{}c@{}}
~\cite{lee2022virtual}
    \end{tabular}
    & 
    \begin{tabular}[t]{@{}l@{}}
        \textbullet\ A metaverse system designed to enhance\\remote education through VR.\\
        \textbullet\ Creating and testing a simulation for\\aircraft maintenance.
    \end{tabular}
                & 
    \begin{tabular}[t]{@{}l@{}}
        \textbullet\ VR\\
        \textbullet\ 3D simulation\\
        \textbullet\ Networking
    \end{tabular}
    & 
    \begin{tabular}[t]{@{}l@{}}
        \textbullet\ Exploring VR's potential in\\enhancing aircraft\\ maintenance education.\\
        \textbullet\ Security of data
    \end{tabular}
    & 
    \begin{tabular}[t]{@{}l@{}}
        \textbullet\ Assessing VR feasibility in technical\\education.\\
        \textbullet\ Exploring the appropriateness of VR\\for technical training across diverse\\engineering disciplines.

    \end{tabular}
    \\
    \hline
    \begin{tabular}[t]{@{}l@{}}
~\cite{gawlik2019experiential}
     \end{tabular}
    & 
    \begin{tabular}[t]{@{}l@{}}
        \textbullet\ Introduce the concept of 3D military\\training.\\
        \textbullet\ Optimizing virtual military training\\assumptions, design approach, and\\proposed training concepts.
    \end{tabular}
                & 
    \begin{tabular}[t]{@{}l@{}}
        \textbullet\ 3D simulation\\
        \textbullet\ VR\\
        \textbullet\ Networking\\
        \textbullet\ Data analysis
    \end{tabular}
    & 
    \begin{tabular}[t]{@{}l@{}}
        \textbullet\ The software may not\\accurately depict\\all military elements\\
        \textbullet\ Some physics experiments\\may be impractical
    \end{tabular}
    & 
    \begin{tabular}[t]{@{}l@{}}
        \textbullet\ Advancing the utilization of VR for\\security and defense applications.\\
        \textbullet\ Exploring advanced training methods\\and the educational potential of the\\virtual world in security and defense.
    \end{tabular}
    \\
    \hline
    \begin{tabular}[t]{@{}l@{}}
~\cite{shen2020study}
    \end{tabular}
    & 
    \begin{tabular}[t]{@{}l@{}}
        \textbullet\ The diverse application of technology in\\military training.\\
        \textbullet\ Advantages of immersive training within\\virtual environments
    \end{tabular}
                & 
    \begin{tabular}[t]{@{}l@{}}
        \textbullet\ VR\\
        \textbullet\ AR\\
        \textbullet\ AI\\
        \textbullet\ Networking\\
        \textbullet\ Data analysis
    \end{tabular}
    & 
    \begin{tabular}[t]{@{}l@{}}
        \textbullet\ Security and privacy\\
        \textbullet\ Presistency and credibility
    \end{tabular}
    & 
    \begin{tabular}[t]{@{}l@{}}
        \textbullet\ Assessing prototypes virtual world's\\feasibility and reliability.\\
        \textbullet\ Examining trust, credibility, and\\user perceptions of security.
    \end{tabular}
    \\
    \hline
    \begin{tabular}[t]{@{}l@{}}
~\cite{jung2022study}
    \end{tabular}
    & 
    \begin{tabular}[t]{@{}l@{}}
        \textbullet\ Immersive simulations and training for\\virtual battles.\\
        \textbullet\ Investigate the application of metaverse\\in military settings, including its concept,\\fundamental types, emerging trends, and\\potential uses.
        
    \end{tabular}
                & 
    \begin{tabular}[t]{@{}l@{}}
        \textbullet\ Communication\\ and networking\\
        \textbullet\ 3D simulation\\
        \textbullet\ VR
    \end{tabular}
    & 
    \begin{tabular}[t]{@{}l@{}}
        \textbullet\ Evaluating credibility via\\subjective assessment.\\
        \textbullet\ Ethical considerations
    \end{tabular}
    & 
    \begin{tabular}[t]{@{}l@{}}
        \textbullet\ Crafting synthetic environments\\for training .\\
        \textbullet\ Integrating cyberspace and battle\\-field via cyber warfare capabilities.\\
        \textbullet\ Integrating the metaverse with AI\\and robotics.
    \end{tabular}
    \\
    \hline
    \cline{1-4}
\end{tabular}
\end{table*}
\subsection{Summary and Lessons Learned}
In this section, we delve into various applications where the metaverse can be employed. These applications span across five distinct domains, including education, healthcare, business and finance, socialization, and defense and critical missions.
\begin{enumerate}
    \item Education in the Metaverse: The metaverse is ushering in an educational revolution, dismantling traditional barriers with immersive learning experiences in virtual classrooms, collaborative projects, and personalized learning paths. This transformative shift offers unparalleled engagement and adaptability, fundamentally reshaping the educational landscape. Within the metaverse, continuous teacher training proves essential for navigating evolving technologies effectively. Prioritizing inclusivity through accessible virtual learning environments emerges as a crucial lesson, ensuring equitable educational opportunities for all. The challenges and recommendations integral to this domain involve addressing the technology integration challenge by establishing comprehensive training programs for educators. Additionally, tackling the digital equity challenge necessitates initiatives to bridge the digital divide and provide universal access. Managing security Concerns requires the development and implementation of stringent protocols, safeguarding student data, and ensuring a secure online learning environment.
    \item Healthcare in the Metaverse: Metaverse applications redefine healthcare, providing unparalleled access to medical expertise, lifelike training simulations, and remote patient monitoring. However, this transformative shift introduces challenges. Striking a delicate balance between innovation and regulatory compliance is paramount. Virtual clinics and telemedicine initiatives bridge gaps, enhancing patient care and medical training. Achieving this delicate balance requires a nuanced approach. Robust data security measures are crucial for building trust and confidence and supporting the responsible integration of metaverse technologies into healthcare. Additionally, navigating the complex landscape of healthcare regulations and ensuring ethical considerations in patient data usage are indispensable aspects.
    \item Business and Finance in the Metaverse: The metaverse revolutionizes corporate landscapes, introducing virtual collaboration spaces and cutting-edge financial modeling tools. Strategic decision-making undergoes a paradigm shift, redefining the dynamics of business and finance in this immersive digital environment. The iterative development process, accompanied by continuous user feedback, proves pivotal in crafting effective business applications within the metaverse. Ensuring the security of financial transactions and safeguarding sensitive corporate data demands rigorous cybersecurity measures. Notable considerations encompass adapting conventional business models to the virtual realm and navigating complex financial regulations within the metaverse. These considerations prompt the need for innovative solutions to facilitate seamless integration. Moreover, optimizing virtual financial processes and addressing potential disparities in financial accessibility pose additional aspects to explore. Suggestions involve substantial investment in cybersecurity infrastructure, collaborative initiatives to establish industry-wide standards, and ongoing efforts to streamline virtual financial procedures and enhance financial inclusivity, all crucial for the continued success of business applications in the metaverse.
    \item Socialization in the Metaverse: The metaverse reshapes social interactions, offering diverse virtual experiences and entertainment options. Users participate in customizable social environments, transcending physical boundaries through global platforms and immersive events. Prioritizing digital well-being and implementing robust community guidelines are paramount. Effective content moderation is essential to counter misinformation and cultivate a positive, inclusive virtual social space. Challenges in metaverse socialization include combating online harassment, safeguarding data privacy, and managing virtual economies. Suggestions involve advancing content moderation tools, enhancing privacy measures, and fostering collaborative initiatives to establish ethical guidelines for virtual interactions.
    \item Industrial Metaverse: The industrial metaverse, with its diverse applications, serves as a powerful tool for optimizing processes across industrial sectors. From collaborative product design to immersive training and efficient manufacturing, the metaverse demonstrates its ability to enhance productivity and innovation. The integration of virtual technologies offers invaluable lessons about the importance of adaptability and collaboration in the face of technological transformations. As industries navigate this new era ushered in by the industrial metaverse, several key lessons emerge. Firstly, the adaptability of workflows and processes is paramount, as the metaverse requires industries to be flexible and responsive to technological advancements. Secondly, collaboration becomes a cornerstone for success, emphasizing the importance of seamless integration and cooperation among diverse facets of industrial operations. Thirdly, a forward-thinking approach is essential, urging industries to embrace innovation and continuously explore ways to leverage the full potential of the metaverse. A dynamic, collaborative, and future-oriented industrial ecosystem is crucial to cultivating the transformative nature of the industrial metaverse.
    \item Defense and Mission-Critical Tasks in the Metaverse: Metaverse applications enhance defense capabilities through realistic simulations, strategic planning tools, and immersive mission execution experiences. Military strategies and preparedness are transformed with virtual war rooms, collaborative scenarios, and advanced training. Maintaining stringent cybersecurity standards is imperative to safeguard sensitive defense information. Continuous ethical scrutiny remains essential to ensure the responsible and accountable use of metaverse technologies in defense applications. Moreover, in the metaverse defense realm, there is a need to address potential vulnerabilities in virtual simulations, ensure secure communication channels, and mitigate the risk of unauthorized access. Strategies for improvement involve investing in cutting-edge cybersecurity measures, conducting regular ethical assessments, and fostering collaborative efforts to establish industry-wide standards for responsible metaverse use in defense applications.
\end{enumerate}
\section{Metaverse Challenges and Future Perspectives}
In this section, drawing inspiration from the aforementioned technological foundations and applications, we present a comprehensive overview of future research directions essential for the practical deployment of the metaverse as a platform. Within each future research direction, we begin by delineating key challenges, followed by an overview of future perspectives, recommendations, and future research directions.
\subsection{Metaverse Infrastructure}
Metaverse development introduces a host of distinctive technical challenges demanding meticulous consideration, especially in anticipation of the upcoming 6G network implementation. A primary challenge revolves around interoperability, a critical concern as diverse platforms employ distinct technologies and protocols, hindering seamless user access to services across the metaverse. To address this, it is strongly advised to establish and embrace metaverse-specific standards and protocols for interoperability. Solutions like Open XR\cite{openxr2023} and MetaOpera\cite{li2023metaopera}, leveraging the advanced capabilities of 6G networks, are recommended to facilitate the smooth transfer of assets and services across diverse platforms. Another pivotal challenge lies in scalability, given the rapid expansion of the metaverse in terms of users, content, and services. This growth has the potential to strain existing infrastructure. Effectively addressing this challenge involves the adoption of decentralized infrastructure and the strategic utilization of 6G networks, coupled with distributed ledger technologies. Solutions such as sharding, sidechains, and off-chain computation, empowered by 6G capabilities, are proposed to significantly enhance performance and reduce latency, meeting the escalating demands of a flourishing metaverse ecosystem. Security surfaces as a major concern due to the increased use of personal and financial data in metaverse applications. It is imperative to leverage the advanced security features of 6G networks. Strong measures, including encryption, multi-factor authentication, and access controls, should be implemented to safeguard user data and assets effectively. Moreover, the development and enforcement of metaverse-specific data privacy and security standards, aligned with the robust capabilities of 6G networks, are deemed essential across platforms and applications. Content creation and curation, predominantly reliant on user-generated content, pose another distinct challenge. To navigate this, the recommendation is to develop and implement metaverse-specific tools and standards for content moderation and curation. Utilizing AI-based content filtering and user reputation systems, aligned with the advanced connectivity and capabilities of 6G networks, is crucial. Creators should be incentivized and supported to produce high-quality and original content, thriving in the enriched connectivity environment of 6G networks. In the expansive realm of the metaverse, the multitude of immersive experiences can pose a challenge, especially for individuals new to virtual environments. To ensure a seamless and engaging experience, such as the user experience (UX), it is imperative to craft interfaces that are not only intuitive but also specifically tailored to the unique dynamics of the metaverse. Providing easily understandable information, along with tutorials designed exclusively for those new to the metaverse, becomes paramount. Continuous research and testing, focusing on the distinct needs of UX in the metaverse, play a vital role in refining the overall experience. Furthermore, leveraging the advanced connectivity and interactive capabilities offered by 6G networks is fundamental. By integrating these enhanced network features, there is a continuous opportunity to enhance interactions and experiences within the evolving landscape of the metaverse. This strategic approach addresses metaverse infrastructure challenges by offering recommendations fine-tuned to the distinctive characteristics of this virtual domain.

\subsection{Security, Privacy, and Trust}
The metaverse, an evolving virtual realm fostering user interactions within simulated environments, holds immense promise in technology and entertainment. However, navigating its complexities entails addressing specific security, privacy, and trust challenges integral to its success. Among the foremost privacy concerns in the metaverse is the safeguarding of personal information, given the substantial data sharing inherent in these platforms. Striking a balance between anonymity and identification poses another challenge, as anonymity, while preserving privacy, can also be exploited for illicit activities. Cybersecurity threats, including hacking and phishing, pose significant risks, potentially resulting in the compromise of personal information and asset theft. On the trust front, verifying user identity proves challenging, with the potential for fake identities and anonymity hindering accurate authentication. Governance gaps pose risks, leading to disputes and eroding user trust. Brand trust is equally critical, with negative incidents capable of driving users away.

To mitigate these challenges, several recommendations are proposed. First, user education initiatives should guide individuals in safeguarding personal information and assets. Second, robust authentication methods must be enforced to limit access to authorized users. Third, the option for users to toggle between anonymity and identification, contextually relevant, aids in preventing illicit activities while preserving privacy. Encrypting data at rest and in transit is a fundamental fourth recommendation to thwart unauthorized access. Regular security audits and vulnerability assessments, as the fifth measure, proactively identify and address potential threats. Sixth, integrating privacy and security considerations into the design and development phases of metaverse platforms ensures a robust foundation. Lastly, collaborative efforts among platforms, security experts, and regulatory agencies can establish standardized metaverse privacy, security, and governance norms, fostering user trust and consistently safeguarding users across platforms.

\subsection{Artificial Intelligence}
The integration of AI within the metaverse presents a landscape rich with opportunities and complexities. Among the significant challenges are data bias, data privacy concerns, and ethical considerations. Data bias is a pivotal concern as biased training data can lead AI algorithms in the metaverse to perpetuate and amplify existing biases, potentially impacting virtual commerce and social interactions. As AI becomes more integral to the metaverse, apprehensions about data privacy are escalating due to the potential misuse or unauthorized access to user data, risking a decline in trust and popularity. Moreover, the advancement of AI in the metaverse necessitates a careful examination of ethical implications. For instance, AI-powered virtual assistants have the potential to influence user opinions and behaviors subtly. Effectively addressing these challenges requires a strategic integration of both Generative AI and conventional AI tailored to the unique demands of the metaverse. Generative AI technologies, including GANs and LLM, offer the ability to create realistic and dynamic virtual content. When applied to the metaverse, Generative AI can counteract biases by leveraging diverse training data, ensuring that the generated content reflects a broad spectrum of perspectives. This minimizes the risk of perpetuating biases in virtual interactions, commerce, and social experiences. Furthermore, Generative AI enhances user engagement and customization, contributing to a more immersive and tailored metaverse experience guided by ethical principles. Moreover, in conjunction with Generative AI, conventional AI algorithms play a crucial role in addressing data privacy and security concerns. Through the implementation of encryption techniques and user-controlled data-sharing settings, conventional AI enhances the protection of user data, mitigating the risk of unauthorized access or misuse. The strengths of conventional AI in structured decision-making and rule-based systems complement the creativity of Generative AI, establishing a comprehensive approach to address the multifaceted challenges in the metaverse. Establishing clear ethical guidelines and standards for AI development, with a specific focus on both Generative AI and conventional AI, ensures that these technologies prioritize user safety and well-being. Regular audits and reviews, encompassing both types of AI systems, have become essential for ongoing adherence to ethical guidelines and the continuous improvement of AI applications in the evolving landscape of the metaverse. Through this collaborative and tailored approach, AI in the metaverse can navigate challenges and unlock its full potential for user-centric and responsible virtual experiences.

\subsection{Metaverse Sustainability}
The rapid development and extensive use of the metaverse introduce substantial sustainability challenges, particularly concerning energy consumption, e-waste generation, carbon footprint, supply chain sustainability, and overall environmental impact. The metaverse's considerable energy consumption raises apprehensions about its long-term sustainability and environmental consequences. The discarded hardware and devices, contributing to e-waste, present challenges in proper disposal and recycling. Additionally, the metaverse's significant carbon footprint adds to climate change concerns, and its intricate supply chain necessitates ethical and sustainable practices to address waste, pollution, and resource depletion. Identifying potential sustainability issues through comprehensive environmental impact assessments becomes crucial for mitigating the metaverse's environmental impact. To tackle these sustainability challenges, metaverse platforms can implement several strategic recommendations. The adoption of energy-efficient technologies and the optimization of data centers can substantially reduce energy consumption. Exploring renewable energy sources like solar, wind, and hydropower is essential for promoting the long-term sustainability of metaverse operations. Developing robust recycling programs, encouraging proper disposal and recycling of e-waste, and designing products with recyclability or reusability in mind are crucial steps to mitigate the environmental impact of discarded hardware and devices. Addressing the metaverse's carbon footprint can involve options such as carbon offsetting through activities like tree planting and investments in renewable energy projects. Ensuring ethical and sustainable practices throughout the supply chain, including fair labor practices, waste reduction, and the use of sustainable materials, is paramount for promoting overall supply chain sustainability. Finally, conducting regular environmental impact assessments helps identify potential sustainability concerns and implement effective strategies to mitigate the metaverse's environmental impact, contributing to a more sustainable future.

\subsection{Metaverse Devices Compatibility}
The intricate landscape of devices within the metaverse presents unique challenges that require precise solutions to ensure seamless integration and enhancement of UX. The incorporation of VR headsets, AR glasses, smartphones, and traditional computing devices introduces complexities associated with hardware disparities and computational intricacies. Heterogeneity in hardware, ranging from high-performance VR systems to ubiquitous smartphones, demands nuanced solutions. Challenges encompass balancing computational requirements across devices, optimizing for diverse processing power and graphics capabilities, and addressing variations in input methods. Ensuring a unified UX entails grappling with network constraints, acknowledging divergent levels of connectivity, and bandwidth considerations among devices prevalent in the metaverse. 

Prospective directions for the metaverse's evolution hinge on collaborative endeavors and innovative solutions meticulously tailored to its distinctive challenges. In the domain of device compatibility, the envisioned pathways involve the advancement of standardization initiatives across hardware specifications and input methods. This necessitates the refinement of development processes to augment interoperability. The integration of adaptive technologies, capable of dynamically adjusting to the computational resources of each device, emerges as a pivotal strategy for sustaining visual fidelity and functional coherence across the diverse platforms intrinsic to the metaverse. Innovations in interface design assume a critical role, envisaging experiences that transcend mere accessibility by conforming to the unique form factors of devices within the metaverse. Noteworthy innovations may encompass the integration of AI to personalize interactions based on the individualized capabilities of each device and user preferences, thereby elevating the overarching UX. Continued user education remains imperative. Empowering users to comprehend the intricacies of device compatibility and promoting best practices constitutes an integral facet for facilitating seamless transitions across varied digital environments within the dynamic metaverse. As the metaverse community embraces standardization, innovation, and ongoing educational initiatives, collective efforts can surmount these challenges, thereby paving the way toward a more accessible, intuitive, and seamlessly integrated digital future within the distinct domain of the metaverse.

\subsection{Metaverse Ethics}
As technology advances and the metaverse becomes more pervasive, it brings forth a set of ethical challenges that demand careful consideration. The metaverse, blurring the boundaries between physical and digital realities, raises key ethical concerns, starting with data privacy. Users' creation and interaction with virtual content generate a wealth of personal data that metaverse platforms must handle responsibly, prioritizing data privacy and implementing measures to protect users' personal information. Security is another critical ethical issue, as the integration of the metaverse into daily life makes it a potential target for cybercriminals. Robust security measures are imperative to safeguard personal information and ensure the platform's uninterrupted operations. Virtual property rights present an additional ethical concern, emphasizing the need for clear guidelines regarding ownership and control over virtual assets. metaverse platforms must establish rules to prevent the exploitation or deprivation of users' virtual property. Concerns related to addiction and social isolation emerge due to the immersive nature of the metaverse. Striking a balance to prevent excessive usage and detachment from real-life interactions is crucial, promoting healthy platform use and preventing addiction. Ethical governance is fundamental to ensuring fairness and equity for all users. Establishing clear content moderation guidelines to prevent illegal or harmful activities and governing the platform ethically contribute to a fair and equitable metaverse for all users. This adaptive approach addresses the unique ethical challenges posed by the evolving landscape of the metaverse.

\subsection{Metaverse Economy}
The emergence and proliferation of the metaverse present unique economic challenges that demand targeted solutions. A paramount concern revolves around virtual asset ownership, where the absence of a clear legal framework leads to disputes and difficulties in enforcing property rights. Establishing a robust legal framework that formally recognizes virtual assets as legitimate property and delineates clear rules for ownership and transfer becomes imperative in addressing this challenge. Uncertainties surround the creation and distribution of value within the metaverse, particularly concerning the valuation of virtual goods and services. To tackle this, there is a need to institute a standardized system that measures the value of virtual assets based on supply and demand dynamics, providing a foundation for a more predictable and stable virtual economy. Monetization stands out as a significant hurdle, with creators and developers struggling to earn revenue from their metaverse content. Enhancing revenue streams can be achieved by introducing more flexible revenue-sharing models and expanding creators' access to a larger user base, thereby providing them with greater incentives to contribute to the metaverse content ecosystem. The exchange of virtual currencies encounters challenges due to the absence of a centralized exchange or regulatory framework. This limitation impacts the liquidity of the virtual economy and exposes users to risks. To alleviate these concerns, the establishment of a centralized virtual currency exchange, accompanied by a robust regulatory framework, is recommended. Such measures ensure improved liquidity, transparency, security, and accountability in virtual currency exchange transactions, fostering a safer and more vibrant economic landscape within the metaverse.
\section{Conclusion} 
This survey undertakes a thorough investigation of the fundamental concepts, prerequisites, and technological underpinnings critical to the development of the metaverse. Initiating with an analysis of the innovative metaverse architecture, we explore its essential characteristics, including immersiveness, spatiotemporal coherence, scalability, heterogeneity, and QoE. The evaluation of metaverse requirements encompasses aspects such as energy efficiency, sustainability, interoperability, decentralization, and security and privacy considerations. Additionally, this survey scrutinizes pivotal technologies within the metaverse, such as interactive experiences, communication and networking, ubiquitous computing, DT, AI, and cybersecurity, elucidating the complex network of elements essential for its seamless operation. Each critical technology is contextualized through an investigation of extant contributions, enabling techniques, and illustrative use cases. Discussion of standards within the metaverse is crucial for promoting a robust, secure, and intuitive virtual environment that can adapt to emerging technologies and evolving user demands. Standardization initiatives typically require collaborative efforts among industry stakeholders, organizations, and regulatory bodies to establish universally accepted norms within the metaverse ecosystem. Furthermore, by delving into the applications of the metaverse across various domains such as education, healthcare, business, finance, socialization, industrial sectors, defense, and mission-critical operations, we elucidate the motivations and challenges pertinent to each area. The survey not only outlines the current landscape but also advances the discourse by offering insightful suggestions for future research and practical recommendations that can be effectively implemented through metaverse applications. Additionally, a comprehensive synthesis of challenges and recommendations in the metaverse domain is presented, addressing issues from infrastructure and cybersecurity to ethical considerations and economic impacts. The necessity for stakeholders to collaborate and coordinate efforts becomes apparent as the metaverse progresses, ensuring interoperability, inclusivity, and the protection of privacy, security, and ethical standards. In anticipation of the metaverse's continuous evolution, this survey serves as an informative guide, poised to elucidate the complex requirements, architectures, challenges, standards, and potential solutions in metaverse applications. Acting as a catalyst for pioneering research in this nascent field, the survey not only captures the present state but also charts a path for the dynamic future of the metaverse, inspiring collaborative endeavors and advancements in pursuit of a responsible and innovative digital frontier.
\section*{Acknowledgements}
\thanks{This work is supported in part by the Microsoft Corp. research gift funds and in part by the DoD Center of Excellence in AI and Machine Learning (CoE-AIML) at Howard University under Contract W911NF-20-2-0277 with the U.S. Army Research Laboratory and Meta/Facebook Research Gift Funds.} 
\balance
\bibliographystyle{IEEEtran}

\end{document}